\documentclass{aa}  

\usepackage{graphicx}
\usepackage{siunitx}
\usepackage{gensymb}
\usepackage{placeins}
\usepackage{multirow}
\usepackage{tablefootnote}
\usepackage{array} % Required for centering columns

\usepackage{soul}
\usepackage{txfonts}
\usepackage{hyperref}

\hypersetup{
    colorlinks=true,
    linkcolor=blue,
    filecolor=blue,      
    urlcolor=blue,
    citecolor=blue
}

\begin{document} 
\defcitealias{hammer}{H18}
\defcitealias{dey}{D23}

   \title{The survey of planetary nebulae in Andromeda (M31) VII.  Predictions of a major merger simulation model compared with chemodynamical data of the disc and inner halo substructures}

   \titlerunning{ }
   
   \author{C. Tsakonas
          \inst{1}
          \and
          M. Arnaboldi \inst{2}
          \and
         S. Bhattacharya   \inst{3, 4}
          \and
          F. Hammer\inst{5}
          \and
          Y. Yang \inst{5}
          \and
          O. Gerhard \inst{6}
          \and
          R. F. G. Wyse \inst{7}
          \and
          D. Hatzidimitriou\inst{1}
          }

   \institute{Department of Astrophysics, Astronomy \& Mechanics, Faculty of Physics, National and Kapodistrian University of Athens, Panepistimiopolis Zografou 15784, Greece\\
              \email{haristsak@phys.uoa.gr}
            \and
             European Southern Observatory, Karl-Schwarzschild-Str. 2, 85748 Garching, Germany            
            \and
            Inter University Centre for Astronomy and Astrophysics, Ganeshkhind, Post Bag 4, Pune 411007, India
             \and
             Centre for Astrophysics Research, Department of Physics, Astronomy and Mathematics, University of Hertfordshire, Hatfield, AL10 9AB, UK
             \and
            GEPI, Observatoire de Paris, PSL Research University, CNRS, Place Jules Janssen, F-92190 Meudon, France
            \and
            Max-Planck-Institut für extraterrestrische Physik, Giessenbachstraße, 85748 Garching, Germany
            \and
            Department of Physics \& Astronomy, The Johns Hopkins University, Baltimore, MD 21218, USA
            }

   \date{Received: 26 November 2024 / Accepted: 16 April 2025}
% \abstract{}{}{}{}{} 
% 5 {} token are mandatory
 
  \abstract
  % context 
   {The nearest giant spiral, the Andromeda galaxy (M31), exhibits a kinematically hot stellar disc, a global star formation episode $\sim$~2~$-$~4~Gyr ago, and conspicuous substructures in its stellar halo that are suggestive of a recent accretion event.}
  % aims 
   {Recent chemodynamical measurements in the M31 disc and inner halo can be used as additional constraints for N-body hydrodynamical simulations that successfully reproduce the disc age-velocity dispersion relation and star formation history as well as the morphology of the inner halo substructures.}
  % methods
   {We combined an available N-body hydrodynamical simulation of a major merger (mass ratio 1 $:$ 4) with a well-motivated chemical model to predict abundance distributions and gradients in the merger remnant at z~$=$~0. We computed the projected phase space and the [M/H] distributions for the substructures in the M31 inner halo, namely, the Giant Stellar Stream (GSS) and the North-East (NE) and Western (W) shelves. We compared the chemodynamical properties of the simulated M31 remnant with recent measurements for the M31 stars in the inner halo substructures. }
  % results 
   {This major merger model predicts (i) multiple distinct components within each of the substructures; (ii) a high mean metallicity and large spread in the GSS and NE and W Shelves, which explain various photometric and spectroscopic metallicity measurements; (iii) simulated phase space diagrams that qualitatively reproduce various features identified in the projected phase space of the substructures in published data from the Dark Energy Spectroscopic Instrument (DESI); (iv) a large distance spread in the GSS, as suggested by previous tip of the red giant branch measurements; and (v) phase space ridges caused by several wraps of the secondary as well as up-scattered main M31 disc stars that also have plausible counterparts in the observed phase spaces. }
  % conclusions 
   {These results provide further strong and independent arguments for a major satellite merger in M31 $\sim$3~Gyr ago and a coherent explanation for many of the observational results that make M31 appear so different from the Milky Way.}


   \maketitle
%
%-------------------------------------------------------------------

\section{Introduction}
As the closest large galaxy, the Andromeda galaxy (M31) was the object of many early groundbreaking studies in extragalactic astronomy (e.g. \citealt{Hubble1929, Rubin1970ApJ}). Due to its large total mass ($\sim$~1.5~$\times$~10$^{12}$~M$_{\odot}$; see \citealt{Bhattacharya2023arXiv230503293B} and references therein) and proximity -773~kpc from the Milky Way, MW; \citealt{conn}- it is an ideal laboratory for galactic archaeology studies through wide-field photometric identification (e.g. \citealt{ibata2001, ferguson2002, Dalcanton2012, williams17}) and resolved spectroscopy (e.g. \citealt{Guhathakurta06, escala2020, escala22, dey}) of red giant branch (RGB) stars.

The most comprehensive imaging survey of M31 is the Pan-Andromeda Archaeological Survey (PAndAS; \citealt{mcconnachie}). It covers a $\sim$400~sq.~degree area and targets the RGB stars within the virial radius of M31. PAndAS revealed an intricate pattern of stellar substructures that populate the inner halo of M31 and whose origin is readily associated with accretion event(s). 

The most striking among the substructures is the Giant Stellar Stream (GSS), a stellar structure extending to more than $\sim$80 kpc \citep{ibata2001, ibata2004, McConnachie03, mcconnachie, Cohen2018} in projected distance from the centre of M31. It is connected to the second brightest substructure in the M31 inner halo, the North-East (NE) shelf, via stellar streams \citep{Merrett2003}. The GSS is the most prominent stellar stream in the Local Group and a clear sign of galactic interaction. It has thus become apparent that, in stark contrast with the MW, which had a quiet accretion history (e.g. \citealt{Wyse, Pillepich2014M, Helmi2020}) over the last $\sim$10 Gyr, M31 seems to have undergone significant accretion events in its recent past \citep{mcconnachie09nature}. 

Following the discovery of the GSS, several simulation attempts were carried out to reproduce its morphology as the result of a minor merger event (i.e. when the mass ratio between the two progenitor galaxies is less than 1~$:$~10, see \citealt{ibata2004, Geehan, font2006, fardal2006, fardal2007, fardal2008, Fardal12, fardal2013, Mori2008, Sadoun2014,Kirihara2017, milosevic2022, milosevic2024}). While differences exist among the various minor merger models, they characteristically consider that a satellite with a total mass of $\sim$10$^{9}$~M$_\odot$ plunged into the gravitational potential of M31 following a radial orbit. Many of these models have been successful at reproducing the morphology of the primary tidal features in the inner halo of the galaxy; namely, the GSS, the NE shelf \citep{ferguson2002, Ibata2005}, and the Western (W) shelf  \citep{fardal2007}. In these models, the GSS appears as a relatively short-lived tidal stream, with the accretion event occurring $\sim$1~Gyr ago. Additionally, they produce wedge-like features in the diagram of the projected distance versus line-of-sight (LOS) velocity (hereafter  R$\rm_{proj}$ versus V$\rm_{LOS}$ diagram) for the satellite stars in these substructures \citep{fardal2007,Kirihara2017, milosevic2024}. Such features in the projected phase space are typical of a radial merger event \citep{Merrifield, Hendel2015MNRAS.454.2472H}. Indeed, coherent kinematic structures (streams, chevrons, wedges\footnote{Wedges and chevrons refer to triangular-shaped patterns in the R$\rm_{proj}$ versus  V$\rm_{LOS}$ diagram of stars in the substructures. Wedges refer to filled triangles and chevrons to empty triangles.}) were later observed in the  R$\rm_{proj}$ versus V$\rm_{LOS}$ diagram of stars in the NE shelf \citep{escala22} and the rest of the inner halo substructures \citep[][hereafter \citetalias{dey}]{dey}, providing additional supporting evidence for a recent merger event. However, such minor merger events typically do not influence the kinematics of the disc of the more massive galaxy \citep{Martig2014}. 

The disc of M31 displays interesting properties that are difficult to explain under the minor merger assumption. \citet{dorman} found that the age-velocity dispersion relation for the stars in the disc of M31 is considerably steeper than that reported (e.g. \citealt{nordstrom2004}) for the MW's disc. The average LOS velocity dispersion at radii larger than R $=$ 10~kpc is an increasing function of stellar age, rising from 30~kms$^{-1}$ for the main sequence stars up to 90~kms$^{-1}$ for $\sim$4~Gyr old RGB stars. A subsequent study from \citet{papaerII} using planetary nebulae (PNe\footnote{PNe are emission line nebulae in a short-lived late stage of stellar evolution for stars with initial masses $\sim$0.8$-$8~M$\odot$ (see the review by \citealt{Kwitter2022PASP..134b2001K} and references therein). The PNe thus span a wide range of parent stellar population ages ($\sim$100 Myr to $\sim$10~Gyr).}) surveyed in M31 \citep{paperI} found 
an increase in the rotational velocity dispersion within the old stellar population, from $\sigma\rm_{\phi} \sim$~61~kms$^{-1}$ for PN progenitor stars with ages less than 2.5~Gyr to 
$\rm \sigma_{\rm\phi} \rm\sim$~101~kms$^{-1}$ for PN progenitors older than 4.5~Gyr, which are associated with the thin disc and the thicker disc, respectively.  The velocity lag displayed by the two populations independently supports the age ordering  \citep{papaerII}. The large amount of kinetic energy input needed to heat the disc to such dispersion requires a merger with a massive companion (see \citealt{hopkins09}).

\citet{paperiv} measured PNe oxygen and argon abundances as a function of the disc radius (out to 30~kpc). Interpreting the PN abundances with tailored galactic chemical evolution models in the oxygen-to-argon ratio versus argon abundance plane, \citet{arna22} showed that the M31 disc experienced an infall of metal-poor gas $\sim$~2~$-$~4~Gyr ago that diluted the interstellar medium (ISM), out of which the subsequently younger stars were formed (see also \citealt{Kobayashi2023} for galactic chemical enrichment models pertaining to this event). Thus, the last significant accretion event in M31 was a star-forming merger in which gas from the satellite and/or the outer M31 gas disc was brought inwards. These results align with the widespread burst of star formation $\sim$2~Gyr ago identified with deep Hubble Space Telescope (HST) observations of the resolved stellar population in the M31 disc \citep{williams17} by the Panchromatic Hubble Andromeda Treasury (PHAT) survey \citep{Dalcanton2012} and in pencil beam observations of the outer regions of the disc of M31 \citep{bernard2012, Bernard15}.

A kinematically hot thick disc such as that in M31 can be produced in a wet major merger event \citep{Brook2004, hopkins09}. The hot thick M31 disc and the burst of star formation $\sim$ 2~$-$~3 Gyr ago together with the intricate morphological characteristics of M31's inner halo were reproduced within the context of a single recent wet galaxy merger event by \citet[][hereafter \citetalias{hammer}]{hammer}. They presented N-body hydrodynamical simulations of a major satellite merger with a mass ratio of 1 $:$ 4 (hereafter major merger for brevity) with significant gas content. The merger resulted from the collision between two disc galaxies, and the coalescence of the nuclei occurred in the time interval 1.8~$-$~3~Gyr ago. The parameters of the simulations were optimised to reproduce M31's inner disc and central (bulge, bar) properties as well as the substructures close to the disc and within the inner halo (see Section~\ref{section general concepts} for further details). The \citetalias{hammer} major merger simulations also predict that the G1-clump substructure observed at the southern edge of M31's disc \citep{ferguson2002} was formed from a substantial perturbation of the outer M31 disc. The kinematics of the PNe co-spatial with the G1-clump were later found to be consistent with that predicted in the major merger scenario \citep{paperVI}. Such a substructure formed out of perturbed host disc material is yet to be observed in minor merger simulations. 

While some of the morphological and kinematic features of the R$\rm_{proj}$ versus V$\rm_{LOS}$ diagram of the GSS and NE and W shelves can be reproduced by minor merger scenarios (e.g. \citealt{fardal2006, font2006}), they fail to reproduce the two kinematically cold components (KCCs) observed in several inner fields of the GSS \citep{Kalirai2006,Gilbert2007, gilbert09} as well as the spread in LOS distance within the stream \citep{conn}. \citetalias{hammer} simulations predict a broad LOS velocity distribution and several kinematically distinct components in the GSS region \citep{paperVI} that are associated with debris generated during the pericentre passages of the satellite (see Section~\ref{ssec:GS} for further details).

Recent observations of RGB stars in the inner halo of M31, including the GSS and NE and W shelves, acquired from the Dark Energy Spectroscopic Instrument (DESI) multi-object spectrograph \citepalias{dey} provide for the first time stellar kinematics and metallicity measurements with a uniform wide-field coverage over tens of square degrees. Combined with metallicity estimates from recent multi-object spectroscopic observations from pencil beam surveys in various regions \citep{Escala2019, escala2020,escala21,escala22,Gilbert2019, Gilbert2020} and additional photometric surveys in the inner halo  \citep{Bernard15, tanaka2010, conn, ogami}, these datasets provide additional chemodynamical constraints against which the major merger simulations of \citetalias{hammer} can be tested. Our investigation aims to extend the comparison to the simulated phase space metallicity distributions of the inner halo substructures with the recent measurements in M31.
%within the context of a major merger for the evolution of M31.}

In the present work, we build on a viable major merger model from \citetalias{hammer} and extend the previous analysis. As there are new observations of the age and abundance distributions for the stellar tracers in the M31 disc and substructures, the major merger model can be tested further. We do so by setting a well-motivated chemical distribution for the two progenitor galaxies at the beginning of the simulation.  We make use of the recent PNe sample measurements in M31's disc from \citet{paperiv} and observational constraints from the stellar mass-metallicity relation (hereafter MZR; see \citealt{maiolino} and references therein) for disc galaxies of similar mass as M31 and observed at redshifts comparable with that of the simulated merger. 
We then follow the disc particle forwards through the merger and compare the chemodynamical properties of the remnant disc and the three prominent inner halo substructures (GSS and NE and W shelves) of M31's simulated analogue with extensive observational data. 

The paper is organised as follows. In Section~\ref{section 2}, we describe the main properties of the simulation selected from \citetalias{hammer}. Section~\ref{section:metallicity in the model} presents the steps undertaken to constrain the initial metallicity distribution in the two progenitor galaxies. The stellar substructures in the inner halo of the simulated galaxy have multiple components that we analyze separately in Section~\ref{section: the multicomponent nature of substructures in the model}. In Section~\ref{chemody}, we compare the  R$\rm_{proj}$ versus V$\rm_{LOS}$ diagram of the GSS and NE and W shelves of our simulated merger remnant with the recent observational results for M31. In Section~\ref{section: metallicity comparison for everything}, we compare the predicted metallicity values with photometric and spectroscopic metallicity measurements in the inner halo substructures. We state our main conclusions in Section~\ref{chpter: conclusions}.

\begin{figure*}[h]
    \centering
    \includegraphics[width=\textwidth]{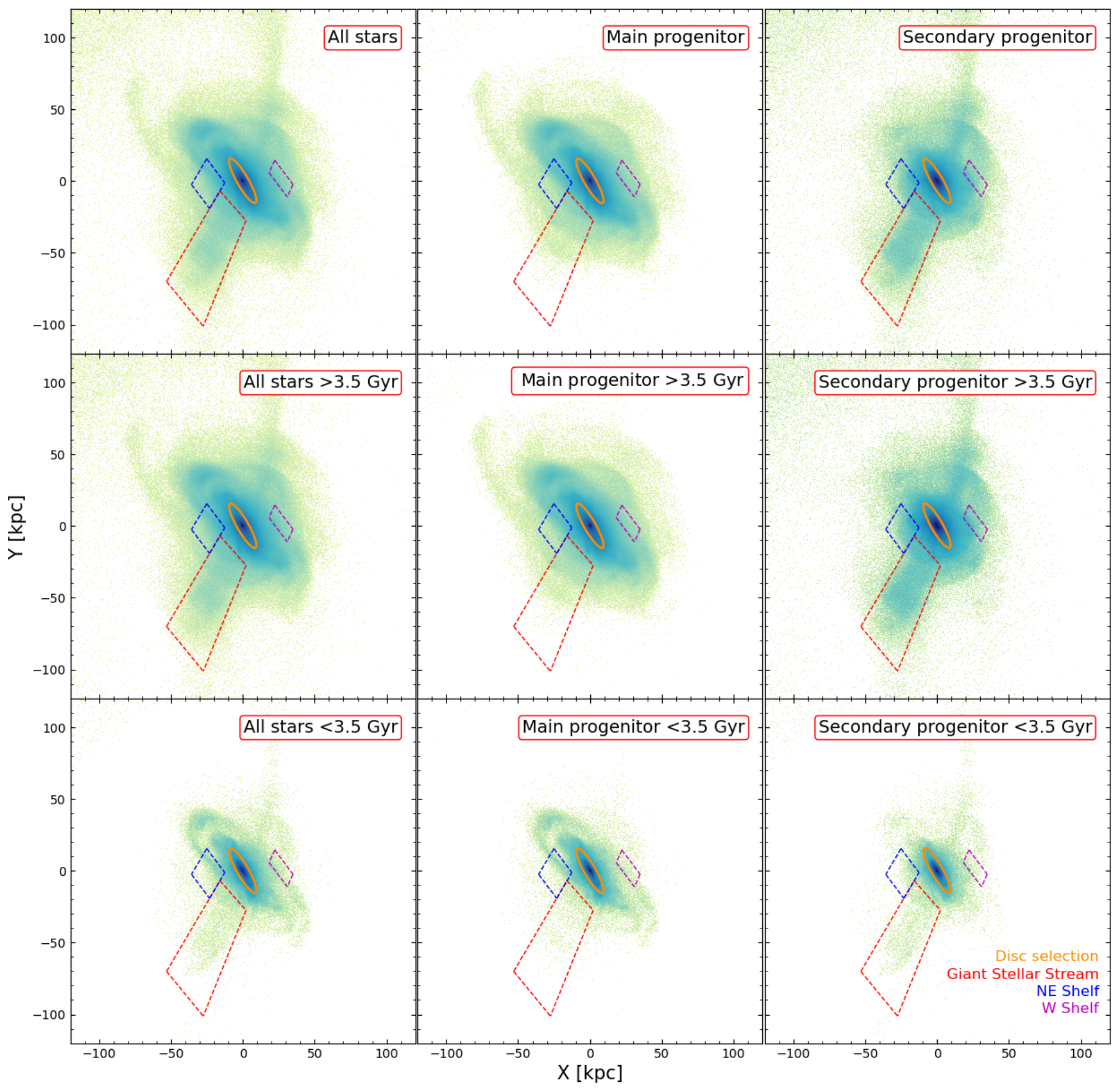}
    \caption{Spatial density map of the star particles at the end of the merger simulation. The galaxy is rotated and projected onto the sky plane according to the inclination and PA of M31. The selected areas for the M31 disc, the GSS, the NE-, and the W-Shelves are indicated with different colours. Upper row: All simulated stellar particles [left], main progenitor stellar particles [middle], and secondary progenitor stellar particles [right] are shown in independent panels. Middle row: Same differentiation, but only for stars older than 3.5 Gyr. Lower row: Same differentiation, but only for stars younger than 3.5 Gyr.}
    \label{fig:modelxy}
\end{figure*}

\section{A major merger simulation for the evolution of M31} \label{section 2}

On the basis of the results obtained from numerical simulations that showed thin discs to be rebuilt after gas-rich galaxy mergers \citep{hammer05, hammer09, hopkins09}, \citetalias{hammer} utilised $\sim$300 high-resolution simulations to investigate and reproduce the morphological properties of M31. These simulations consisted of two bulgeless disc galaxies, both with substantial gas discs, in addition to the stellar discs. \citetalias{hammer} have let free most of the parameters within a large range (see their Table~1), such as the mass ratio from 1~$:$~5 to 1~$:$~2. However, \citetalias{hammer} did not consider minor merger cases (including mass ratio significantly larger than 1~$:$~10), conversely to the works of \cite{fardal2013} and \cite{Kirihara2017} for instance, where the mass of the secondary galaxy is assumed to be $\sim$5~$\times$~10$^{9}$~M$_\odot$. The rationale for the adopted strategy is illustrated in \citetalias{hammer} and in the introduction of the current work. To recap, only a major merger can reproduce the observed properties of M31's disc: a strong star forming episode 2-3.5 Gyr ago (see \citealt{bernard2012, Bernard15, williams17}), the steep age-velocity dispersion relation  \citep{dorman}, and the 10~kpc ring \citep{Lewis2015ApJ...805..183L}. 
The mass merger ratio (constrained from 1~$:$~4 to 1~$:$~5) is independently found in the investigation carried out by \cite{papaerII}, who used the relation between the disc scale height (H$\rm_z$) and the satellite-to-disc-mass ratio
(M$\rm_{sat}$/M$\rm_{disc}$) described by \cite{Hopkins08}, to explain the dynamically hot 4.5~Gyr-old population in M31's disc (with an H$\rm_{z~4.5 Gyr} \simeq$~0.86~kpc; \citealt{Dalcanton15}).

In addition to the observed properties of the M31 disc, there are compelling observational constraints on the timing of the merger event in M31 and its first pericentre passage occurring $\sim$ 7$-$8 Gyr ago, coming from the clustering of the quenching time of dwarf spheroidal galaxies  (10$^{5}$ M$_\odot$ < M$_{*}$ < 10$^{7}$ M$_\odot$)  in M31; see \citet{Souza2021,savino2025}.

We choose model \#336, one of the best M31 analogues and the highest resolution simulation of \citetalias{hammer}, with 20  million particles, as it did the best job of matching the observations at the time it was published.  The mass resolution is 13750 M$_{\odot}$ and 55000 M$_{\odot}$ per baryon and dark matter particle respectively. By utilizing the major merger model, we aim to make predictions for the chemodynamical phase space structure of the resulting M31 analogue, with particular emphasis on the prominent substructures in the inner stellar halo, including the GSS (Section~\ref{ssec:GS}).

The simulation is deployed for a total time length of 7.5~Gyrs. Each progenitor is initially modelled by two components, a dark matter halo (comprising 80\% of the total mass) and a thin disc that includes both stars and gas. The scale length of the gas discs is three times that of the stellar discs and the scale height of all discs (stellar and gaseous) is equal to 1$/$10 of the scale length of the stellar discs (i.e. the gas/stellar discs of each progenitor are assumed to be thin and have the same scale height). Both discs have flat rotation curves, which reach V$\rm_{max}$ $\simeq$~220 kms$^{-1}$ for the main and V$\rm_{max}$ $\simeq$~140 kms$^{-1}$ for the secondary disc.

\subsection{General properties and time evolution of the M31 simulations} \label{section general concepts}

\citetalias{hammer} used the N-body hydrodynamical code GIZMO to simulate the collision between the two bulgeless, thin discs with a mass ratio of 1~$:$~4. The primary features of the dynamical evolution of the merger are as follows. The first pericentric passage occurs at a lookback time of $\sim$7~Gyr. Due to their high relative velocity, the two discs reside far from each other for the next $\sim$3.5~Gyr, until the second pericentric passage occurs. After the second pericentric passage, the two galaxies do not drift far apart from each other. The third pericentric passage occurs $\sim$1~Gyr later, followed by a full-fledged merging process. Multiple subsequent pericentric passages at short time intervals follow and the satellite progenitor is completely disrupted. The coalescence of the nuclei takes place 1.8 $-$ 3~Gyr ago, and from a lookback time of  $\sim$2~Gyr, the two galaxies are fully merged into a single galaxy.

Following the second pericentric passage $\sim$3.5~Gyr ago, the subsequent pericentric passages of the secondary release tidally distorted material (stellar debris from both progenitors) in the inner halo of the remnant. This tidal debris shows up as overdensities along the plane of the infalling satellite's orbit, resulting in overlapping loops that are destined to merge with M31. The simulation predicts that the superposition of these loops along the LOS and their projections on the sky produce the observed properties of the inner halo substructures in velocity and position.

During the collision, the ensuing gravitational torques within the gas brought by the satellite and the gas from the main progenitor resulted in the ignition of star formation in the main progenitor disc $\sim$~2~$-$~3~Gyr ago, as was observed by \citet{williams17} and \citet{bernard2012, Bernard15}. During the final stages of the merger, many short pericentric passages of the secondary heat the main progenitor's pre-existing stellar distribution, resulting in the formation of the kinematically hot thick disc in the remnant. The thick disc is mostly comprised of stars older than the collision event (ages $>$4~Gyr). A residual core from the secondary progenitor is absent in the remnant disc, as the central parts of the secondary progenitor completely dissolve during the numerous short pericentric passages, and its - initially less bound - stellar outskirts are deposited into inner halo substructures (GSS and NE and W shelves). 

Overall, the major (1 $:$ 4) mass merger model for the evolution of M31 successfully predicts the steep age-velocity dispersion relation \citep{dorman,papaerII}, the star formation burst in the disc \citep{williams17}, the chemodynamics of the G1 clump \citep{paperVI} and the clustering of the quenching time of the dwarf spheroidal galaxies $\sim$8 Gyr ago \citep{Souza2021}, associated with the first pericentre passage of the merging satellite.

\subsection{Morphology of the merger remnant}\label{subsec:remnant}

The distribution of the stars in the remnant at the completion of the simulation is portrayed in Figure~\ref{fig:modelxy}. Star particles are demarcated separately for each progenitor (all particles [left], main progenitor particles [middle], and secondary progenitor particles [right]). Middle and lower rows further differentiate stars into two age groups, with 3.5~Gyr being the defining age; this specific choice is discussed within the context of the star formation history (SFH) of the disc of M31 in Appendix~\ref{sfh appendix}. Given the timing of the merger (the second pericentric passage occurred $\sim$3.5~Gyr ago), stars older than 3.5~Gyr formed before the strong interactions, while younger stars formed from gas mainly in the rebuilt thin disc (see Section \ref{section general concepts}). From the beginning of the simulation, stellar and gaseous particles of each progenitor are labelled as main or secondary galaxy particles and retain this label throughout the simulation (even if a gaseous particle is converted into a star within the remnant disc). 

Plotted as a 2D histogram, Figure~\ref{fig:modelxy} encapsulates the predicted stellar distribution at the end of the simulation (z~$=$~0). The orange ellipse shows the boundary of the remnant galaxy disc. We introduce this selection based on the estimated length of the major (30~kpc; \citealt{papaerII}) and minor  (3.99~kpc for an inclination of 77$^{\circ}$; \citealt{Geehan}) axes of the galaxy. For the appropriate rotation of the modelled disc, we assume a position angle (PA) of 30$^{\circ}$ (instead of the commonly cited PA = 38$^{\circ}$), which properly reproduces the observed morphology of the disc. We note that the thin stellar disc in the simulation has a scale length of 3.6~kpc \citepalias{hammer}. \citet{yin} estimate a disc scale length of 6.08~kpc for M31. We use this measurement as a baseline, and we scale our ellipsoidal selection for M31's disc accordingly (scaling down the major and minor axes in the model accordingly). The selection of the inner halo substructures (GSS and NE and W shelves) is based on their projected sky positions relative to the centre of M31, which reproduce the locations of their observed counterparts \citep{mcconnachie}; our demarcations are almost identical to the spatial selections made for the GSS, the NE-, and the W-Shelves in \citet{paperVI}.  For an extensive discussion on other prominent features evident in Figure~\ref{fig:modelxy} (like the stream at the north-west part of the remnant, dubbed the Counter GS), we refer the reader to \citetalias{hammer} and \citet{paperVI}. 

In Figure~\ref{fig:modelxy} [upper, middle], stellar particles from the more massive progenitor populate mainly the central parts of the galaxy, including the inner and outer parts of the disc, and the inner halo. In Figure~\ref{fig:modelxy} [upper, right], particles from the secondary galaxy form the tidal streams and shell-like substructures in the halo of the remnant galaxy, with the GSS being mostly formed from stellar debris of the secondary galaxy. Separate illustrations of the two age groups are given in the middle and lower rows. These distributions illustrate that the bulk of the inner halo is composed of old stars. 
In the lower panel, the distribution of young stars in the model traces the recent episode of star formation confined to the inner disc.

\section{Predicting oxygen abundances and metallicity gradients in the M31 disc}\label{section:metallicity in the model}
 
The \citetalias{hammer} simulations compute the enrichment of gas particles due to star formation, in addition to computing the orbits of both star and gas particles in the time-dependent gravitational potential. Implementing a suitable set of assumptions, one can then compute and compare the stellar metallicity (expressed in terms of [M/H] or [Fe/H] for $\alpha$/Fe = 0.0) and oxygen abundance in various regions of the disc, substructures, and inner halo of the simulated remnant with the abundance measurements of the oxygen/iron content from stellar and gaseous probes. 

The metallicity of each gas and stellar particle at the beginning of the simulation is a free parameter. The N-body hydro-dynamical models compute i) the metallicity of the star particles formed out of the gas distribution during the simulations and ii) the enrichment of the gas particles due to star formation. For those stars associated with the initial stellar discs of the progenitors, there is no further enrichment. These stars maintain the value assigned ab initio throughout the entire simulation. Star particles formed from gas that is initially in the secondary progenitor maintain their label (i.e. secondary particle), even if the star formation episode occurred within the remnant disc.

\subsection{The enrichment process of the interstellar medium in the \citetalias{hammer} simulations} \label{subsection: enrichment process}

The approach to compute the chemical enrichment process of the ISM in the simulation is described in detail in \citet{cox}. It assumes a yield of 0.02 per solar mass of stars formed. The enriched material is instantly recycled and depends only on the instantaneous star formation rate of each gas particle. The star formation rate is determined by the local gas density and is calculated individually for each gas particle, with metal enrichment carried out within such particles. The metal enrichment is smoothed over the smoothing kernel of the corresponding gas particles. Specifically for model 336 \citepalias{hammer} used in this paper, the smoothing kernel of star-forming particles has a median kernel size of 0.16~kpc (ranging from 0.05 to 0.18~kpc). Metal diffusion is not explicitly modelled as a separate process, but estimates of the properties of gas particles (enrichment, among others) are smoothed over the smoothing kernel, effectively mimicking the effects of metal diffusion \citep{Okamoto2005,Tornatore2007,Wiersma2009,Roca-Fabrega2021}. Each gas particle carries its metal content, and when new star particles form, they inherit the metallicity from their parent gas particles at that point in time. Once a star is born, its metal content remains unchanged.

This particular enrichment recipe is akin to the enrichment carried out through core-collapse supernovae (SN~$\rm{II}$); no delayed enrichment,  analogous to supernovae type Ia (SN~$\rm{Ia}$), is implemented in the code.

We convert, when needed, metallicity to oxygen abundance given the solar oxygen abundance (12~+~log(O/H)$_{\odot}$~=~8.69; \citealt{Asplund2009}) and, assuming that the solar elemental abundance is universal \citep{salaris}, through the following equation:

\begin{equation}
   \rm [M/H]= 12 + \log(O/H) - 8.69.
    \label{eq:m/h}
\end{equation}

We also express the stellar and gas metallicity in terms of iron abundance: [M/H]~=~[Fe/H], assuming $\alpha$/Fe~=~0.0 (solar alpha enrichment).

\subsection{Constraints on the metallicity distribution of the initial stellar and gaseous discs}\label{sec:initialdiscs}

Each galaxy, representing the main and secondary progenitor, consists of two distinct structures, that is a stellar and a gaseous disc, which have different disc scale lengths.  Thus, the initial oxygen abundance (or, equivalently, [M/H]) distribution must be set for each of the four discs at the beginning of the simulation. We adopt a linear relationship between radius and oxygen abundance values, with a negative gradient. Hence, the general functional form of the initial oxygen distribution in any of the discs at the beginning of the simulation is

\begin{equation}
      (12+\rm log(O/H))_{ini,R} = \nabla~(12+log(O/H))~\times R~+~(12+\rm log(O/H))_{R=0},
      \label{eq:initial oxygen distribution}
   \end{equation}
where $\rm\nabla$~(12+log(O/H)) is the adopted (negative) gradient, R is the distance of each particle (stellar or gaseous) from the galaxy centre in the plane of the disc, and (12+log(O/H))$\rm_{R=0}$ is the oxygen abundance at R$=$0. Hence, we need to specify i) the gradient $\rm \nabla$~(12+log(O/H)) and ii) the central (12+log(O/H))$\rm_{R=0}$ values for each of the four discs. 

To this end, we used the MZR \citep{maiolino} and the reported oxygen abundance from a large sample of PNe \citep{paperI, paperIII} in the M31 disc \citep{paperiv} to constrain the initial metallicity distribution for the stellar and gas discs in the simulation. As an additional consistency check, the abundance gradients in the simulated M31 disc at z~=~0 are compared with the measured abundance gradients determined for different generations of stars \citep{Gregersen2015,Saglia2018,Gajda2021, paperiv}. 

\begin{table}
 \caption{Initial conditions for the simulation of \citetalias{hammer} for the main and the secondary progenitor.}
    \centering
    \footnotesize
    \setlength\tabcolsep{5pt}
    \begin{tabular}{c c c } 
    
 \hline
 & & \\
 Parameter & Main & Secondary \\
  & & \\
 \hline \hline
   & & \\
 M$_{\star}$ [M$_\odot$]& 6.63 $\times$ $10^{10}$ & 0.66 $\times$ $10^{10}$ \\ 
 log (M$_{\star}$/M$_{\odot}$) & 10.82 &  9.82 \\ 
 h$\rm _{s}$ [kpc] &  2.8  &  2.8 \\ 
h$\rm _{g}$  [kpc] &   8.4 &  8.4\\  
 f$\rm _{gas}$  & 0.5& 0.8 \\   
 (12+ log(O/H))$_{\rm <1~R_{e}}$ [dex]&     8.706 $\pm $ 0.15&  8.518 $\pm$ 0.15\\ 
 $\nabla\rm (log(O/H))$ [dex/kpc]  &  -0.1 $\pm$ 0.05 & -0.1 $\pm$ 0.05\\
%Gas (12+log(O/H))$_{\rm <1~R_{e}}$ [dex] &         8.706 $\pm 0.15$ & 8.518 $\pm$ 0.15\\  
  & & \\
  \hline
    \end{tabular}
    \tablefoot{M$_{\star}$ is the stellar mass of each galaxy. h$\rm _{s}$ is the scale length of the stellar discs,  h$\rm _{g}$ is the scale length of the gas discs, and f$\rm _{gas}$ is the fraction of gas mass to the total baryonic mass in each progenitor (the baryonic mass is 20$\%$ of the total mass). (12+log(O/H))$_{\rm <1~R_{e}}$ are the values from the MZR measured for galaxies at z$  \simeq 1 $ \citep{rodriguez} and for a stellar mass equal to the mass of each progenitor respectively. The uncertainties reported are calculated as the $ 1 \times \sigma$ dispersion within the relevant redshift bin. The spread in the oxygen gradient values is derived from Figure~9 of \citet{curtigrad}.}
    \label{tab:ics}
\end{table}

\subsubsection{Constraints from the stellar mass-metallicity relation and observed and simulated metallicity gradients}

The MZR links the average of oxygen abundance, within the effective radius (R$\rm_{e}$), of the galaxy's gas-phase metallicity and its stellar mass (see \citealt{tremonti,maiolino,curti}). The simulation by \citetalias{hammer} starts at a lookback time of 7.5~Gyr (z $\sim$ 1). Since the effective radii and stellar masses of the two progenitors are known quantities in the simulation, we calibrate the average oxygen abundance within the R$\rm{_e}$, that is (12+log(O/H))$_{\rm <1~R_{e}}$, based on the oxygen abundance of similar stellar mass discs at z~$\simeq$~1 using the MZR. 

In Table~\ref{tab:ics}, we list the masses, scale lengths, and gas fractions for the two disc progenitors in the simulation. We note that the MZR depends on the adopted method to infer the gas-phase oxygen abundance. The direct (Te) abundance determination using auroral lines, which favour lower (12+log(O/H))$_{\rm <1~R_{e}}$ values, is considered the most accurate (see review by \citealt{maiolino} and references therein). However, no Te-based MZR is yet available at z~$\sim$~1. To account for this observational incongruity, we utilise the study from \citet{rodriguez}, where, although they use a strong-line calibration to infer the oxygen abundance of their sampled galaxies, they attain relatively low (12+log(O/H))$_{\rm <1~R_{e}}$ values due to their better estimate of the extinction and the underlying Balmer absorption. The (12+log(O/H))$_{\rm <1~R_{e}}$ values for the given progenitors' stellar masses based on this study are also noted in Table~\ref{tab:ics}. 

Concerning setting the initial metallicity gradient $\nabla$~(12+log(O/H)) of the two progenitor galaxies,  \citet{curtigrad} find a significantly wide spread in the measured values for the metallicity gradient with redshift (see their Figure~8). %Such a large spread may originate from observational uncertainties. 
In our current study, we then proceed by adopting the same $\nabla$~(12+log(O/H)) value equal to -0.1 $\pm$ 0.05 for the four discs (see the parameters adopted in Table~\ref{tab:derived parameters}). The arguments in support of this choice are presented in detail in Appendix~\ref{gradient appendix}.

The (12+log(O/H))$\rm_{R=0}$ abundance values are set independently of the gradient values using the constraints set from the (12+log(O/H))$_{\rm <1~R_{e}}$ MZR values at z~$=$~1 (see Table~\ref{tab:derived parameters}) within the errors. Therefore, the (12+log(O/H))$\rm_{R=0}$ values are different for the main and secondary. The metallicity of each star/gas particle is then assigned at the beginning of the simulation according to Equation~\ref{eq:initial oxygen distribution}, with a value for the gradient in the range -0.1 $\pm$ 0.05, see Table~\ref{tab:ics}. The abundance distributions at z~=~0 for the different stellar populations in the M31 remnant are then compared with the oxygen abundance distributions for PNe in the M31 disc (see Section \ref{section PNe}).

\subsubsection{Constraints from oxygen abundance distribution of M31 planetary nebulae} \label{section PNe}

    In what follows, we verify that our assumptions for the initial metallicity gradient make predictions which are consistent with measurements of the oxygen abundances from stellar population tracers of different ages in the M31 disc. The survey of PNe in M31, including the disc region from R$\rm_{GC}$ = 3 $-$ 30 kpc \citep{paperI, papaerII, paperiv}, provided age estimates and oxygen abundance distributions for stellar progenitors within the M31 disc. PNe with parent stellar populations older than $\sim$4.5 Gyr are associated with the kinematically hotter, thicker disc, while those that are $\sim$2.5 Gyr old are associated with the kinematically colder, thin disc \citep{papaerII}. Additionally, these two PN samples also had distinct oxygen abundance distributions \citep{paperiv}.

We thus compared the oxygen abundance distribution of these two age samples with that of their model counterparts. Based on the SFH of the modelled particles versus that determined for the observed M31 disc \citep{williams17}, we link the older and younger PN population with that of modelled particles having ages $>$3.5~Gyr and $<$3.5~Gyr, respectively. The milestone of 3.5~Gyr for the observed PNe which then separates between thin and thick disc PNe was set independently of the merger timescales in the simulations. The selection of this reference milestone, in conjunction with the disc SFH, is discussed in detail in Appendix~\ref{sfh appendix}.

\begin{figure}[t]
    \centering
    \includegraphics[width=0.5\textwidth]{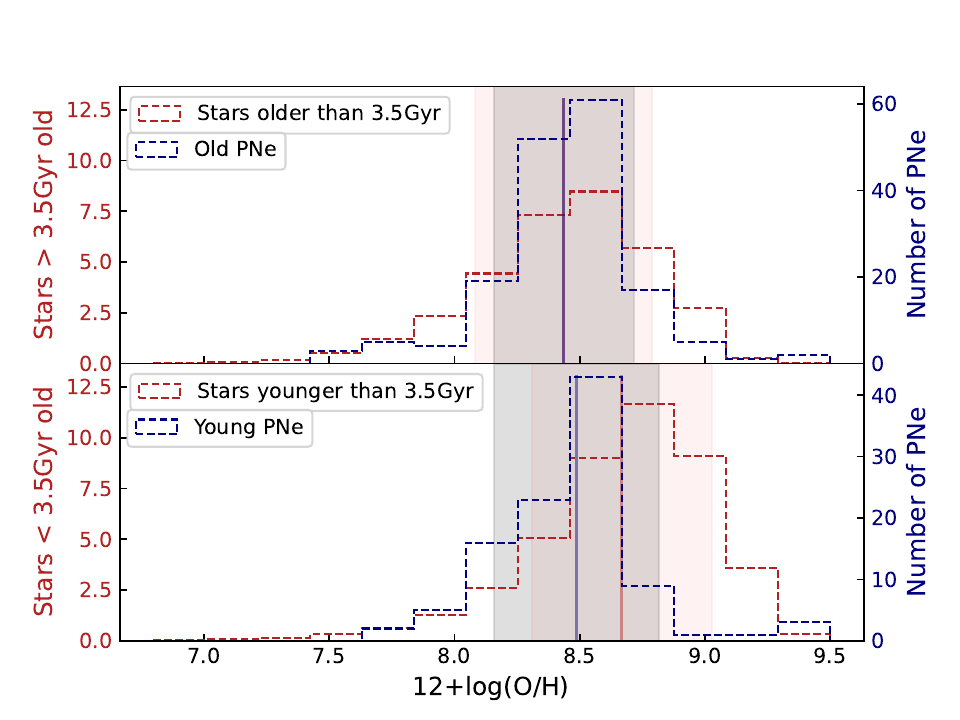}
    \caption{Upper panel: Comparison of the resulting oxygen abundance of old stars ($>$3.5 Gyr) in the modelled disc with the oxygen abundance of old PNe from \citet{paperiv}. Each vertical line specifies the mean values of the two datasets. Shaded regions represent the standard deviation ($1 \times \sigma$) of each dataset. The y-axis [left] is the mass percentage of stars older than 3.5~Gyr in the model. The PNe are plotted according to their number counts in the survey (y-axis [right]). Lower panel: Same as the upper panel but reporting the comparison between younger stars ($<$ 3.5~Gyr) in the model and young PNe.}
    \label{fig:oxcombined}
\end{figure}

\begin{table}
\caption{Adopted and derived parameters of the major merger model and observational constraints.}
    \centering
    \footnotesize
    \begin{tabular}{c c c } 
    
     & & \\
       \multicolumn{3}{c}{Adopted and derived parameters of the model at z $=$ 1}  \\
    
 \hline
 Parameter & Main & Secondary \\

 \hline

(12+log(O/H))$_{\rm R=0}$  [dex]&   9.0  &   8.75  \\  
$\nabla$(12+log(O/H))  [dex/kpc]  &  -0.1 &  -0.1 \\  
<12+log(O/H)>$_{\rm Stellar, <1~R_{e}}$ [dex]&    8.73  &  8.48 \\ 
<12+log(O/H)>$_{\rm Gas, <1~R_{e}}$ [dex] &        8.70  & 8.45 \\  
\hline
    \\
   \multicolumn{3}{c}{Derived parameters of the model at z~$=$~0} \\
    
 \hline
 Parameter & Old stars & Young stars \\

 \hline 
<12 + log(O/H)> [dex]& 8.47   & 8.67  \\
$\sigma$ (12 + log(O/H)) [dex] & 0.35 & 0.36\\
   (12+log(O/H))$_{\rm R=0}$  [dex]&   8.52  &   8.77  \\  
$\nabla$(12+log(O/H)) [dex/r$_{d}$] &  -0.06 &  -0.075  \\  
\hline
\\
   \multicolumn{3}{c}{PNe survey \citep{paperiv}}  \\
  \hline

 Parameter & Old PNe & Young PNe \\
\hline
<12 + log(O/H)> [dex]& 8.48 $\pm$ 0.02  & 8.57 $\pm$ 0.03\\
<12 + log(Ar/H)> [dex]& 6.25 $\pm$ 0.02  & 6.32 $\pm$ 0.03\\
$\sigma$ (12 + log(O/H)) [dex] & 0.21 & 0.28\\
$\sigma$ (12 + log(Ar/H)) [dex] & 0.20 & 0.29\\
   (12+log(O/H))$_{\rm R=0}$  [dex]&   8.31 $\pm$ 0.05  &   8.6 $\pm$ 0.08  \\
    (12+log(Ar/H))$_{\rm R=0}$  [dex]&   6.3 $\pm$ 0.05  &   6.51 $\pm$ 0.08  \\ 
$\nabla$(12 + log(O/H)) [dex/r$_{d}$] & 0.036 $\pm$ 0.018  & -0.079 $\pm$ 0.036\\
$\nabla$(12 + log(Ar/H)) [dex/r$_{d}$] & -0.03 $\pm$ 0.018  & -0.109 $\pm$ 0.036\\
  \hline

\\
   \multicolumn{3}{c}{RGB survey \citep{Gregersen2015}}  \\
  \hline

 Parameter & 4 Gyr old RGBs &  \\
 \hline
 $\nabla$[M/H] [dex/kpc] &  -0.02 $\pm$ 0.004&   \\
 & & \\

    \end{tabular}
    \tablefoot{(12+log(O/H))$\rm_{R=0}$ and $\nabla$~ (12+log(O/H)) are the adopted values for the linear functional form (Equation~\ref{eq:initial oxygen distribution}) of the oxygen abundance in the initial discs, that is the central value of the oxygen abundance and its gradient along the galactocentric radius. <12+log(O/H)>$_{\rm Stellar,\, <1~R_{e}}$ and  <12+log(O/H)>$_{\rm Gas,\, <1~R_{e}}$ are the oxygen abundance values we compute within one R$\rm_{e}$ for the initial discs at z~$=$~1. The derived parameters for the remnant disc at z~$=$~0 are the average oxygen abundance (< 12+log(O/H)>) for the two families of modelled stars, older and younger than 3.5~Gyr and its standard deviation ($\sigma$), their central abundance ((12+log(O/H))$_{\rm R=0}$), and the oxygen abundance gradient ($\nabla$(12+log(O/H))). The same parameters for oxygen and argon abundances are reported for the two families (old: >4.5~Gyr; young: $\sim$2.5~Gyr) of PNe from \citet{paperiv} . The metallicity gradient reported in \citet{Gregersen2015} is also noted.  }
    \label{tab:derived parameters}
\end{table}

    We carry our tests whereby the central value is fixed as in Table~\ref{tab:ics}, while we adjust the gradient from the shallowest (-0.05 dex/kpc) to the steepest possible values (-0.15 dex / kpc). The resulting (at z~=~0) <12~+~log(O/H)> of the older stars\footnote{ Stars in the simulated remnant M31 disc were selected  within an elliptical region (as described in Section~\ref{subsec:remnant}). A central 3~kpc aperture is masked in the spatial selection of the disc region since the PNe survey does not sample regions of high surface brightness in the centre of M31.} with ages $>$3.5~Gyr in the simulation is compared to the measured value for the older PNe.  We find a good agreement for $\nabla$~(12+log(O/H))~=~-0.1, while for the steeper/shallower values of the gradients, the average <12~+~log(O/H)> values of the abundance distributions of the older stars are off by more than 1~$\times$~$\sigma$. The comparison of the predicted oxygen abundance distributions of the simulated stars with those of old and young PNe for a $\nabla$~(12+log(O/H))~=~-0.1 gradient is shown in Figure~\ref{fig:oxcombined}.

Figure~\ref{fig:oxcombined} [upper panel] shows that the oxygen abundance distribution for the older stars\footnote{The oxygen abundance and metallicity ([M/H]) will be used interchangeably to probe the stellar metallicity as well using equation~\ref{eq:m/h}.} is in very good agreement with that of the old PNe. We recall that the cumulative stellar mass of stars older than 3.5~Gyr account for $\sim$ 85\% of the total stellar mass at z~=~0. Figure~\ref{fig:oxcombined} [lower panel] shows the corresponding quantities for the younger stars that contribute 15\% of the total stellar mass at z~=~0. The oxygen abundance distribution of the simulated young stars is broader and skewed towards higher oxygen abundance values than that measured for PNe with young stellar progenitors. Quantitatively, the <12+log(O/H)> of the young simulated stars is about $\sim$0.1~dex more metal-rich than that measured for the young PNe. Such an offset is comparable to the standard deviation values (Table~\ref{tab:derived parameters}). In summary, our approach reproduces the oxygen abundance distribution for the majority of stars (85\%) and within 1$\sigma$  for the remaining 15\%.

At this stage, we achieved a good agreement between the simulated old stars (ages $>$3.5 Gyr) and the oxygen abundance distribution of the PN with the older progenitors. This is an important result, as the substructures in the inner halo (GSS and NE and W shelves) comprise mainly old stellar populations \citep{brown2006, brown2007}, and we would test the model predictions for their projected phase space and metallicity distributions in Sections ~\ref{chemody} and ~\ref{section: metallicity comparison for everything}.

For the simulated younger stars, their oxygen distribution turns out to be slightly ($\sim$0.1~dex) more metal-rich than that measured for young PNe. Given the limitations of the chemical enrichment treatment as implemented in the \citetalias{hammer} simulations, we consider the current setup to provide a satisfactory approximation to within 1$\sigma$ for the chemical properties of the younger stars in the disc.

\subsection{Simulated radial abundance gradients for the different stellar populations in the M31 disc}\label{sec:radial gradients}

\begin{figure}[t]
    \centering
    \includegraphics[width=0.5\textwidth]{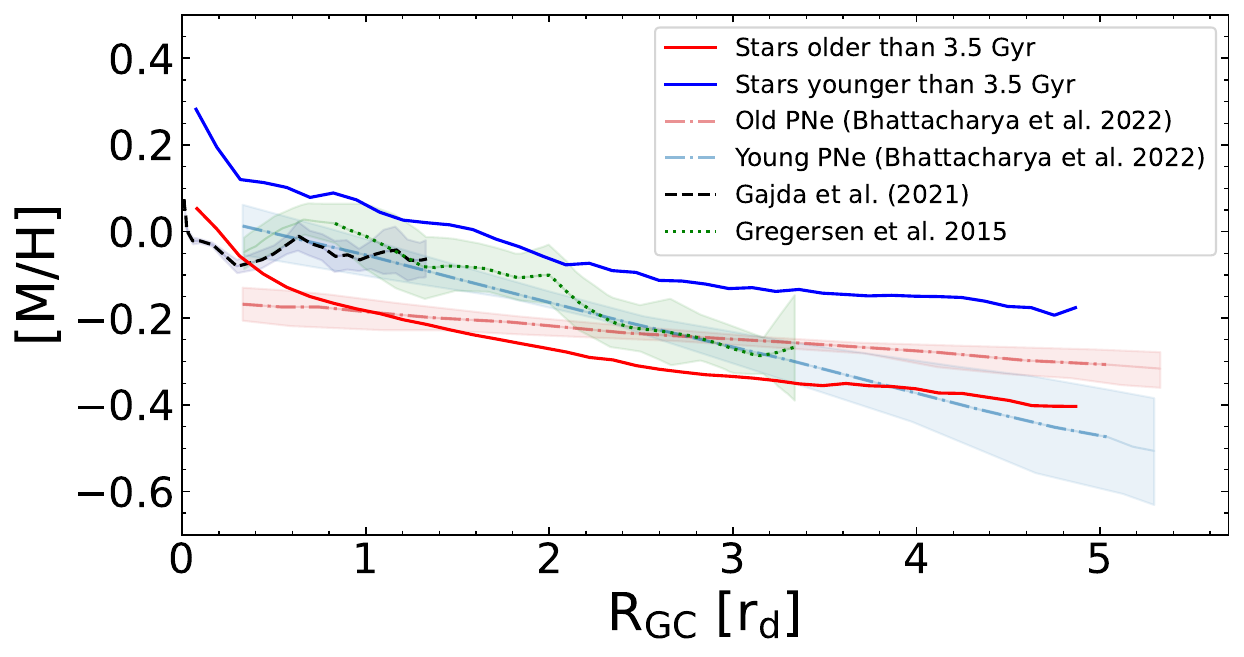}
    \caption{Comparison of the resulting metallicity along the galactocentric radius for old (red solid curve) and young (blue solid curve) stars in the model with corresponding observations, expressed in units of disc scale length (r$\rm_{d}$). For the old and young PNe sample from \citet{paperiv}, the 12+log(Ar/H) gradient are shown.}
    \label{fig:gradient}
\end{figure}

The chemical evolution of discs of spiral galaxies is reflected in their radial abundance gradient \citep[see][and references therein]{maiolino}. In this study, we implemented a procedure to predict the radial oxygen gradients for stellar populations for the two different ages from the \citetalias{hammer} merger simulations for the M31 remnant disc. Using Equation ~\ref{eq:m/h}, we convert the oxygen abundances to stellar metallicity gradients, and we utilise different observational data sets as possible benchmarks for comparison. 

In \citet{Saglia2018}, the authors derived the metallicity gradient from IFU observations of the M31 central regions, within R$\rm_{GC}$~$\sim$~4~kpc, and measured a [M/H] gradient of 0.0~$ \pm $~0.03 dex/kpc. The mean [M/H] for their disc population is 0.03~$\pm$~0.03 dex, without postulating an age distinction for the stellar populations under study.  In the model, within R$\rm_{GC}$~$\sim$~4~kpc, we report a mean [M/H] of -0.16 dex for all stars, a mean [M/H] of 0.03 dex for stars younger than 3.5 Gyr, and a mean [M/H] of -0.2 dex for stars older than 3.5 Gyr. 

Constrained by the aforementioned IFU observations, made-to-measure modelling of M31 by \cite{BlanaDiaz2018} was utilised by \cite{Gajda2021} to determine the [M/H] radial profile out to $\sim $~10 kpc (Figure \ref{fig:gradient}). \citet{Gregersen2015} measured a metallicity gradient (see Table~\ref{tab:derived parameters}) for R$\rm_{GC}$~$\simeq$~4~$-$~20 kpc in the M31 disc using stars from the PHAT survey (also, Figure \ref{fig:gradient}). They assumed solar [$\alpha$/Fe] value and a constant age of 4~Gyr for the RGB stars. \citet{paperiv} measured oxygen and argon radial abundance gradients for the two age groups of PN progenitors. Within the radial range of 3~$-$~30 kpc, their reported oxygen abundance gradients are shown in Table~\ref{tab:derived parameters} and plotted in Figure \ref{fig:gradient}, separately for their two PNe groups.

Figure~\ref{fig:gradient} also presents the mean metallicity in the modelled stars for our two distinct age groups as a function of galactocentric radius. The fitted parameters for the radial oxygen abundance (akin to [M/H]) distribution of these two age groups are noted in Table~\ref{tab:derived parameters}. For the young modelled stars, the gradient is consistent with the negative gradient from the young PNe, but with a higher (12+log(O/H))$_{\rm R=0}$ value. For the older modelled stars, we find a negative gradient, different from the positive gradient found for older PNe, but with a higher (12+log(O/H))$_{\rm R=0}$ value that leads to a consistent <12 + log(O/H)>. As shown in Figure ~\ref{fig:gradient}, the [M/H] as a function of the galactocentric radius for the two groups of modelled stars brackets the observed distributions from stellar probes of different ages.
 
As a general remark, we also point out that old stars in the simulation model show a flatter gradient relative to younger stars, as expected in galaxies that experienced mergers (e.g. \citealt{Zinchenko2015ApJ...806..267Z, Tissera2019MNRAS.482.2208T}).

\subsection{Radial redistribution of gas particles during the major merger event} 

The metallicity distributions of stellar and gaseous particles in the remnant M31 disc at the end of the merger simulation are the results of the i) initial assignment of the oxygen abundance gradients in the stellar and gaseous discs of both main and secondary progenitors, and contributions from either ii) the triggered star formation or iii) the radial redistribution of the stellar and gaseous particles during the merger event.

\begin{figure}[t]
    \centering
    \includegraphics[width=0.5\textwidth]{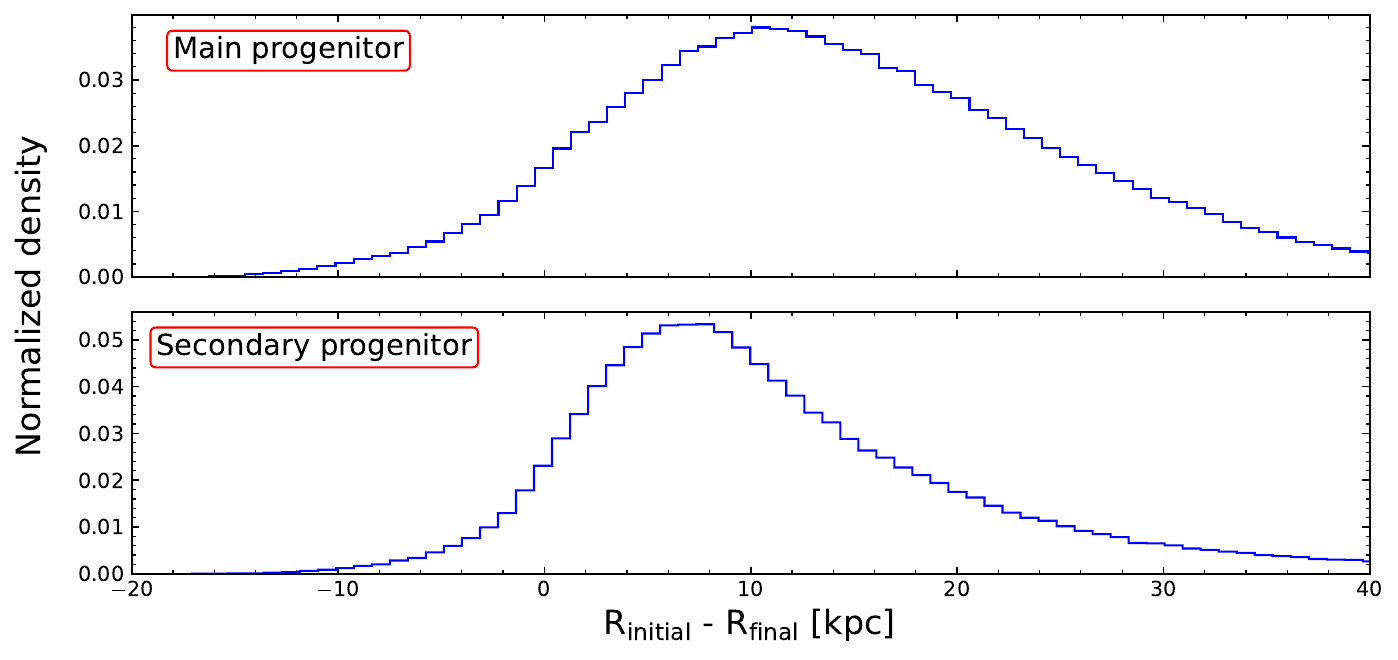}
    \caption{Normalized density of the distribution of the difference between the initial (R$\rm_{initial}$) and final (R$\rm_{final}$) galactocentric distances of young ($<$3.5 Gyr) stars in the two progenitor galaxies. Here, R$\rm_{initial}$ is the distance of each gas particle from the centre of its progenitor (main or secondary) at the beginning of the simulation (7.5~Gyr ago), whereas R$\rm_{final}$ is the galactocentric distance of the stellar particle that spawned from its gaseous progenitor and is situated in the remnant disc. Upper panel: Stars younger than 3.5~Gyr from the main progenitor. Lower panel: Stars younger than 3.5~Gyr from the secondary progenitor.}
    \label{fig:ini1}
\end{figure}

To assess the interplay between the adopted initial abundance gradients and the enrichment due to the star formation in the simulation, we estimate the radial redistribution of young stars in the disc from the initial radial distance of the gas particles in the progenitors' discs. In Figure~\ref{fig:ini1}, we show the $\Delta$R (i.e., R$\rm_{initial}$ - R$\rm_{final}$) of stars younger than 3.5~Gyr separated based on their initial progenitor's association (we remind the reader that star/gas particles retain their initial progenitor identity regardless whether they are converted into stellar particles within the remnant disc).

A preliminary assessment of the spatial rearrangement of stars within the final remnant is as follows: As seen in Figure~\ref{fig:ini1}, 93\% of young stars ($<$3.5~Gyr old), from both secondary and main progenitors, have moved inwards ($\Delta$R~$>$~0) from their initial galactocentric radius. We note again, that since we deal with stars younger than 3.5~Gyr, these particles were initially gaseous and were converted into stars during the last 3.5~Gyr of the simulation. Hence, the majority of young stars in the model came initially from the gas particles residing in the outskirts of the progenitors. The initially negative (-0.1 dex/kpc) metallicity gradient led to their lower metal content.

The in-depth assessment of the impact of different mechanisms (e.g. the stellar bar, the energy injection from the secondary, the spiral arms, etc.) on the observed radial redistribution of disc stars caused by the major merger event, will be the topic of a future study. 

\section{New predictions of the multidimensional phase space structure of the Z~=~0 remnant in the simulation} 
\label{section: the multicomponent nature of substructures in the model}

\begin{figure*}[h]
    \centering
    \includegraphics[width=\textwidth]{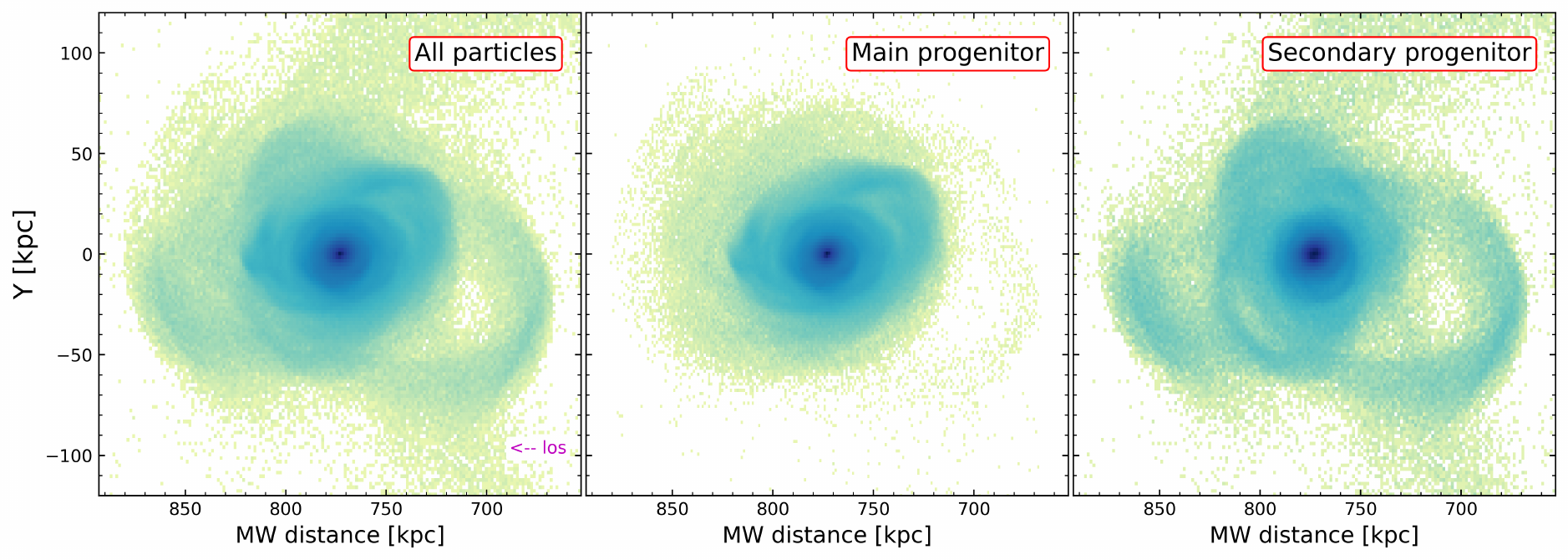}
    \caption{Two-dimensional histogram of the LOS (Z, Y) plane view of the M31 remnant. The centre of the galaxy is fixed at 773 kpc from the MW \citep{conn}. All simulated particles [left], main progenitor particles [middle], and secondary progenitor particles [right] are plotted separately. The Z$-$axis of the simulation denotes our LOS and is designated as MW distance (our vantage point is marked on the bottom right of the leftmost panel). The Y-axis corresponds to the distance from the centre of M31 in the direction from north (up) to south (down) in units of kiloparsecs.}
    \label{fig:modelzy}
\end{figure*}

In a minor accretion event scenario, the substructures in the M31 inner halo (i.e. GSS and NE and W shelves), are single leading (NE shelf) and trailing (GSS, W shelf) orbital wraps of tidally disrupted stars (e.g. \citealt{fardal2013,kirihara2014,Kirihara2017}) which were formed in the last 1 Gyr. Hence, the shell patterns for the NE and W shelves are predicted to correspond to successive orbital passages of GSS-related tidal debris \citep{escala22}. The phase space distributions (projected radius R$\rm_{proj}$ versus  LOS velocity V$\rm_{LOS}$) of their stellar components are therefore smooth (see, for example, Fig.~10 in \citealt{escala22}). Over-densities or ridges may be the "smoking gun" for the possible presence of an intact core of the secondary progenitor, at the location of the NE shelf (see again Fig.~10 in \citealt{escala22} featuring the simulated projected phase space diagram for the NE shelf in case of the survival of the secondary progenitor's intact core).

In a major merger, the substructures acquire complex configurations in 3D space as a consequence of subsequent pericentric passages. The released tidal debris stars in the distinct loops produce over-densities and ridges in the observed phase space and complex structures along the LOS distance \citepalias{hammer}.  This is particularly interesting given the presence of the KCCs in the GSS \citep{Kalirai2006, gilbert09}, recently confirmed by the DESI observations of the inner halo of M31 \citepalias{dey}.  In addition to the KCCs, the distributions of the LOS distances versus photometric metallicity estimates, from the resolved stellar populations studies carried out by  \citet{conn} and \citet{ogami}, show broad probability distributions for the LOS distances along the GSS with secondary-peaks, and $\sigma$[M/H]$\simeq$1~dex for the metallicity distributions in the GSS and inner halo substructures. See also the work by \citetalias{dey} for an illustration of the coherent kinematic structures (streams, wedges, and chevrons) in the positions and velocities distribution of individual stars at different locations in the M31 inner halo, as well as their metallicity distribution.  

Using the \citetalias{hammer} simulations, we can now explore the network along the LOS, phase space, and metallicity distributions of the GSS and NE and W shelves analogues in the M31 simulated remnant. In Figure~\ref{fig:modelzy}, we show the LOS particle distribution of the simulated galaxy, plotted in the (Z,Y) plane, to illustrate its spatial properties. The composite nature of each structure emerges from the several loops seen in projection, mostly associated with the several pericentric passages along the infalling orbit of the secondary galaxy (see the rightmost panel in Figure~\ref{fig:modelzy}). Their association with the GSS and NE and W shelves over-densities was previously presented in \citetalias{hammer}. Here we focus on their LOS distance structure, projected phase space properties, and metallicity distributions. 

To group particles within each substructure of the inner halo of the simulated galaxy, we adopted Density-based Spatial Clustering of Applications with Noise (DBSCAN; \citealt{dbscan}) which catalogues the multiple components within the GSS, the NE-, and W-Shelves. This code utilises the 3D position of the stars rather than simply over-densities in the projected number density distribution on the sky. It recognises distinct clusters of points in high-density regions of any given parameter space. Our adopted parameters and their specific implementation in the current study are described in detail in Appendix~\ref{appendix:dbscan}. 

Once particles are grouped in each of the three substructures (GSS and NE and W shelves) of the M31 simulated inner halo, we study the morphology of their R$\rm_{proj}$ versus  V$\rm_{LOS}$ phase space diagrams, and their metallicity distribution.  In Section~\ref{section: metallicity comparison for everything}, we shall carry out a one-to-one comparison of the [M/H] average values and standard deviations from relevant regions in the simulated M31 remnant with metallicity measurements for the same spatial regions of M31 itself. 

In Figures~\ref{fig:desi_gs}, \ref{fig:desi_ne} and \ref{fig:desi_ws}, we identify any stream-, chevron-, and wedge-like over-density patterns in each component of the resulting substructures in the simulated M31 remnant. We used the orange-dashed and black-dotted lines for ridge and shell pattern identification in the main and secondary particles, respectively. Whenever the same feature arose in the phase space distributions of stellar particles originating from either progenitor, we plot the lines in red. These morphological features are to be compared with the observed ones in the R$\rm_{proj}$ versus  V$\rm_{LOS}$ diagrams recently acquired from the DESI observations in the inner halo of M31 \citepalias{dey}.

In the phase space diagrams shown in the following sections, we plot the (V$\rm_{LOS}$ - V$\rm_{sys}$) versus R$\rm_{proj}$, where V$\rm_{sys}$ is the systemic velocity of M31 equal to -300 kms$^{-1}$ \citep{Watkins2013} and R$\rm_{proj}$ is the projected linear distance (in kpc) from the centre of the simulated M31 remnant, computed as R$\rm_{proj}$ = $\rm\sqrt{(X^{2} + Y^{2})}$.

\begin{figure*}[htp]
\centering
    \includegraphics[width=\textwidth]{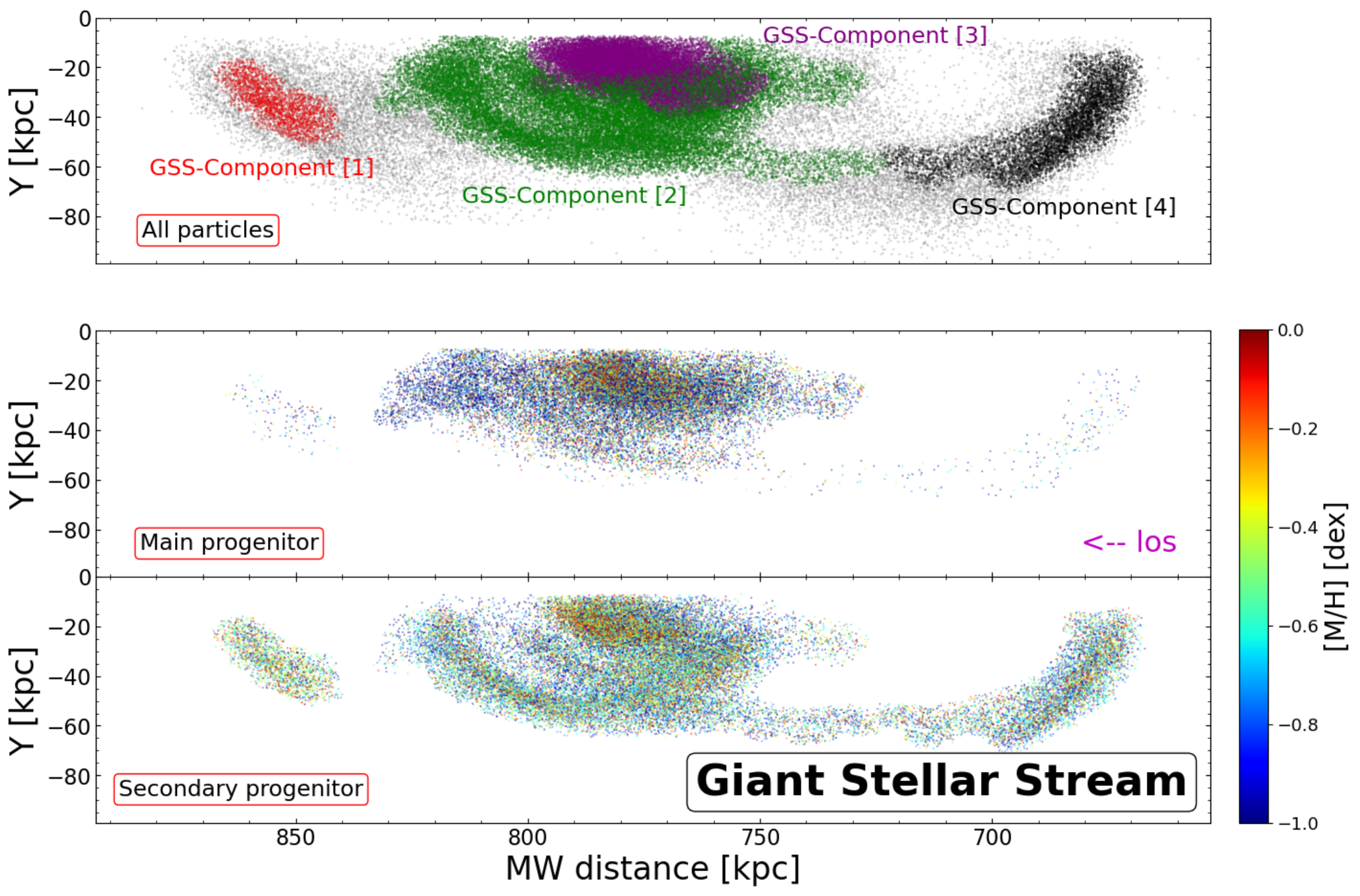}
    \caption{Upper panel: Illustration of the distribution along the LOS of all the stellar particles of the spatially selected GSS (see Figure ~\ref{fig:modelxy}), grouped into four numbered components by DBSCAN. Grey points are stellar particles that are not associated with any group of particles according to the DBSCAN clustering algorithm. The entire spatial distribution of the modelled GSS stars along the LOS is compared to independent observations in Figure~\ref{fig:ogami_gs}. Particles in the upper panel are colour-coded according to their association with a DBSCAN component, as detailed in the legend. Middle panel: Particles originating from the main progenitor that belong to the identified components in the GSS. They are colour-coded by their metallicity. The grey particles shown in the upper panel are not included. Lower panel: Same as the middle panel, but for stellar particles which originate from the secondary progenitor. }
    \label{fig:3d_gs}
\end{figure*}

\subsection{The Giant Stellar Stream }\label{ssec:GS}
The PAndAS survey \citep{mcconnachie09nature} showed that the GSS is the dominant contributor to the number of RGB stars in the M31 inner halo. It covers at least $\sim$6$^{\circ}$ on the sky, and it reaches out to $\sim$80 kpc \citep{McConnachie03, conn, Cohen2018} in projected distance. The proximity of M31 allows observers to study the resolved RGB population in the stream (e.g. \citealt{ibata2004,Guhathakurta06, Ferguson2016,Wojno2023}), and use them to estimate the LOS distances from the tip of the RGB \citep{McConnachie03, conn,ogami}, in addition to obtaining the radial velocities of RGB stars (e.g. \citealt{Guhathakurta06,gilbert09}; \citetalias{dey}) and PNe \citep{paperVI}. 

\begin{figure*}[h]
    \centering
    \includegraphics[width=\textwidth]{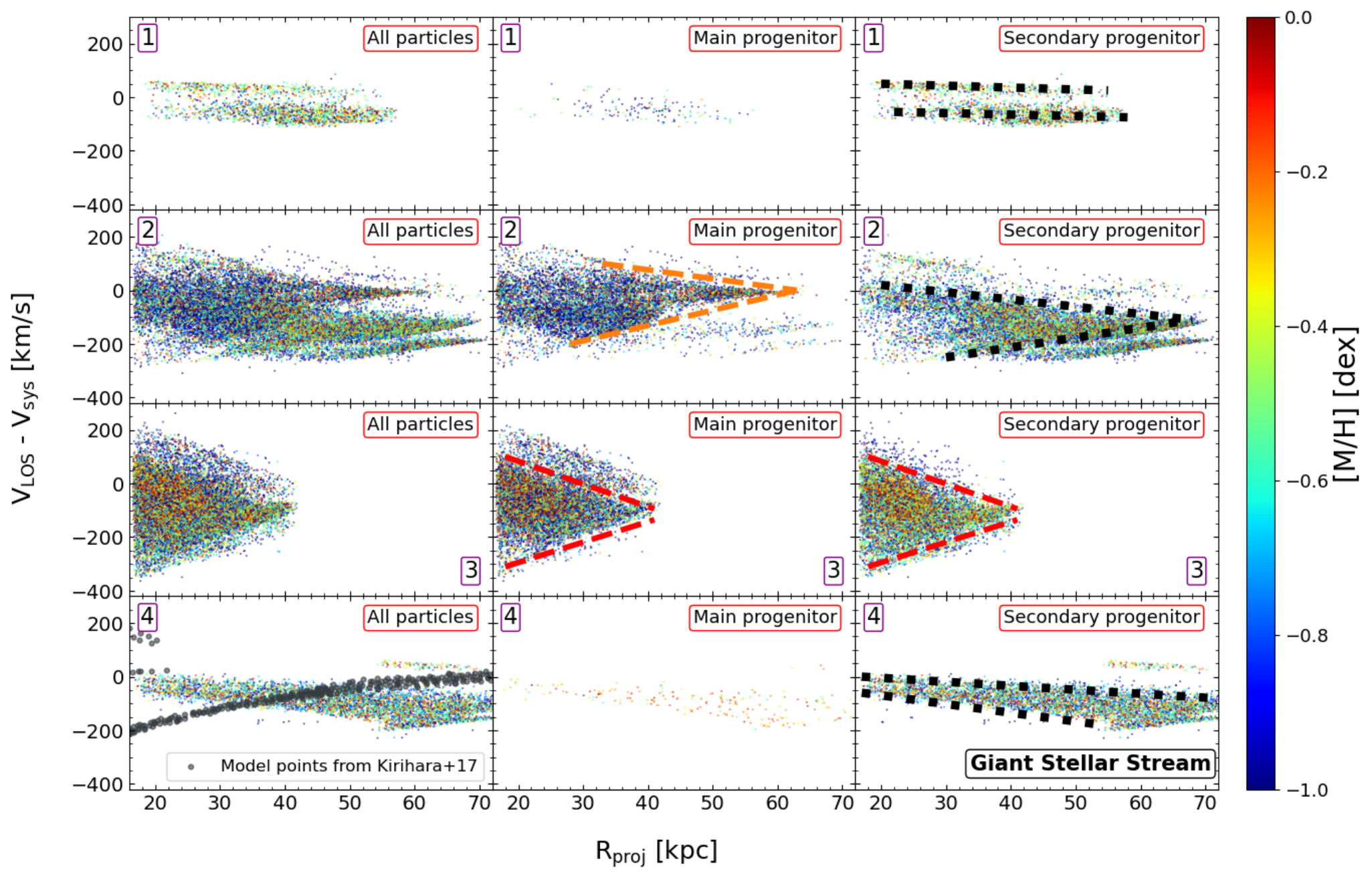}
    \caption{Phase space diagram (V$\rm_{LOS}$ - V$\rm_{sys}$) versus R$\rm_{proj}$. Here, V$\rm_{sys}$ is the systemic velocity of M31 equal to -300 kms$^{-1}$ \citep{Watkins2013}, and R$\rm_{proj}$ is the projected linear distance in kiloparsecs to the centre of the simulated M31 remnant for the modelled stars in the GSS. The four panels show the projected phase space for each GSS-component ([1] to [4]; see Figure ~\ref{fig:3d_gs}) identified by DBSCAN. All the stellar particles are colour-coded by their metallicity. All simulated particles [left column], main progenitor particles [middle column], and secondary progenitor particles [right column] are separately illustrated. Ridges from the main progenitor are marked with orange-dashed lines while those from the secondary are marked with black-dotted lines. When the same morphological feature is identified in both progenitors, then it is marked by red-dashed lines in both panels.  In the leftmost panel of the GSS-component [4] row, the grey circles are simulated stars from the satellite's trailing tail, reproduced from \cite{Kirihara2017}.}
    \label{fig:desi_gs}
\end{figure*}

In Figure~\ref{fig:3d_gs}, we show the (Z, Y) density distribution of the stellar particles in the region of the GSS, that is all the star particles extracted from the region enclosed by the red-dashed polygon of Figure~\ref{fig:modelxy} in the projected  (X, Y) view on the sky of the M31 simulated remnant galaxy. We plot the star particles in the GSS region colour-coded according to the grouping identified by DBSCAN in the 3D particle distribution  (X, Y, Z) of the M31 remnant. DBSCAN identifies four components of the GSS regions; they are labelled GSS-components [1] to [4] in the uppermost panel of Figure~\ref{fig:3d_gs}. In the middle and lower panels of Figure~\ref{fig:3d_gs}, we show the particles in groups [1] to [4] according to their origin, either main (middle panel) or secondary (lower panel). 

The distribution in the  (Z, Y) plane and the DBSCAN grouping clearly show that the S-components [1], [2], and [4] include the loops farthest away from the center. These loops are dominated by star particles originating from the secondary progenitor. The plane that contains the loops is seen nearly edge-on and the extended GSS structure is then the result of the superposition of GSS-components [1] to [4] along the LOS. GSS-component [3] is dominated instead by stellar particles from the main progenitor and lies closer to the M31's disc.

In the middle and lower panels of Figure~\ref{fig:3d_gs}, star particles are colour-coded according to their metal content. There is a definite metallicity gradient as a function of radial distance along the simulated GSS, with the more metal-rich star particles found closer to the centre of M31. The lower panel shows that the more metal-rich particles in the secondary are found in the DBSCAN components at smaller distances. In the middle panel, the star particles in the GSS-component [3], which originate from the main progenitor, show a radial metallicity gradient as well, with more metal-rich star particles found at smaller galaxy radii. 

During a radial merger event, stars from the less massive companion become gravitationally unbound, and tidal debris are deposited in a leading and a trailing tail \citep{ferguson2002, Hendel2015MNRAS.454.2472H}. As in any galactic disruption event, the least bound stars in the disrupted satellite are deposited at the largest distances in the host potential (e.g. \citealt{Amorisco2017}). As stars approach the apocentre of their orbits in the deeper potential well, they reverse their path, which leads to an enhancement in stellar density at the apocentre \citep{Merrifield}. Their morphology will resemble a shell, and a V-shaped, chevron- or wedge-like feature will appear in their R$\rm_{proj}$ versus V$\rm_{LOS}$ diagram. %We expect to detect shell-like features in all three substructures.

The R$\rm_{proj}$ versus V$\rm_{LOS}$ phase space diagrams for all the star particles for the GSS-components [1] to [4] are plotted in Figure~\ref{fig:desi_gs}, leftmost panels. We illustrate the different regions of the star particle distribution in panels [1] to [4], and then we further showcase whether star particles originate from the main (central panel) or secondary (right panel) progenitor. We outline wedge patterns in GSS-components [2] and [3]. Streams and tail features appear in GSS-components [1] and [4], which are populated by stars originating from the secondary progenitor. The morphology of the GSS-components arises since the orbit of the centre of mass of the secondary progenitor becomes more radial as it sinks to the centre of the deeper potential well (as described in \citealt{Amorisco2017}). Therefore, stellar debris removed during the early stages of the merger produces streams/arms (i.e. like GSS-components [1] and [4]), while stellar debris released later on generates wedges and chevrons (i.e. GSS-components [2] and [3]).

Specifically, the morphology of the phase space distribution of the GSS-component [1] displays a stream-like feature populated by particles from the secondary progenitor (Figure ~\ref{fig:desi_gs}, panel [1]). 
 
The phase space distribution for the GSS-component [2] displays two distinct V-shaped chevrons, populated by star particles from the main and the secondary particles (Figure~\ref{fig:desi_gs}, panel [2]). These two chevrons reach different projected distances, with the chevron populated by the secondary star particles reaching larger distances and negative velocities, with its turnaround point at V$\rm_{LOS} - $ V$\rm_{sys} \simeq $ 100 kms$^{-1}$ at R$\rm_{proj} \simeq $ 70 kpc. Overall, the GSS-component [2] is a prominent feature of the simulated GSS. A wedge pattern is seen in the phase space distribution of the GSS-component [3] with similar density of star particles and ranges in radius and velocities for both progenitors; this feature is shown with a red dashed line to emphasise that we deal with the same feature in both progenitors (Figure~\ref{fig:desi_gs}, panel [3]). The GSS-component [4] is a tail/stream extended over large projected distances from the main body of the galaxy, towards the MW along the LOS. This tidal tail exhibits the largest projected velocity and distance from the centre of M31 among the GSS components. A stream-like feature is seen in its phase space distribution (Figure ~\ref{fig:desi_gs}, panel [4]). 

To summarise, the major merger simulations would lead to the presence of distinct kinematic components in the GSS region, which were previously identified in observations \citep{Kalirai2006, gilbert09, dey}. Such stars originate from either the main or the secondary progenitor \citep{paperVI}. Hence, in the \citetalias{hammer} merger simulation, the GSS appears to be a composite structure comprising loops of stars at different radial distances/energies, with significant structural differences from the smooth trailing stream of a minor accretion event. The complex kinematic structure of the GSS can only be reproduced by a major merger, since a minor merger would have produced a single trailing tail; see the grey points of \citet{Kirihara2017} simulation reproduced in the leftmost panel of Figure~\ref{fig:desi_gs}, GSS-component [4], all particles.

\subsection{The North-East shelf}\label{ssec:NE}

\begin{figure}[htp]
    \centering
    \includegraphics[width=0.5\textwidth]{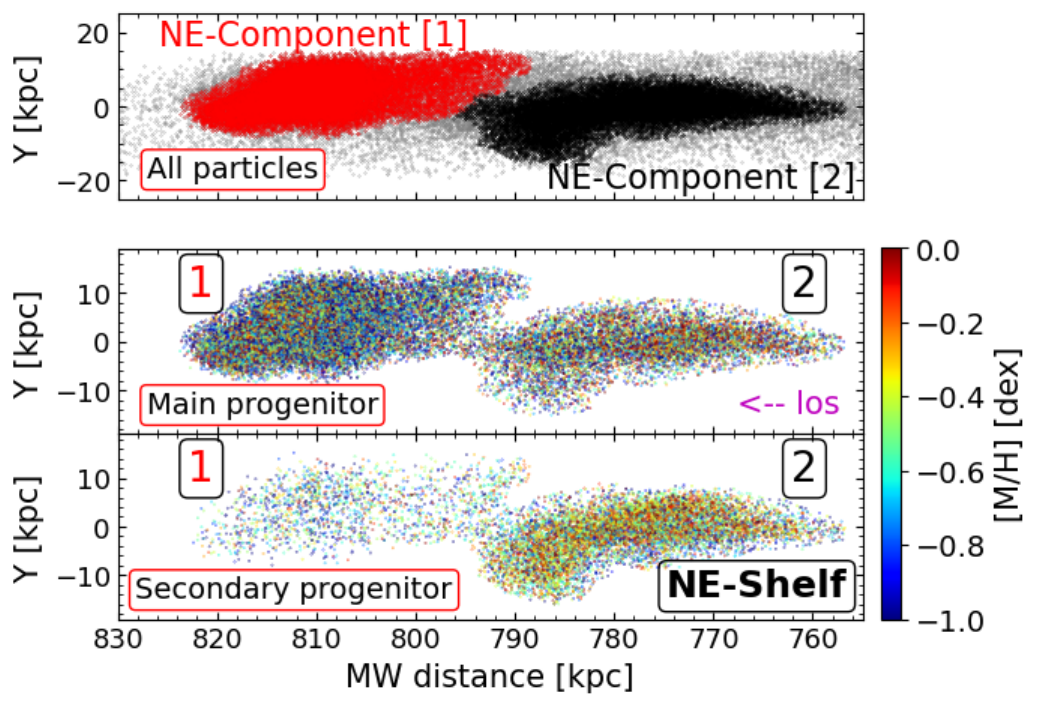}
    \caption{ Illustration of the distribution along the LOS of all the stellar particles of the spatially selected NE shelf (see Figure ~\ref{fig:modelxy}) grouped into two numbered components by DBSCAN (upper panel: red and black points labelled in the legend). Grey points are stellar particles that are not linked to any group of particles according to the DBSCAN clustering algorithm. The entire spatial distribution of the modelled NE shelf stars along the LOS is compared to independent observations in Figure~\ref{fig:ogami_ne}. Middle and lower panels: Same as Figure~\ref{fig:3d_gs} but for the particles in the NE shelf region. }
    \label{fig:3d_ne}
\end{figure}

\begin{figure*}[htp]
    \centering
    \includegraphics[width=0.9\textwidth]{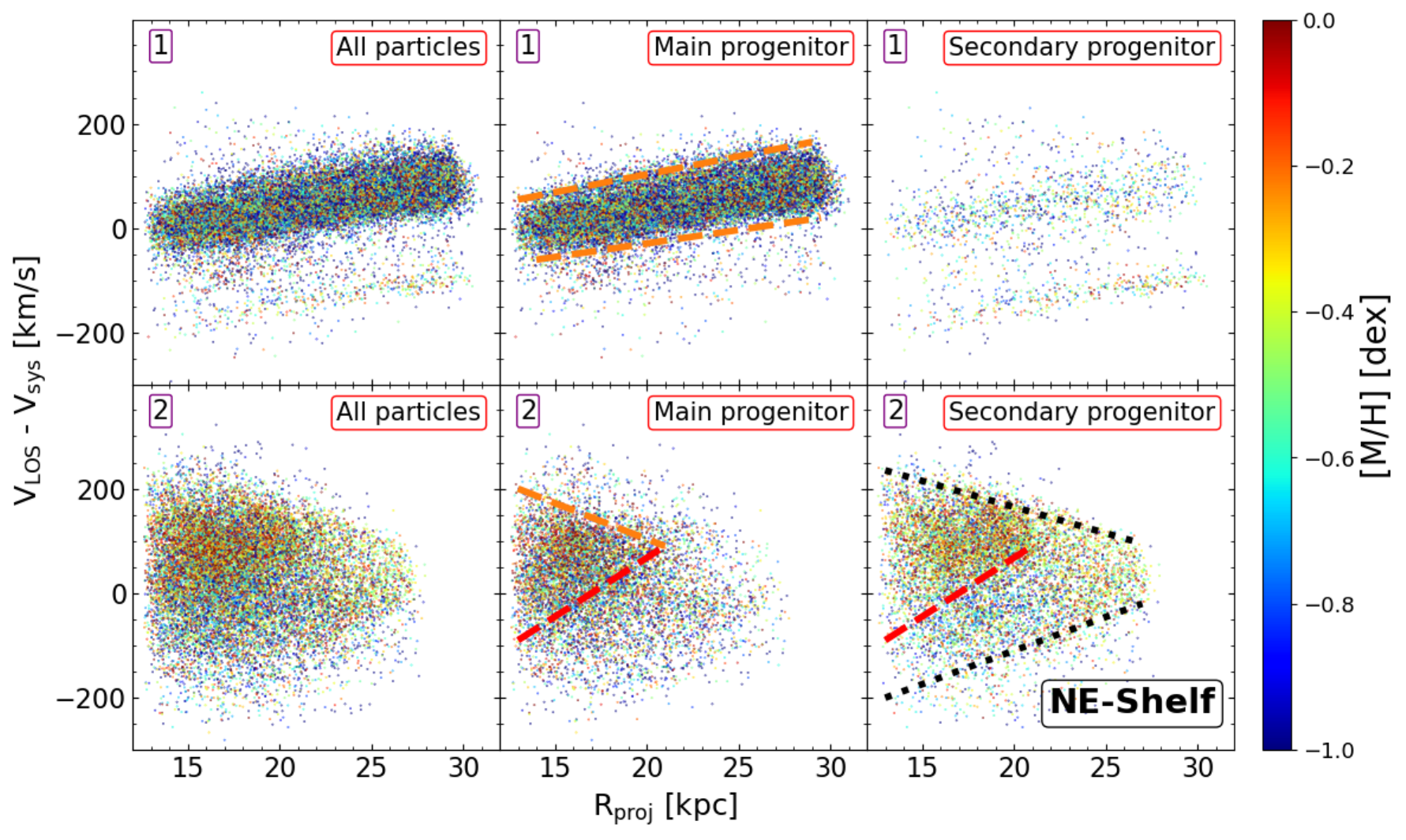}
    \caption{Same as Figure \ref{fig:desi_gs} but for the NE shelf with two components.}
    \label{fig:desi_ne}
\end{figure*}

The NE shelf indicates a region exhibiting an abrupt increase in surface stellar density in the north-east region of the inner halo of M31 \citep{ferguson2002}. The NE shelf is the outermost shell according to \citet{escala22}, extending to R$\rm_{proj} \simeq $ 31~kpc and spanning $\sim $400 kms$^{-1}$ in velocity.

DBSCAN is used to identify clusters in the (X, Y, Z) particle distribution at the NE location in the M31 remnant that emerge as stellar overdensities in their 3D unfolding. Figure ~\ref{fig:3d_ne} reveals two components in the (Z, Y) plane of the selected region. These two clusters have different LOS distances to the MW, with NE-component [1] in the radial range of 820 to 800 kpc, and NE-component [2] in the 790 to 760 kpc range.

In Figures~\ref{fig:3d_ne} and ~\ref{fig:desi_ne}, we present the LOS structure of the NE shelf in the M31 remnant, and the R$\rm_{proj}$ versus  V$\rm_{LOS}$ distribution for each component, respectively. In both figures, it is apparent that NE-component [1] comprises mostly particles from the main progenitor. In the R$\rm_{proj}$ versus V$\rm_{LOS}$ diagram of Figure~\ref{fig:desi_ne}, there is a single, extended feature in this component indicating that these stars exhibit disc kinematics. We also note that the star particles in component [1] show a wide spread of metallicity, with a large contribution from the range [M/H]~$<$~-0.6 dex. The assessment of the stellar ages within the rotationally supported NE-component [1] resulted in $\sim$23\% of stars being younger than 5~Gyr old. \citet{Bernard15} report that their "disc-like" fields in the NE shelf regions have had one-quarter of their mass formed in the last 5 Gyr. We then interpret component [1] as an extension of the rotationally supported M31 disc, whose younger metal-poor stars are formed from the outermost gas in the main progenitor. For the contamination of the M31 disc stars in the NE shelf, see also the discussion in \citet{escala22}.

Next, we turn to the NE-component [2]. This cluster of stars comprises wedges that are populated by stars originating either from the main or the secondary progenitor.  

The leftmost panel of Figure~\ref{fig:desi_ne} [lower row] featuring all the particles within NE-component~[2], gives a vivid illustration of the broad metallicity spread predicted in this study for the NE shelf, with distinctly higher metallicity than for the rotationally supported NE-component  [1]. The cluster of stellar particles in NE-component~[2] of the NE shelf can be decomposed into two wedge patterns (see~Figure ~\ref{fig:desi_ne}), with an inner wedge appearing mainly in the R$\rm_{proj}$ versus V$\rm_{LOS}$ diagram of particles from the main progenitor, and an outer wedge populated by particles from the secondary. 
The lower part of the inner wedge appears in both distributions, so we show it as a red-dashed contour in both panels. The radial extent of the wedge in the NE shelf populated by particles from the secondary covers a range of radii and velocities very similar to the ranges for these quantities determined from the wedge distribution of the NE shelf RGB stars published by \citet{escala22}.

We note that a difference in the maximum R$\rm_{proj}$ of the wedges suggests distinct apocentre points for the underlying coherently moving group of stars. Stars from the pre-merger disc (which have an apocentre closer to the centre of M31) slow to a halt before reversing their path sooner than particles from the secondary. This is expected since stars with different initial orbits reach distinct apocentre radii \citep{Merrifield}.

A possible cause for NE-component [2] to include particles from the main is that, as the satellite passes through the pre-merger disc of the galaxy, stars are dragged along the orbits of the secondary. Therefore, finding particles in each stellar substructure in the inner halo coming from the pre-merger disc of M31 is expected in a major merger scenario, due to the substantial gravitational pull of the more massive secondary galaxy.

\subsection{The Western shelf}\label{ssec:WShelf}

\begin{figure}[h]
    \centering
    \includegraphics[width=0.5\textwidth]{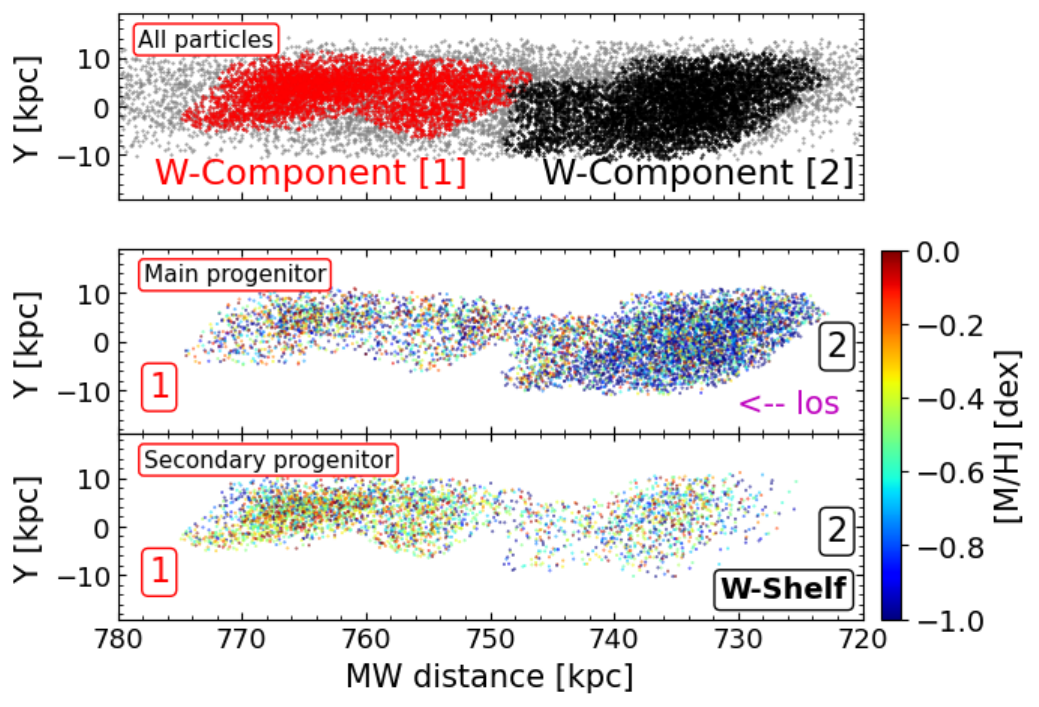}
    \caption{Illustration of the distribution along the LOS of all the stellar particles of the spatially selected W shelf (see Figure~\ref{fig:modelxy}), grouped into two numbered components by DBSCAN (upper panel, red and black points labelled in the legend). Grey points are stellar particles that are not linked to any group of particles according to the DBSCAN clustering algorithm. The entire spatial distribution of the modelled W shelf stars along the LOS is compared to independent observations in Figure~\ref{fig:ogami_ws}. Middle and lower panel: Same as Figure \ref{fig:3d_gs} but for the W shelf. }
    \label{fig:3d_ws}
\end{figure}

\begin{figure*}[h]
    \centering
    \includegraphics[width=0.9\textwidth]{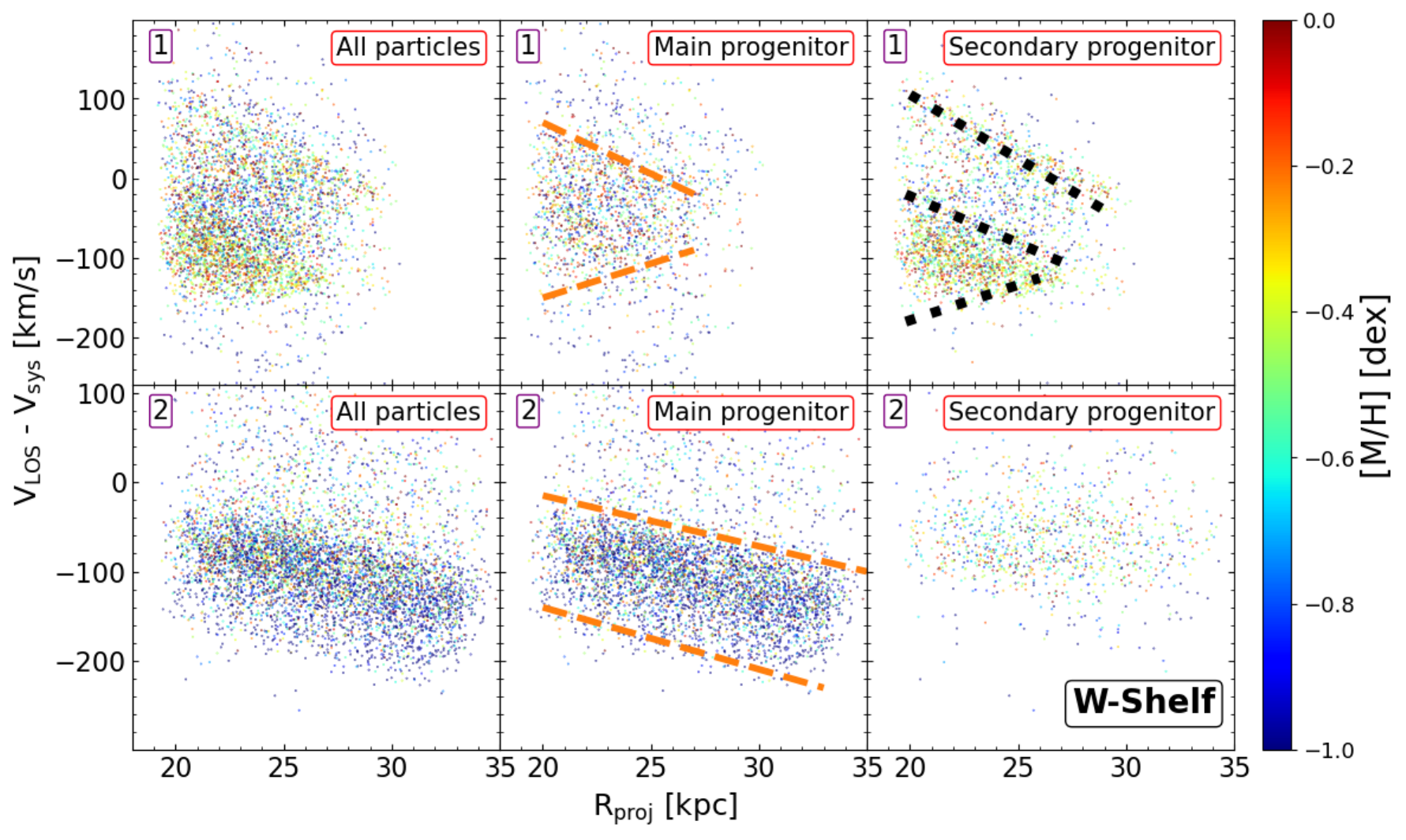}
    \caption{Same as Figure \ref{fig:desi_gs}, but for the W shelf with two components. }
    \label{fig:desi_ws}
\end{figure*}

The W shelf in M31 is a stellar overdensity signaled by an increased surface brightness, embedded in the low-surface brightness, smooth halo. It extends from the western side of M31's disc \citep{fardal2007, Ferguson2016}.  

Using DBSCAN, we identify two clusters in the (X, Y, Z) distribution of particles in the western region of the M31 remnant: W-component~[1] and W-component~[2], shown in Figure~\ref{fig:3d_ws} [upper panel] in different colours. These two clusters have different LOS distances to the MW, with W-component~[1] in the radial range of 770 to 750~kpc, and W-component~[2] in the 750 to 730~kpc range. Both progenitors contribute to these clusters, with the secondary being more prominent in W-component~[1], and the main progenitor in W-component~[2]. In Figure~\ref{fig:3d_ws}, star particles in the middle and lower panels are colour-coded according to their metallicity; their coloured distribution shows that star particles from the secondary in W-component~[1] have a higher metal content than those in W-component~[2] originating from the main progenitor. 

In Figure~\ref{fig:desi_ws}, we show the phase space distributions of the stellar particles in the W shelf components [1] and [2]. Two wedge-like patterns are evident in the W-component [1] (see Figure ~\ref{fig:desi_ws}, upper panels); one is populated by particles from the main progenitor, and another pattern appears in the distribution of particles from the secondary. The two wedges span almost the same projected distance (R$\rm_{proj,max} \simeq$~26~kpc) and velocity ranges. A stream-like patch of stars is also identified in particles from the secondary. The presence of wedges from main and secondary stars and their colour-coded metallicity indicate a large width of the metal abundance distribution for this component as well.

The phase space distribution of the stellar particles in the W-component [2] shows that these star particles (coming from the pre-merger disc) exhibit disc-like kinematics, reaching a negative V$\rm_{LOS}$ consistent with the rotation of the M31 disc on its western side and have lower metallicity than the stellar particles populating W-component~[1]. 

\section{R$\rm_{proj}$ versus V$\rm_{LOS}$ diagrams for the Giant Stellar Stream and the North-East and Western shelves from the 1~$:$~4 mass merger simulation}\label{chemody}

The simulations by \citetalias{hammer} are optimised to reproduce numerous observed properties of M31's disc and the morphology of its inner halo. Our investigation aims at extending the comparison with M31 to the phase space and metallicity distributions of the halo substructures, including the recently acquired chemodynamical data from the wide-field spectroscopic sample of \citetalias{dey}. %within the context of a major merger for the evolution of M31. 

In this comparison, it is important to take into account the principal limitations of the model (see also Appendix~\ref{different kinetic angle}): (i) The \citetalias{hammer} simulation is viewed at a specific snapshot during an ongoing evolution; (ii) while a large number of initial parameters was sampled to optimise previously available observations, the sampling is still coarse and important assumptions had to be made; (iii) the final viewing direction also has remaining uncertainties which influence the $V_{\rm LOS}$ towards the MW; (iv) the chemical model is based on a simplified enrichment scheme during the simulation, as well as on a posteriori comparison with limited data. Since the substructures in the halo of M31 are inherently time-dependent, these uncertainties can have significant effects: features seen in the data may not all be simultaneously found in the model.  Consequently, we would not expect high levels of accuracy in comparing observed phase space features and their metallicity properties quantitatively with similar features in the model. It is therefore remarkable that multiple features seen in the data are reproduced qualitatively, and in some cases even quantitatively.

Over and under densities in the  R$\rm_{proj}$ versus V$\rm_{LOS}$ diagram of the GSS stars were evident from previous spectroscopic surveys \citep{Merrett2006, Gilbert2007, gilbert09, Fardal12,escala2020, escala22, paperVI}, and were already reproduced in simulations with different levels of success (e.g. \citealt{Fardal12, Kirihara2017, milosevic2022, milosevic2024}).  Recently, the DESI survey (\citetalias{dey}) provided a catalogue of $\sim$~7000 sources residing in the inner halo of M31, selected according to criteria shown in their Figure~1, with measured radial velocities and projected distances, along with spectroscopic metallicity estimates. This is the most uniform spectroscopic coverage of the inner halo of M31 to date.

\begin{figure*}[h]
    \centering
    \includegraphics[width=\textwidth]{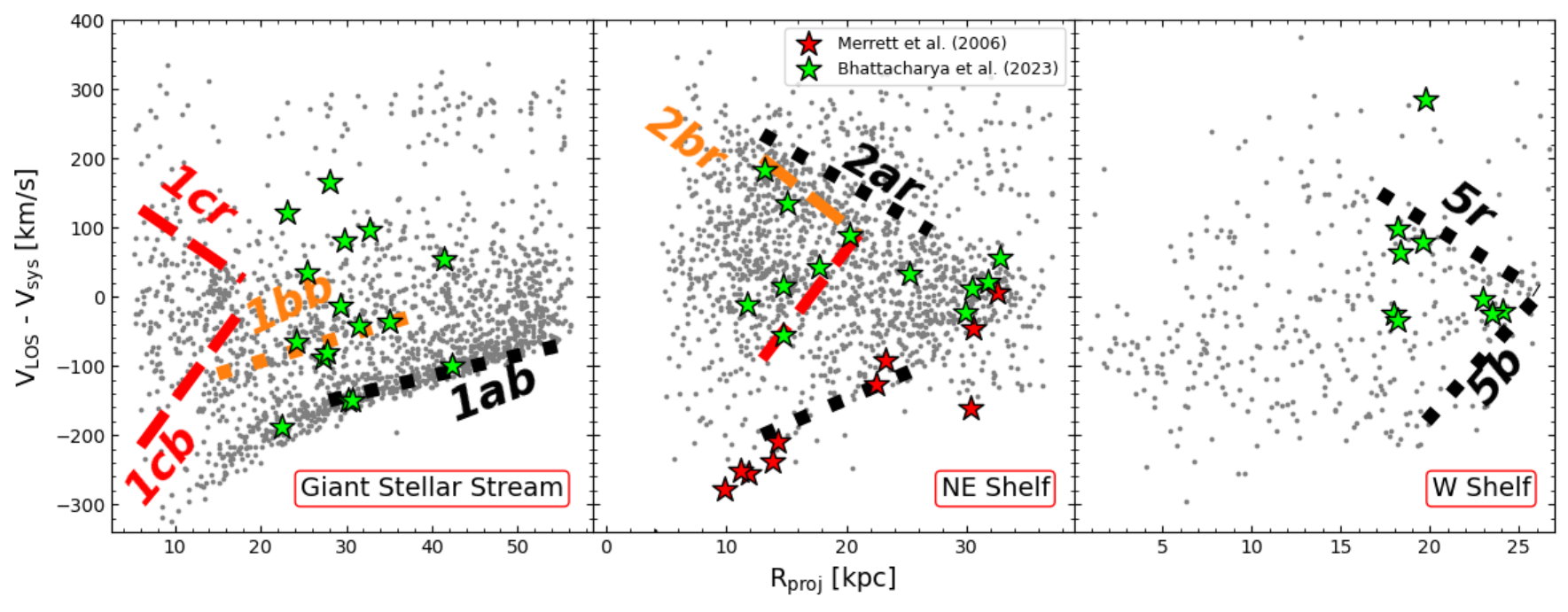}
    \caption{Position versus (V$\rm_{LOS}$ - V$\rm_{sys}$) velocity diagrams for three areas in the DESI survey. Each grey point is a star identified in the DESI survey. Each selected area encloses the substructure denoted by the bottom right label (GSS [left], NE shelf [middle], and W shelf [right]). Linear features (streams, wedges, and chevrons) from DESI are overplotted, with the labels used by \citetalias{dey}. Orange dashed and black dotted lines designate particles from the main and the secondary progenitor, respectively,  of  analogous model features in Figs.~\ref{fig:desi_gs}, \ref{fig:desi_ne}, \ref{fig:desi_ws}. Whenever the same feature appears in the phase space of stars from both progenitors, it is sketched with red-dashed lines. The PNe data from \citet{paperVI} are overplotted for each substructure as green stars. The PNe, classified as outliers by \citet{Merrett2003, Merrett2006}, are overplotted in the NE shelf region as red stars. } 
    \label{fig:desi_phase_space}
\end{figure*}

In Figure~\ref{fig:desi_phase_space}, we reproduce the R$\rm_{proj}$ versus  (V$\rm_{LOS}$~-~V$\rm_{sys}$) distributions from the DESI survey (grey points) extracted in the regions co-spatial with GSS and NE and W shelves. On these distributions, we overplot the R$\rm_{proj}$ and (V$\rm_{LOS}$~-~V$\rm_{sys}$) measurements for the PNe in the same regions as red and green stars, from \citet{Merrett2003, Merrett2006}, and \citet{paperVI}, respectively. To guide the eye, the wedge features already identified and labelled in \citetalias{dey} are reproduced in the plots. They are colour- and line-coded according to similar features in the M31 model specified in Section~\ref{section: the multicomponent nature of substructures in the model}, Figures~\ref{fig:desi_gs}, \ref{fig:desi_ne}, \ref{fig:desi_ws}. Some of the overdensities 
identified here are not labelled or identified in \citetalias{dey}. Specifically, the lower envelopes of the two wedge-like patterns that emerge in the phase space diagram of the NE shelf are not labelled in \citetalias{dey} but are identified as such in this study. We discuss the comparison between the observed and the simulated phase space for each substructure in the inner halo of M31 in turn in the following sections. 

\subsection{Giant Stellar Stream phase space comparison}\label{GSps}
        
In the leftmost panel of Figure~\ref{fig:desi_phase_space}, we examine the DESI RGBs \citepalias{dey} and PNe \citep{paperVI} phase space at the location of the GSS. \citetalias{dey} identified a chevron at a low projected distance of R$_{\rm proj} \leq$~20~kpc (the red-dashed lines  in Figure~\ref{fig:desi_phase_space} labelled ``1cr, 1cb''). A similar chevron structure is evident in the simulated GSS (see the phase space of GSS-component [3] in Figure~\ref{fig:desi_gs}), with the same morphology for stellar particles from both progenitors. The tip of the wedge occurs at R$\rm_{proj} \simeq $~40~kpc in the simulation model.  

Another feature identified in the observed GSS phase space as ‘1ab’ by \citetalias{dey}, was initially reported by \citet{gilbert09} to be a kinematically cold component, that is a stream of stars whose velocity varies between $\sim$~-300~kms$\rm^{-1}$ for R$\rm_{proj}$~$\sim$~10~kpc to $\sim$-100~kms$\rm^{-1}$ for R$\rm_{proj}$~$\sim$~55~kpc.  In Figure~\ref{fig:desi_gs}, we find an elongated wedge pattern in the secondary particles of GSS-component [2], whose lower edge closely resembles the observed feature labelled ‘1ab’ in the DESI phase space.  

In addition, there is the overdensity labelled ‘1bb’ by \citetalias{dey}. We note that ‘1bb’ and ‘1ab’ are the counterparts in the DESI phase space of the GSS KCC \citep{Kalirai2006,Gilbert2007}. When examining the simulated phase space in Figure~\ref{fig:desi_gs},  the lower edges of the distributions of the main and secondary stellar particles in the GSS-component [2] are reminiscent of these features, as already noted in \citet{paperVI}.

We are guided by a similarity in R$\rm_{proj}$ and V$\rm_{LOS}$  ranges to associate GSS-component [2] and [3] with the prominent overdensity features of the DESI phase space diagram for the RGBs in the GSS. Differently, we do not find a matching structure for GSS-component [1] and [4], seen in Figure~\ref{fig:desi_gs}. We acknowledge that GSS-component [1] exhibits low surface brightness; therefore, it may not be detectable. The GSS-component [4] is placed at a particularly large LOS distance and is most affected by projection effects. Furthermore, as discussed in \citetalias{hammer}, a small shift in the initial conditions strongly affects its kinematics\footnote{The specific temporal choice made in model \#336 for the entire z~$=$~0 snapshot is also relevant. If one allows the model to evolve further in time, it leads to the accretion of an additional loop surrounding the post-merger disc, while in reality, it has not reached the galaxy yet.
}.

In summary, we find that wedges, chevrons, and KCCs in the GSS can be produced by several wraps of the merging secondary galaxy in the model. A moderate contribution from the main progenitor stars along the orbit of the merging secondary is also expected. 

\subsection{North-East shelf phase space comparison}\label{ssec:NEps}

We next examine the  R$\rm_{proj}$ versus V$\rm_{LOS}$ diagram of stars in the DESI survey at the location of the NE shelf; see the middle panel of Figure~\ref{fig:desi_phase_space}. 
\citetalias{dey} identified as over-densities the linear structures labelled ‘2br’ and ‘2ar’. We also note that the edge at negative velocities in the radial range of R$\rm_{proj}$ from 10 to 30~kpc was prominent in the NE shelf phase space in Figure~10 of \citet{escala22}, and is populated by the PN outliers in \citet{Merrett2003, Merrett2006}. In Figure~\ref{fig:desi_phase_space} [middle], PNe data from \citet{paperVI} are plotted as green stars, while PNe outliers from \citet{Merrett2003, Merrett2006} are plotted as red stars. 

In Figure~\ref{fig:desi_ne}, the phase space of the simulated NE-component~[2] showed clearly two V-shaped patterns: an inner wedge populated by particles from the main galaxy, and an outer wedge populated by particles from the secondary. The inner chevron of NE-component~[2] is strongly reminiscent of the feature ‘2br’ in \citetalias{dey} (orange-dotted line in Fig.~\ref{fig:desi_phase_space}). The inner wedge in the simulated NE shelf matches the observed one in both the R$\rm_{proj}$ and V$\rm_{LOS}$ positions. 
    
To reproduce the outermost wedge pattern that appears in DESI phase space for the NE-Shelf, with its tip at $\sim$34~kpc, we demarcate on Figure~\ref{fig:desi_phase_space} the upper and lower black dotted lines from secondary particles of the NE-component~[2]. The upper line traces the upper envelope of the observed wedge, while the lower line traces the corresponding observed lower envelope. 

We also note here that the NE shelf phase space predicted from N-body models simulating a minor merger event predicts overdensities in the NE shelf suggestive of an intact core of the secondary progenitor \citep{fardal2007, escala22}. Such overdensities are morphologically very different from the inner wedge associated with ‘2br’ in \citetalias{dey}. The double wedge structure emerges in \citetalias{hammer} simulations at the location of the NE shelf as particles from the main progenitor are pulled by the passage of the secondary. A sufficiently massive object is required. Hence, the agreement with the DESI features of the chevron- and wedge-like overdensities in the projected phase space diagram for the NE shelf gives additional support to the major merger scenario for the origin of the M31 inner halo substructures. 

\subsection{Western shelf phase space comparison}\label{ssec:Wshelfps}

In the DESI phase space at the location of the W shelf (Figure~\ref{fig:desi_phase_space} [right]), the labelled overdensities are ‘5r’ and ‘5b’, delimiting the upper and lower envelopes of the DESI phase space wedge distribution. The linear outer edges in the W-component~[1] in the upper panels of Figure~\ref{fig:desi_ws} are reminiscent of the wedge in the DESI phase space, which extends over the very same radial and velocity ranges. The PN data in the W shelf region from \citet{paperVI} (green stars) also trace the outer contour of the wedge pattern.

We note that while W-component~[2] is clearly evident in Figure~\ref{fig:desi_ws}, it is not visible in the \citetalias{dey} observed phase space (Figure~\ref{fig:desi_phase_space}). This can be understood in terms of target selection bias in the DESI survey which required a metallicity of [M/H] $>$ -0.5~dex (their Figure~1) while the W-component~[2] is more metal poor ([M/H] $<$ -0.6~dex). 

In our model, the wedge at the location of the W shelf is populated by star particles from both main and secondary progenitors. This property has implications for the abundance distribution in the W shelf. This holds also for the GSS and NE shelf, where similar superpositions along the LOS occur. This topic is investigated further in Section~\ref{section: metallicity comparison for everything}. 

As a final remark, we note that PNe seem to trace the densest regions delineating the chevrons and wedge patterns in the NE and W shelves, and in the GSS in Figure~\ref{fig:desi_phase_space}. PNe, being discrete tracers of light and stellar kinematics, allow the overdensities in phase space to be traced, which are the chevrons and wedge patterns in the M31 substructures in this work. A similar property is shown by the PN overdensity tracing the crown in M87, as discussed in \citet{longobardi15}. 

\section{Abundance distributions in the inner halo substructures: Predictions and comparison with observations}\label{section: metallicity comparison for everything}

\begin{table*}
\caption{Median metallicity ([M/H]) values and their standard deviation estimates from \citet{escala2020}, \citet{escala22}, and \citetalias{dey}, and the predicted values for the equivalent regions in the simulated M31 remnant.}
   \centering
    \resizebox{\textwidth}{!}{
    \begin{tabular}{c |c| c| c| c| c| c| c| c| c| c} 
     \hline
     \multirow{3}{*}{Region} & \multirow{3}{*}{\citetalias{dey}} & \multirow{3}{*}{\citet{escala2020, escala22}} & \multicolumn{2}{c|}{Component 1 [C1]} & \multicolumn{2}{c|}{Component 2 [C2]} & \multicolumn{2}{c|}{Component 3 [C3]} & \multicolumn{2}{c}{Component 4 [C4]} \\ 

     \cline{4-11}
     & & & Main & Sec & Main & Sec & Main & Sec & Main & Sec \\
     \hline
     \\
    Giant Stellar Stream & -0.37$\pm$0.72 & -0.52$\pm$0.45 & -0.79$\pm$0.52 & -0.49$\pm$0.31  &  -0.78$\pm$0.59 & -0.61$\pm$0.36 &  -0.56$\pm$0.50 & -0.53$\pm$0.38  & -0.74$\pm$0.37 & -0.60$\pm$0.33 \\ 
    NE Shelf &-0.35$\pm$0.75 & -0.44$\pm$0.37 & -0.65$\pm$0.49 & -0.60$\pm$0.41 & -0.50$\pm$0.48 & -0.45$\pm$0.36 & - & - & - & - \\ 
    W Shelf & -0.43$\pm$0.90 & - & -0.56$\pm$0.51 & -0.45$\pm$0.36 & -0.89$\pm$0.55 & -0.55$\pm$0.39 & - & - & - & -  \\ 
    \\
     \hline
    \end{tabular}
    }
    
    \tablefoot{For \citet{escala2020}, we used Equation 1 from \citet{paperIII} to convert [Fe/H] values into [M/H] taking into account the relative $\alpha$-enhancement within the field. For \citet{escala22} we calculated the average photometric metallicity (and its standard deviation) of the six fields within the NE-Shelf. For modelled stars, the median and the standard deviation values for each component of the substructures are separately presented for main and secondary stellar particles.}  \label{tab:spectro_metallicity_comparison}
\end{table*}

\begin{figure*}[h]
    \centering
    \includegraphics[width=\textwidth]{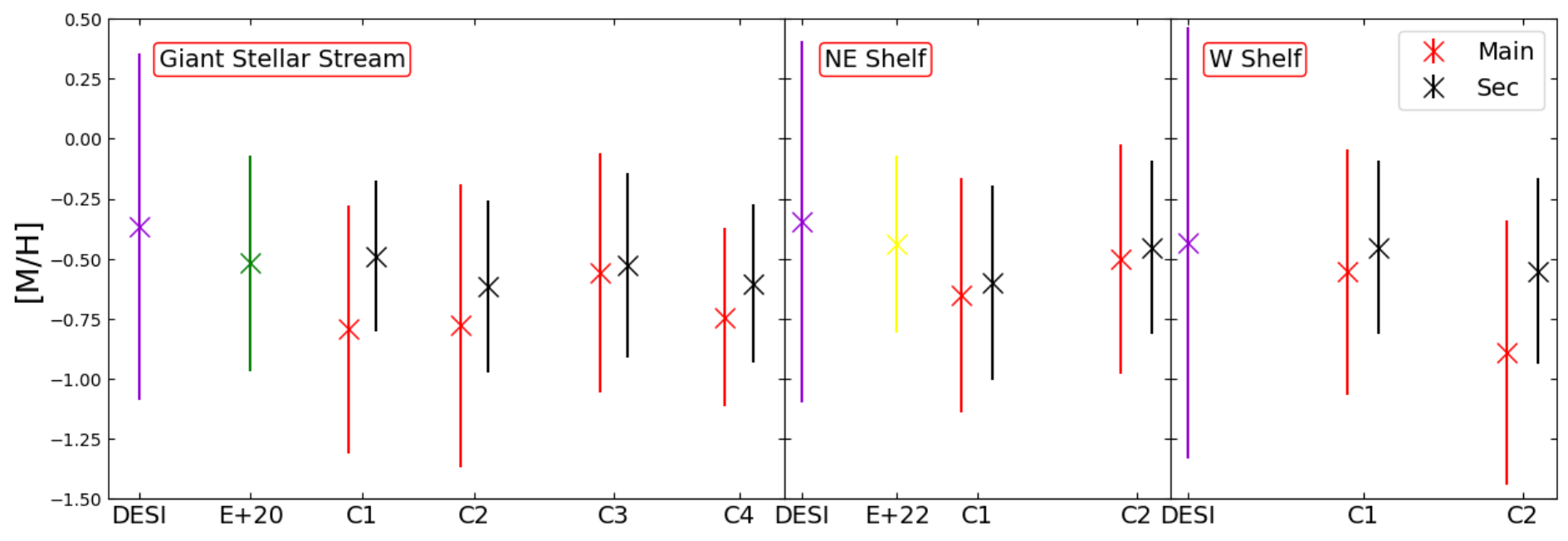}
    \caption{Summary overview of the median metallicity values within GSS, NE and W shelves of the model to be compared with spectroscopic metallicity estimates from the DESI survey, \citealt{escala2020} (for the GSS; labelled as E+20 in the figure), and \citet{escala22} (for the NE shelf; labelled as E+22 in the figure). For each substructure, we report the median values from their components identified in the simulations by DBSCAN and labelled as C1, C2, etc. The error bars denote the standard deviation both for the data distributions and the distributions obtained from the simulation model.}
    \label{fig:desi_metal_compa}
\end{figure*}

We now turn to examining the median metallicity of each component of the three substructures in the model, distinguishing between main and secondary particles. We also compare with the median spectroscopic metallicity of the equivalent region in \citetalias{dey}. For the GSS and the NE shelf, we include in our comparison the metallicity estimates from the SPLASH survey (\citealt{escala2020} and \citealt{escala22}\footnote{\citet{escala22} have acquired spectra for stars in the NE shelf region. However, they estimate the photometric metallicity after they make a kinematic selection to pin down NE shelf stars in order to minimise the contamination from the M31 disc. }, respectively). 
Table~\ref{tab:spectro_metallicity_comparison}  and 
Figure~\ref{fig:desi_metal_compa}
report available spectroscopic metallicity measurements and the corresponding quantities predicted for each component of the simulated M31 inner halo substructures. The error bars denote the standard deviation of the metallicity in the simulation and the observations. 

We note that the selection method for M31 stars implemented in the DESI survey is biased towards redder sources in (g $-$ i)  colour and therefore may primarily sample the metal-rich and older RGB, as well as the redder AGB stars; hence, they do not sample the metal-poor regions well \citepalias{dey}. Consequently, we expect our model-predicted metallicity values to be slightly metal-poorer on average. 

Apart from pencil-beam spectroscopic measurements at specific locations, estimates of the metallicity from photometry for the resolved stellar populations in the GSS were carried out over an extended area. Using a two-fold (stellar magnitude and colour) algorithm implemented on the PAndAS resolved RGB photometric dataset \citep{mcconnachie09nature}, \citet{conn} constrained distances from the MW through the TRGB method and photometric metallicities along the GSS. The measurements by \citet{conn}  show variations in LOS distances along the GSS, a variation in the [M/H] average values along the stream,  with the most metal-rich section at 1~degree ($\simeq$~40~$-$~50~kpc) along the stream, and a significant spread of $\sigma$[Fe/H], up to 1~dex, for the metallicity distribution of the RGB population of GSS stars. 

More recently, \citet{ogami} utilised a novel approach to account for the severe contamination from the MW's stellar halo\footnote{MW dwarf stars occupy the same region in the colour-magnitude diagram as the  M31 giants.} and obtained a sample of $\sim$90\% accuracy in selecting M31 stars. They provide the photometric metallicity distribution, projected distances from the M31 center, and distances from the MW. This survey covers the regions of the GSS, the NE and W shelves. \citet{ogami} confirm the variation of [M/H] along the GSS, and a large $\sigma$[M/H] $\simeq$ 1~dex, which implies the presence of solar and super-solar metallicity stars in the GSS. \citet{ogami} find that stars in the NE and W shelves have average values of their metallicity distribution of $\simeq$ -0.5~dex with $\sigma$[M/H] $\simeq$~0.7$-$0.9~dex, which are also significantly broad distributions. Photometric metallicity estimates for the resolved stellar population in the NE shelf are also reported in \citet{escala22} and \citet{Bernard15}, and for the W shelf in \citet{tanaka2010}.  The metallicity distributions for each substructure are shown in Figures~\ref{fig:ogami_gs}, \ref{fig:ogami_ne}, and \ref{fig:ogami_ws}, and a comparison with observations is reported below.

\subsection{Line-of-sight distance and metallicity distributions for the Giant Stellar Stream}\label{metallicity Comparison for the Giant Stream}

    \begin{figure*}[htp]
    \centering
    \includegraphics[width=\textwidth]{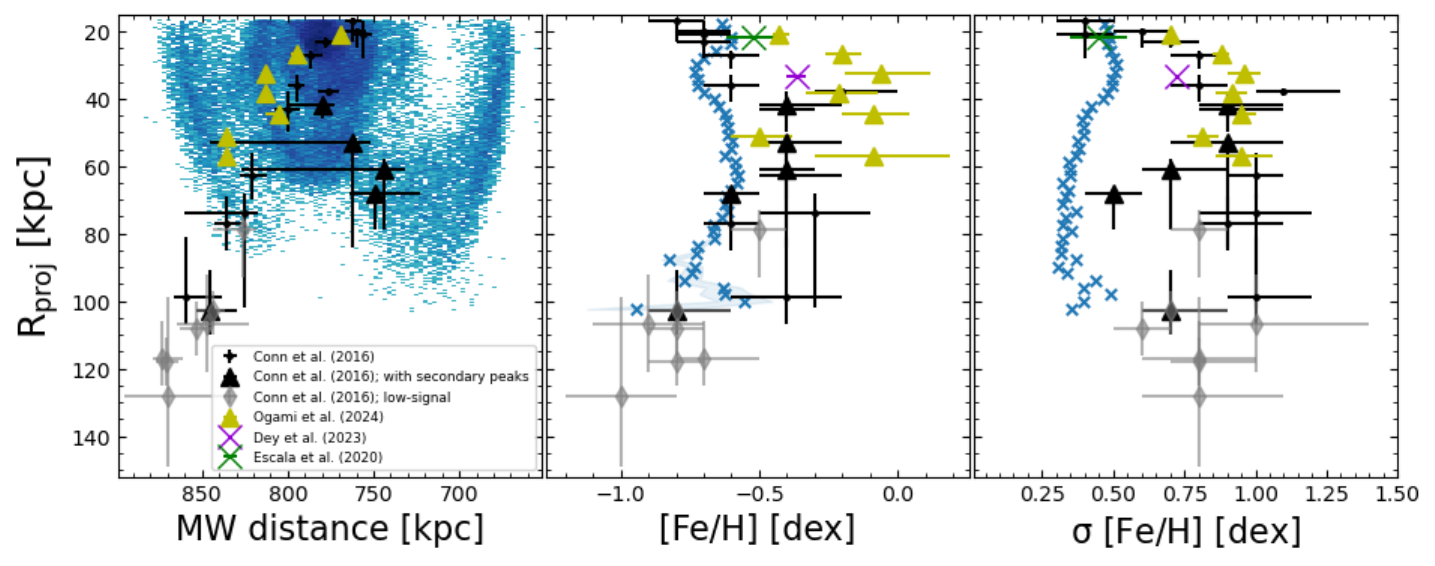}
    \caption{Left panel: M31 R$\rm_{proj}$ and LOS distance for various regions in the GSS, from photometric measurements of \citet{conn} and \citet{ogami}. For \citet{conn} GSS sub-fields, small black circles denote distances and metallicities inferred from the single highest peak of their PDF, black triangles for quantities inferred by taking into account a secondary peak in the PDF, and grey diamonds for the fields with low signal-to-noise and high contamination fraction. 
    \citet{ogami} data are shown as yellow triangles. In blue we show the surface number density of star particles in the model. Middle panel: Comparison of the GSS's average metallicity values from \citet{conn}, \citet{escala2020}, \citetalias{dey} and \citet{ogami} with the equivalent values predicted by the model (indicated by blue crosses), binned in $\sim$2~kpc bins for clarity. The error bars for the [M/H] estimated by \citet{ogami} represent the 68\% uncertainty. \citet{ogami} estimate their total systematic error in the estimated MW distance to be $\sim$27~kpc.  Error bars for the data of \citet{conn} represent their 1$\sigma$ (68.2 percent) uncertainties. \citet{conn} assume a fixed age of 9~Gyr for the GSS, while \citet{ogami}  assume a fixed age of 13~Gyr. The critical assumption of constant age breaks down in the GSS part which is close to the M31 disc (see for example the age spread in the resolved stellar population in \citealt{Bernard15}).
    Right panel: Comparison of the $\sigma$[Fe/H]  for stars in the model with equivalent measurements from observed datasets of \citet{conn}, \citet{escala2020}, \citetalias{dey} and \citet{ogami}.}
    \label{fig:ogami_gs}
    \end{figure*}
    
In Figure~\ref{fig:desi_metal_compa} [left] and Table~\ref{tab:spectro_metallicity_comparison},   we report the metallicity values estimated for the GSS components, from the simulated major merger combined with a well-motivated chemical model. These values are generally consistent with the inferred spectroscopic metallicity data from \citet{escala2020}, while they are slightly metal-poorer than those from DESI, as anticipated based on their selection bias \citepalias{dey}. The widths of the metallicity distributions determined via spectroscopy of RGB stars in the GSS component by \citet{escala2020} and \citetalias{dey} are narrower than those determined photometrically for the RGB population located in the GSS regions. 

In the model, the lower metallicities of GSS-components~[1] (main [M/H]~$\simeq$~-0.8 dex, secondary [M/H]~$\simeq$~-0.5) and [4] (main [M/H]~$\simeq$~-0.74 dex, secondary [M/H]~$\simeq$~-0.6, see Table~\ref{tab:spectro_metallicity_comparison})  are consistent with being populated by stars located in the outermost parts of their progenitors' discs and, thus, with lower metallicity.

In Figure~\ref{fig:ogami_gs}, we show the comparison between our simulation-inferred i) M31 LOS distance versus  R$\rm_{proj}$, ii) mean metallicity [Fe/H]  versus  R$_{\rm proj}$, and iii) standard deviation of the metallicity $\sigma$[Fe/H] versus R$_{\rm proj}$ of stars in the GSS, and compare them with the equivalent quantities inferred from photometric measurements of resolved stellar populations in \citet{conn} and \citet{ogami}\footnote{ The reader should bear in mind that the stellar ages assumed in the independent photometric estimates of the metallicity distribution of the stars in GSS and NE and W shelves are different; \citet{conn} assume a fixed age of 9~Gyr for the GSS, while \citet{ogami} and \citet{escala22} assume a fixed age of 13~Gyr and 12~Gyr respectively. This critical assumption breaks down in the GSS and the other substructures closest to the M31 disc; see for example the age spread of targets in \cite{Bernard15}}. The metallicity values and their standard deviation from \citet{escala2020} and \citetalias{dey} are also included. The modelled stars selected and plotted in Figure~\ref{fig:ogami_gs} included all stars in the spatial selection of the GSS in Figure~\ref{fig:modelxy} along the entire LOS.

\citet{conn} divide the GSS into spatially adjacent sub-fields and estimate the probability distribution function (PDF) for the heliocentric distance, the average photometric metallicity [M/H], and the RGB metallicity spread of each field. For some of the fields that comprise the GSS, they found double peaks in the PDFs. These are independent evidence for several distinct components in the GSS along the LOS. In Figure~\ref{fig:ogami_gs}, the larger, black triangles show the quantities when secondary peaks of the PDFs are present. We further comment that the agreement of the model's estimation of the LOS distance for stars in the GSS is sufficiently good out to an R$\rm_{proj}$~$\sim$~80~kpc. The outermost fields observed by \citet{conn} are plotted in grey, due to the low signal available and a very high contamination fraction ($\sim$90\%) from foreground dwarf stars in the MW halo in those regions. 
Contrary to the \citet{conn} data, the model GSS does not extend to R$\rm_{proj}$~$>$~80~kpc. 
As discussed already, some discrepancies are expected, as the major merger simulation was not tailored to reproduce these observations. We note that \citetalias{hammer} feature an alternative model (\#255) that reproduces the 3D structure of the GSS more closely. This model features a shorter radial distance for the first pericentre passage of the two progenitor galaxies, which defines the time elapsed between two passages. Since GSS comprises tidal debris that is retracted to the gravitational potential of the remnant after the merger, the time between the passages determines its morphology.

An important success of the model is the reproduction of the metallicity trend along the GSS from \citet{conn} observations, which features a kink at a projected distance of $\sim$ 50~$-$~70~kpc. The model also predicts a broad RGB metallicity distribution ($\simeq$~0.3~$-$~0.5~dex). We discuss these in turn below. 

The kink at R$\rm_{proj}\simeq$~40~$-$~50~kpc arises naturally in the major merger simulation as superposition along the LOS of star particles in wedges of different radial extensions from main and secondary stars. Hence, we do not need to assume a steep initial metallicity gradient (see \citealt{milosevic2022, milosevic2024}, where an initial metallicity gradient of -0.3 dex/kpc is adopted for the secondary galaxy). The multiple superposed loops and wedges that build the GSS along the LOS in the simulation here lead to the observed metallicity patterns and curvature as a function of R$\rm_{proj}$.

We also compare the average values of [M/H] and the metallicity spread ($\sigma$[M/H]) for GSS regions in the model with equivalent quantities from \citet{ogami}. Their photometric average metallicities are higher than those of the model and also of the measured values by \citet{conn}. A possible bias in the observed stars is discussed in Appendix~\ref{metallicity young stars}. Appendix~\ref{metallicity young stars} shows that a better agreement with the mean photometric metallicity in \citet{ogami} can be reached by selecting the younger ($<$3.5~Gyr) stellar population of modelled stars. 

In Figure~\ref{fig:ogami_gs} rightmost panel, we show the $\sigma$[M/H] from the RGB width estimated by \citet{conn} (black), \citet{ogami} (yellow), \citet{escala2020} (green), and \citetalias{dey} (purple symbols) compared with the model predictions. While the model predicts a $\sigma$[M/H] in the range 0.3~$-$~0.6, the photometrically inferred $\sigma$[M/H] are larger, in the range 0.5~$-$~1.0 dex.

We note that the super-solar metallicities of the stars in the KCC of the GSS from \citetalias{dey}, the high average metallicity values obtained by \citet{ogami}, and the significant width of the RGB population ($\sim$1 dex) for a large region of the GSS \citep{conn,ogami} all point to a massive progenitor galaxy with an extended star formation history. This is favourable for reaching such high values of stellar metallicity as per the MZR (\citealt{maiolino}) and producing a wide spread in LOS distances and metallicities.

\subsection{Line-of-sight distances and metallicity distribution for the North-East shelf}

\begin{figure}[t]
   %\centering
   \includegraphics[width=0.5\textwidth]{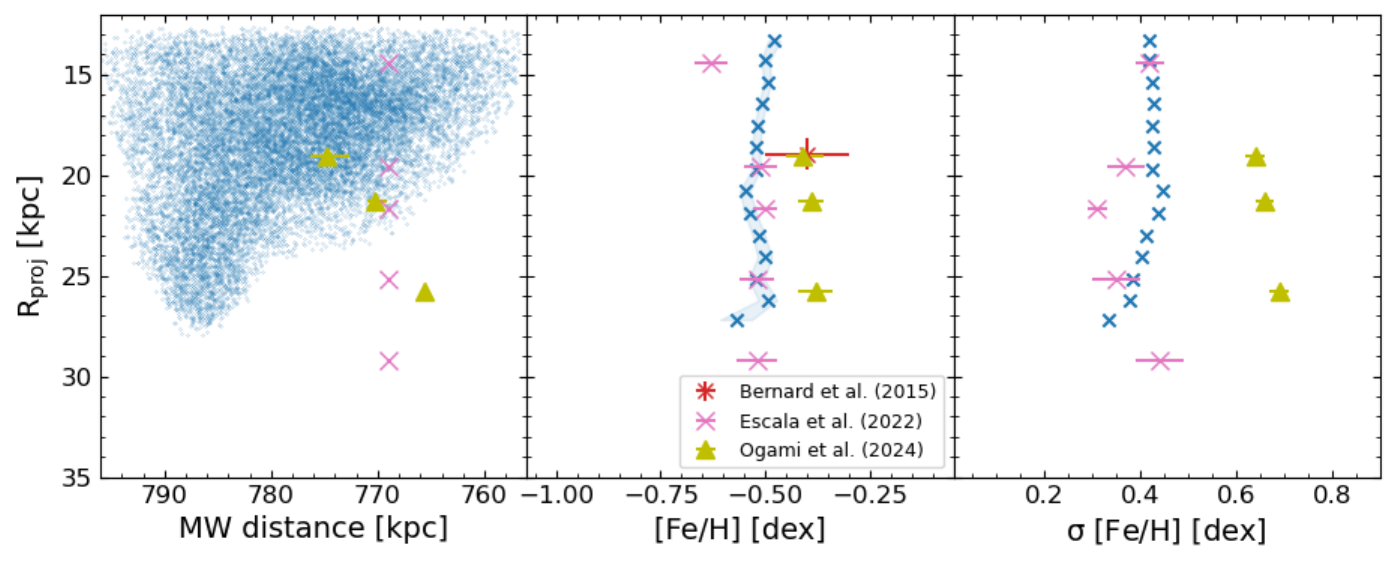}
    \caption{Line-of-sight distribution, average metallicity and metallicity spread for NE shelf component[2], blue symbols. \citet{ogami} data are shown as yellow triangles. The metallicity values and error bars from \citet{escala22} are shown in pink. Whenever they report two components in an observed field, their mean is plotted. The same MW distance of 769~kpc is assumed for all their NE shelf fields. \citet{ogami} and \citet{escala22} assume a fixed age of 13~Gyr and 12~Gyr respectively. This critical assumption breaks down for distances near the M31 disc; see for example the age spread in \cite{Bernard15}. The \citet{Bernard15} measurement is shown in red. For the projected distance, we have taken the average of the two fields for the NE shelf, while the relevant error corresponds to the size of each field.}
    \label{fig:ogami_ne}
\end{figure}

In Figure~\ref{fig:desi_metal_compa} [middle] and Table~\ref{tab:spectro_metallicity_comparison}, we list the estimated metallicity average values for the NE shelf in our simulation model.
We note that for this comparison, the computed [M/H] and $\sigma$[M/H] are evaluated for NE-component~[2] only, since also the spectroscopic and photometric metallicity estimates  (\citealt{Richardson2008}, \citealt{Bernard15}, \citealt{paperIII}, \citealt{escala22}) are derived for the shelf population only, while the contributions of the M31 disc plus halo, and MW halo are removed in the observations. As expected, the average metallicity values from the model are slightly metal-poorer than \citetalias{dey}'s spectroscopic metallicity values; see Figure~\ref{fig:desi_metal_compa} [middle].

In Figure~\ref{fig:ogami_ne},  we show the surface number density of stellar particles versus LOS distances [left], the average metallicity values in our model binned in 2~kpc bins [middle], and the $\sigma$[Fe/H] [right]. In these plots, we also show the photometric metallicity, LOS distances, and metallicity width estimates from \citet{ogami}, \citet{Bernard15}, and \citet{escala22}.

In our model, the broad RGB width arises naturally from the superposition along the LOS of wedges in the NE shelf phase space, which are populated by main and secondary star particles (see NE shelf component[2] in Figure~\ref{fig:desi_ne}).
The $\sigma$[Fe/H] of the model is in excellent agreement with the equivalent values for the NE shelf as measured by \citet{escala22}. Instead \citet{ogami} obtain a higher [M/H] average and larger $\sigma$[Fe/H] values, possibly from the contribution of other M31 components (e.g. the disc or the halo) superposed on the NE shelf.

\subsection{Line-of-sight distances and metallicity distribution for the Western shelf}

\begin{figure}[t]
   \includegraphics[width=0.5\textwidth]{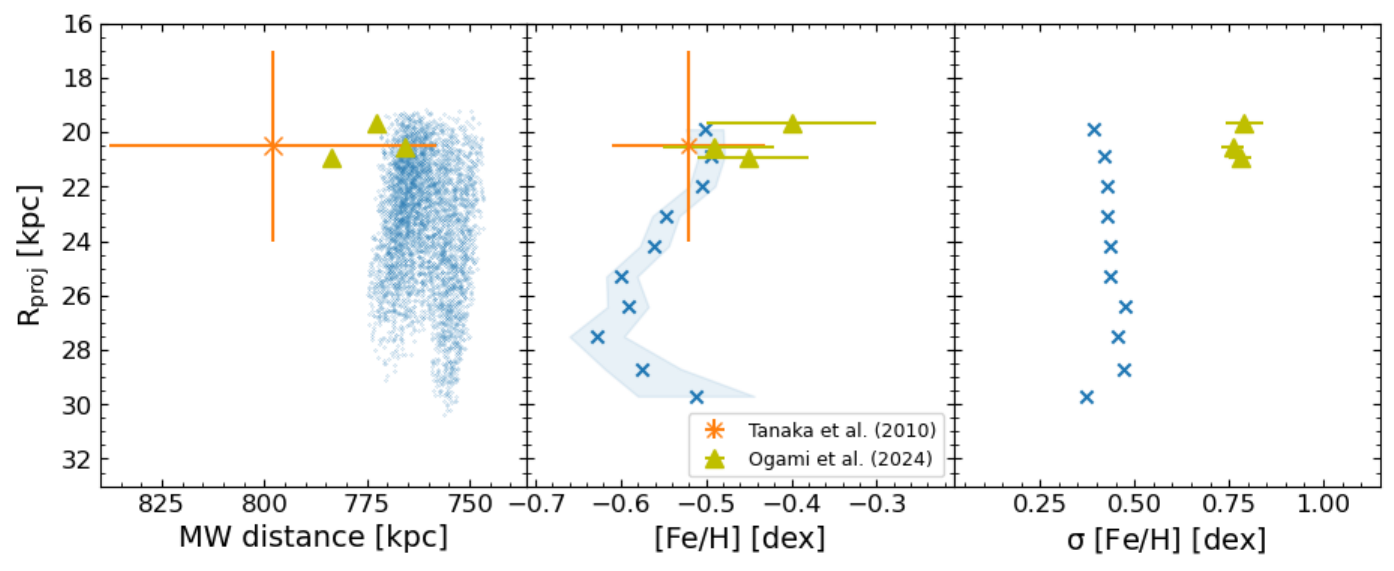}
    \caption{Line-of-sight distribution, average metallicity and metallicity spread for W shelf component[1], blue symbols. \citet{ogami} data are shown as yellow triangles. The plotted [M/H] and its relevant error from \citet{tanaka2010} are obtained from their measured $\alpha$ and [Fe/H] abundances as described in \citet{paperIII}; they are presented in orange in the figures. For their R$_{\rm proj}$ we estimated the average of the projected distance of their surveyed fields, and the error was calculated as the distance of this average from the field edges.}
    \label{fig:ogami_ws}
\end{figure}

In Figure~\ref{fig:desi_metal_compa} [right] and Table~\ref{tab:spectro_metallicity_comparison}, we show that our simulated metallicity values for the W-component [1] are consistent with the spectroscopic metallicity reported by the DESI survey for this substructure. 

In Figure ~\ref{fig:ogami_ws}, we also report the predicted average [M/H] and RGB width ($\sigma$[M/H]) values for the simulated stellar particles for the W shelf component[1] in Figure~\ref{fig:desi_ws}, together with the metallicity average values,  $\sigma$[M/H], and distance estimates from \citet{tanaka2010} and \citet{ogami}.  LOS distances and average [M/H] values agree very well with the observed values. The $\sigma$[M/H] values reported in \citet{ogami} are $\simeq$~0.8~dex while in the simulation model, we obtain 0.3~$-$~0.5~dex. The larger spread in \citet{ogami} may be related to superposed contributions of other M31 components like the disc and the halo, in addition to W shelf stars. 
 
\section{Conclusions} \label{chpter: conclusions}

The giant Andromeda galaxy offers a unique opportunity for a close-up study of the hierarchical mass assembly of galaxies predicted by the $\Lambda$CDM model \citep{White1978, Bullock&Johnston2005}.

We utilised an N-body hydrodynamical simulation of a major merger between two disc galaxies with a mass ratio of 1~$:$~4 and combined it with a well-motivated chemical model to determine the chemodynamical properties of the M31 merger remnant. The combined model shows (i) a high metallicity and a large spread in the GSS and NE and W shelves, thus explaining the various photometric and spectroscopic measurements; (ii)  a large distance spread in the GSS suggested by previous TRGB measurements; and (iii) phase space ridges, wedges and chevrons for all three substructures generated by several distinct pericentre passages of the secondary progenitor together with up-scattered main M31 disc stars, that also have plausible counterparts in the observed phase spaces.

These comparisons provide further independent and strong arguments for a major satellite merger in M31 $\sim$3~Gyr ago and a coherent explanation for many observational results that make M31 appear so different from the MW. The wide spread in metallicities and LOS distances of the stars comprising the GSS serves as evidence for a complex and extended star formation history in the satellite galaxy and is suggestive of a relatively massive galaxy.  

We compared the emerging chemodynamical properties of the substructures in the inner halo of the M31 simulated analogue, the  GSS and the two shelves, with the available data in the literature. Our main results are as follows: 

\begin{itemize}
          
    \item We observed a multi-component nature for the NE and W shelves and especially for the GSS. In a major merger, the GSS contains multiple overlapping components along the LOS stemming from consecutive wraps of the satellite. The stellar kinematics of the GSS in the merger remnant are consistent with observations initially reported by \citet{Kalirai2006} and \citet{Gilbert2007, gilbert09} and even more so with the results from the wide-field coverage of the inner halo by the DESI survey \citepalias{dey}.

     \item The  R$\rm_{proj}$ versus V$\rm_{LOS}$ diagram for each of the three main substructures in the inner halo of the modelled M31 reveals coherent wedges, chevrons, and stream-like patterns. These features are echoed in the corresponding phase space of resolved stars observed by \citetalias{dey}. The two distinct wedge patterns that appear in the phase space of the observed NE and W shelves stars also emerge for modelled stars within the equivalent regions. This two-fold wedge pattern (with a different apocentre for each wedge) does not appear in minor merger models (e.g. \citealt{fardal2007, Kirihara2017, milosevic2024}). This is justified since a sufficiently massive secondary galaxy is required to drag stellar particles from the pre-merger disc and deposit them in tidal shells.

     \item The metallicity gradient reported by \citet{conn} and \citet{ogami} along the GSS displaying an increment of the metallicity at 40~$-$~50 kpc distance for the GSS is qualitatively reproduced by the model.  The model predicts a spread of $\sigma$[M/H] in the range of 0.3 $-$ 0.5~dex which is consistent with the widths of metallicity distributions from spectroscopic measurements but narrower than $\sigma$[M/H] values inferred from photometric metallicities (\citealt{conn}; \citealt{ogami}).

    \item The major merger model showcases average [M/H] values for the M31 inner halo substructures that are in agreement with spectroscopic metallicity observations (\citealt{escala2020}; \citetalias{dey}). Photometrically estimated \citep{ogami, conn, Bernard15, escala22, tanaka2010} LOS distances, projected distances, and metallicities of stars within the three inner halo substructures are also echoed by the model. In addition, the simulation revealed the effect of the stellar age distribution, which is not accounted for in the current measurements of the metallicity from colour-magnitude diagrams of resolved stellar populations. A known bias of photometric surveys is to select the brightest stars within the GSS and the rest of the substructures, which lie close to the TRGB and thus tend to be younger and more metal rich. The basic assumption of a single age for stellar populations of the satellite galaxy may not be applicable for a massive satellite with an extended star formation history. In particular, the assumption of a single age for the resolved stellar population may break down for the substructures closest to the M31 disc. 
  \end{itemize}

\noindent
Apart from the reproduction of various traits of the M31 stellar halo, the gas-rich major merger model features a reconstructed stellar disc. We obtained the following results for the M31 disc remnant:
\begin{itemize}

     \item The model qualitatively reproduces the observed metallicity gradient of the M31 disc. The resulting gradients for the two age groups of modelled stars, younger and older than 3.5~Gyr, bracket the metallicity gradients measured for the thin and thicker disc stellar populations.
     \item The assessment of the radial redistribution of disc stars revealed that the majority of young stars in the model initially came from the gas particles residing in the outskirts of the progenitors. The initially negative (-0.1 dex/kpc) metallicity gradient led to their lower metal content. Hence, the model provides a viable explanation for the moderately metal-poor young stars in the outer disc of M31 (see \citealt{Bernard15} and \citealt{arna22}).
     \item Future surveys with samples that effectively account  for MW contamination without introducing  metallicity and ages biases - unlike the selection implemented by \citetalias{dey} to reduce the contamination from MW halo stars - will reveal the younger metal poor outer disc in the W shelf and in the other substructure fields closer to the M31 disc.
\end{itemize}

Upcoming facilities and instrumentation will focus even more on M31, which is the ideal testbed for resolved stellar population studies of galaxies affected by a recent major merger. The inauguration of facilities with a wide field of view, such as the Roman Space Telescope (see \citealt{Dey2023romandromeda}), and instrumentation, including the Prime Focus Spectrograph \citep{Subarupfs} on the Subaru Telescope will also allow for resolved stellar population studies in galaxies outside the Local Group. Since mergers are regular events in the evolution of galaxies, the extensive study and juxtaposition against simulated analogues of a major merger remnant such as M31 may be helpful as a guideline for dealing with the upcoming data sets and future modelling efforts.

\begin{acknowledgements}
      The authors wish to thank the referee, Prof. Masao Mori, for his useful comments. CT was supported by COST (European Cooperation in Science and Technology) Action CA18104 - Revealing the Milky Way with Gaia (MW-GAIA) during his visit at the Paris Observatory. CT wants to thank the European Southern Observatory (ESO) for the opportunity to visit ESO, Garching, Germany under the Early-Career Scientific Visitor scheme. CT wants to thank IUCAA and especially Prof. Kanak Saha for the hospitality during his 6-month visit to its premises. CT acknowledges the use of the High-Performance Computing facility Pegasus at IUCAA, Pune. SB is funded by the INSPIRE Faculty Award (DST/INSPIRE/04/2020/002224), Department of Science and Technology (DST), Government of India. SB and MAR thank ESO, Garching, Germany for supporting SB's visits through the 2021 and 2023 SSDFs. This research was co-supported by the Excellence Cluster ORIGINS which is funded by the Deutsche Forschungsgemeinschaft (DFG, German Research Foundation) under Germany´s Excellence Strategy – EXC-2094 – 390783311. This work was co-supported by the DAAD under the Australia-Germany joint research program with funds from the German Federal Ministry for Education and Research. The analysis made use of the following packages: NumPy \citep{numpy}, Pandas \citep{pandas}, and Matplotlib \citep{matplotlib}. 
\end{acknowledgements}
\bibliographystyle{aa} 
\bibliography{ref.bib}

\begin{thebibliography}{108}
\expandafter\ifx\csname natexlab\endcsname\relax\def\natexlab#1{#1}\fi

\bibitem[{{Amorisco}(2017)}]{Amorisco2017}
{Amorisco}, N.~C. 2017, \mnras, 464, 2882

\bibitem[{{Arnaboldi} {et~al.}(2022){Arnaboldi}, {Bhattacharya}, {Gerhard}, {Kobayashi}, {Freeman}, {Caldwell}, {Hartke}, {McConnachie}, \& {Guhathakurta}}]{arna22}
{Arnaboldi}, M., {Bhattacharya}, S., {Gerhard}, O., {et~al.} 2022, \aap, 666, A109

\bibitem[{{Asplund} {et~al.}(2009){Asplund}, {Grevesse}, {Sauval}, \& {Scott}}]{Asplund2009}
{Asplund}, M., {Grevesse}, N., {Sauval}, A.~J., \& {Scott}, P. 2009, \araa, 47, 481

\bibitem[{{Bernard} {et~al.}(2012){Bernard}, {Ferguson}, {Barker}, {Hidalgo}, {Ibata}, {Irwin}, {Lewis}, {McConnachie}, {Monelli}, \& {Chapman}}]{bernard2012}
{Bernard}, E.~J., {Ferguson}, A. M.~N., {Barker}, M.~K., {et~al.} 2012, \mnras, 420, 2625

\bibitem[{{Bernard} {et~al.}(2015){Bernard}, {Ferguson}, {Richardson}, {Irwin}, {Barker}, {Hidalgo}, {Aparicio}, {Chapman}, {Ibata}, {Lewis}, {McConnachie}, \& {Tanvir}}]{Bernard15}
{Bernard}, E.~J., {Ferguson}, A. M.~N., {Richardson}, J.~C., {et~al.} 2015, \mnras, 446, 2789

\bibitem[{{Bhattacharya}(2023)}]{Bhattacharya2023arXiv230503293B}
{Bhattacharya}, S. 2023, arXiv e-prints, arXiv:2305.03293

\bibitem[{{Bhattacharya} {et~al.}(2019{\natexlab{a}}){Bhattacharya}, {Arnaboldi}, {Caldwell}, {Gerhard}, {Bla{\~n}a}, {McConnachie}, {Hartke}, {Guhathakurta}, {Pulsoni}, \& {Freeman}}]{papaerII}
{Bhattacharya}, S., {Arnaboldi}, M., {Caldwell}, N., {et~al.} 2019{\natexlab{a}}, \aap, 631, A56

\bibitem[{{Bhattacharya} {et~al.}(2022){Bhattacharya}, {Arnaboldi}, {Caldwell}, {Gerhard}, {Kobayashi}, {Hartke}, {Freeman}, {McConnachie}, \& {Guhathakurta}}]{paperiv}
{Bhattacharya}, S., {Arnaboldi}, M., {Caldwell}, N., {et~al.} 2022, \mnras, 517, 2343

\bibitem[{{Bhattacharya} {et~al.}(2021){Bhattacharya}, {Arnaboldi}, {Gerhard}, {McConnachie}, {Caldwell}, {Hartke}, \& {Freeman}}]{paperIII}
{Bhattacharya}, S., {Arnaboldi}, M., {Gerhard}, O., {et~al.} 2021, \aap, 647, A130

\bibitem[{{Bhattacharya} {et~al.}(2023){Bhattacharya}, {Arnaboldi}, {Hammer}, {Yang}, {Gerhard}, {Caldwell}, \& {Freeman}}]{paperVI}
{Bhattacharya}, S., {Arnaboldi}, M., {Hammer}, F., {et~al.} 2023, \mnras, 522, 6010

\bibitem[{{Bhattacharya} {et~al.}(2019{\natexlab{b}}){Bhattacharya}, {Arnaboldi}, {Hartke}, {Gerhard}, {Comte}, {McConnachie}, \& {Caldwell}}]{paperI}
{Bhattacharya}, S., {Arnaboldi}, M., {Hartke}, J., {et~al.} 2019{\natexlab{b}}, \aap, 624, A132

\bibitem[{{Bla{\~n}a D{\'\i}az} {et~al.}(2018){Bla{\~n}a D{\'\i}az}, {Gerhard}, {Wegg}, {Portail}, {Opitsch}, {Saglia}, {Fabricius}, {Erwin}, \& {Bender}}]{BlanaDiaz2018}
{Bla{\~n}a D{\'\i}az}, M., {Gerhard}, O., {Wegg}, C., {et~al.} 2018, \mnras, 481, 3210

\bibitem[{{Brook} {et~al.}(2004){Brook}, {Kawata}, {Gibson}, \& {Freeman}}]{Brook2004}
{Brook}, C.~B., {Kawata}, D., {Gibson}, B.~K., \& {Freeman}, K.~C. 2004, \apj, 612, 894

\bibitem[{{Brown} {et~al.}(2007){Brown}, {Smith}, {Ferguson}, {Guhathakurta}, {Kalirai}, {Rich}, {Renzini}, {Sweigart}, {Reitzel}, {Gilbert}, \& {Geha}}]{brown2007}
{Brown}, T.~M., {Smith}, E., {Ferguson}, H.~C., {et~al.} 2007, \apjl, 658, L95

\bibitem[{{Brown} {et~al.}(2006){Brown}, {Smith}, {Ferguson}, {Rich}, {Guhathakurta}, {Renzini}, {Sweigart}, \& {Kimble}}]{brown2006}
{Brown}, T.~M., {Smith}, E., {Ferguson}, H.~C., {et~al.} 2006, \apj, 652, 323

\bibitem[{{Bullock} \& {Johnston}(2005)}]{Bullock&Johnston2005}
{Bullock}, J.~S. \& {Johnston}, K.~V. 2005, \apj, 635, 931

\bibitem[{{Cohen} {et~al.}(2018){Cohen}, {Kalirai}, {Gilbert}, {Guhathakurta}, {Peeples}, {Lehner}, {Brown}, {Bianchi}, {Barger}, \& {O'Meara}}]{Cohen2018}
{Cohen}, R.~E., {Kalirai}, J.~S., {Gilbert}, K.~M., {et~al.} 2018, \aj, 156, 230

\bibitem[{{Conn} {et~al.}(2016){Conn}, {McMonigal}, {Bate}, {Lewis}, {Ibata}, {Martin}, {McConnachie}, {Ferguson}, {Irwin}, {Elahi}, {Venn}, \& {Mackey}}]{conn}
{Conn}, A.~R., {McMonigal}, B., {Bate}, N.~F., {et~al.} 2016, \mnras, 458, 3282

\bibitem[{{Cox} {et~al.}(2006){Cox}, {Jonsson}, {Primack}, \& {Somerville}}]{cox}
{Cox}, T.~J., {Jonsson}, P., {Primack}, J.~R., \& {Somerville}, R.~S. 2006, \mnras, 373, 1013

\bibitem[{{Curti} {et~al.}(2020{\natexlab{a}}){Curti}, {Maiolino}, {Cirasuolo}, {Mannucci}, {Williams}, {Auger}, {Mercurio}, {Hayden-Pawson}, {Cresci}, {Marconi}, {Belfiore}, {Cappellari}, {Cicone}, {Cullen}, {Meneghetti}, {Ota}, {Peng}, {Pettini}, {Swinbank}, \& {Troncoso}}]{curtigrad}
{Curti}, M., {Maiolino}, R., {Cirasuolo}, M., {et~al.} 2020{\natexlab{a}}, \mnras, 492, 821

\bibitem[{{Curti} {et~al.}(2020{\natexlab{b}}){Curti}, {Mannucci}, {Cresci}, \& {Maiolino}}]{curti}
{Curti}, M., {Mannucci}, F., {Cresci}, G., \& {Maiolino}, R. 2020{\natexlab{b}}, \mnras, 491, 944

\bibitem[{{Dalcanton} {et~al.}(2015){Dalcanton}, {Fouesneau}, {Hogg}, {Lang}, {Leroy}, {Gordon}, {Sandstrom}, {Weisz}, {Williams}, {Bell}, {Dong}, {Gilbert}, {Gouliermis}, {Guhathakurta}, {Lauer}, {Schruba}, {Seth}, \& {Skillman}}]{Dalcanton15}
{Dalcanton}, J.~J., {Fouesneau}, M., {Hogg}, D.~W., {et~al.} 2015, \apj, 814, 3

\bibitem[{{Dalcanton} {et~al.}(2012){Dalcanton}, {Williams}, {Lang}, {Lauer}, {Kalirai}, {Seth}, {Dolphin}, {Rosenfield}, {Weisz}, {Bell}, {Bianchi}, {Boyer}, {Caldwell}, {Dong}, {Dorman}, {Gilbert}, {Girardi}, {Gogarten}, {Gordon}, {Guhathakurta}, {Hodge}, {Holtzman}, {Johnson}, {Larsen}, {Lewis}, {Melbourne}, {Olsen}, {Rix}, {Rosema}, {Saha}, {Sarajedini}, {Skillman}, \& {Stanek}}]{Dalcanton2012}
{Dalcanton}, J.~J., {Williams}, B.~F., {Lang}, D., {et~al.} 2012, \apjs, 200, 18

\bibitem[{{Dey} {et~al.}(2023{\natexlab{a}}){Dey}, {Najita}, {Filion}, {Han}, {Pearson}, {Wyse}, {Thob}, {Anguiano}, {Apfel}, {Arnaboldi}, {Bell}, {Beraldo e Silva}, {Besla}, {Bhattacharya}, {Bhattacharya}, {Chandra}, {Choi}, {Collins}, {Cunningham}, {Dalcanton}, {Escala}, {Foote}, {Ferguson}, {Gibson}, {Gnedin}, {Guhathakurta}, {Hawkins}, {Horta}, {Ibata}, {Kallivayalil}, {Koch}, {Koposov}, {Lewis}, {Macri}, {McKinnon}, {Nidever}, {Olsen}, {Patel}, {Petersen}, {Petric}, {Price-Whelan}, {Rich}, {Riley}, {Saha}, {Sanderson}, {Sharma}, {Sohn}, {Soraisam}, {Steinmetz}, {Valluri}, {Vivas}, {Williams}, \& {Wojno}}]{Dey2023romandromeda}
{Dey}, A., {Najita}, J., {Filion}, C., {et~al.} 2023{\natexlab{a}}, arXiv e-prints, arXiv:2306.12302

\bibitem[{{Dey} {et~al.}(2023{\natexlab{b}}){Dey}, {Najita}, {Koposov}, {Josephy-Zack}, {Maxemin}, {Bell}, {Poppett}, {Patel}, {Beraldo e Silva}, {Raichoor}, {Schlegel}, {Lang}, {Meisner}, {Myers}, {Aguilar}, {Ahlen}, {Allende Prieto}, {Brooks}, {Cooper}, {Dawson}, {de la Macorra}, {Doel}, {Font-Ribera}, {Garc{\'\i}a-Bellido}, {Gontcho A Gontcho}, {Guy}, {Honscheid}, {Kehoe}, {Kisner}, {Kremin}, {Landriau}, {Le Guillou}, {Levi}, {Li}, {Martini}, {Miquel}, {Moustakas}, {Nie}, {Palanque-Delabrouille}, {Prada}, {Schlafly}, {Sharples}, {Tarl{\'e}}, {Ting}, {Tyas}, {Valluri}, {Wechsler}, \& {Zou}}]{dey}
{Dey}, A., {Najita}, J.~R., {Koposov}, S.~E., {et~al.} 2023{\natexlab{b}}, \apj, 944, 1

\bibitem[{{Dorman} {et~al.}(2015){Dorman}, {Guhathakurta}, {Seth}, {Weisz}, {Bell}, {Dalcanton}, {Gilbert}, {Hamren}, {Lewis}, {Skillman}, {Toloba}, \& {Williams}}]{dorman}
{Dorman}, C.~E., {Guhathakurta}, P., {Seth}, A.~C., {et~al.} 2015, \apj, 803, 24

\bibitem[{{D'Souza} \& {Bell}(2021)}]{Souza2021}
{D'Souza}, R. \& {Bell}, E.~F. 2021, \mnras, 504, 5270

\bibitem[{{Escala} {et~al.}(2022){Escala}, {Gilbert}, {Fardal}, {Guhathakurta}, {Sanderson}, {Kalirai}, \& {Mobasher}}]{escala22}
{Escala}, I., {Gilbert}, K.~M., {Fardal}, M., {et~al.} 2022, \aj, 164, 20

\bibitem[{{Escala} {et~al.}(2020){Escala}, {Gilbert}, {Kirby}, {Wojno}, {Cunningham}, \& {Guhathakurta}}]{escala2020}
{Escala}, I., {Gilbert}, K.~M., {Kirby}, E.~N., {et~al.} 2020, \apj, 889, 177

\bibitem[{{Escala} {et~al.}(2021){Escala}, {Gilbert}, {Wojno}, {Kirby}, \& {Guhathakurta}}]{escala21}
{Escala}, I., {Gilbert}, K.~M., {Wojno}, J., {Kirby}, E.~N., \& {Guhathakurta}, P. 2021, \aj, 162, 45

\bibitem[{{Escala} {et~al.}(2019){Escala}, {Kirby}, {Gilbert}, {Cunningham}, \& {Wojno}}]{Escala2019}
{Escala}, I., {Kirby}, E.~N., {Gilbert}, K.~M., {Cunningham}, E.~C., \& {Wojno}, J. 2019, \apj, 878, 42

\bibitem[{{Fardal} {et~al.}(2006){Fardal}, {Babul}, {Geehan}, \& {Guhathakurta}}]{fardal2006}
{Fardal}, M.~A., {Babul}, A., {Geehan}, J.~J., \& {Guhathakurta}, P. 2006, \mnras, 366, 1012

\bibitem[{{Fardal} {et~al.}(2008){Fardal}, {Babul}, {Guhathakurta}, {Gilbert}, \& {Dodge}}]{fardal2008}
{Fardal}, M.~A., {Babul}, A., {Guhathakurta}, P., {Gilbert}, K.~M., \& {Dodge}, C. 2008, \apjl, 682, L33

\bibitem[{{Fardal} {et~al.}(2007){Fardal}, {Guhathakurta}, {Babul}, \& {McConnachie}}]{fardal2007}
{Fardal}, M.~A., {Guhathakurta}, P., {Babul}, A., \& {McConnachie}, A.~W. 2007, \mnras, 380, 15

\bibitem[{{Fardal} {et~al.}(2012){Fardal}, {Guhathakurta}, {Gilbert}, {Tollerud}, {Kalirai}, {Tanaka}, {Beaton}, {Chiba}, {Komiyama}, \& {Iye}}]{Fardal12}
{Fardal}, M.~A., {Guhathakurta}, P., {Gilbert}, K.~M., {et~al.} 2012, \mnras, 423, 3134

\bibitem[{{Fardal} {et~al.}(2013){Fardal}, {Weinberg}, {Babul}, {Irwin}, {Guhathakurta}, {Gilbert}, {Ferguson}, {Ibata}, {Lewis}, {Tanvir}, \& {Huxor}}]{fardal2013}
{Fardal}, M.~A., {Weinberg}, M.~D., {Babul}, A., {et~al.} 2013, \mnras, 434, 2779

\bibitem[{{Ferguson} {et~al.}(2002){Ferguson}, {Irwin}, {Ibata}, {Lewis}, \& {Tanvir}}]{ferguson2002}
{Ferguson}, A. M.~N., {Irwin}, M.~J., {Ibata}, R.~A., {Lewis}, G.~F., \& {Tanvir}, N.~R. 2002, \aj, 124, 1452

\bibitem[{Ferguson \& Mackey(2016)}]{Ferguson2016}
Ferguson, A. M.~N. \& Mackey, A.~D. 2016, Substructure and Tidal Streams in the Andromeda Galaxy and its Satellites, ed. H.~J. Newberg \& J.~L. Carlin (Cham: Springer International Publishing), 191--217

\bibitem[{{Font} {et~al.}(2006){Font}, {Johnston}, {Guhathakurta}, {Majewski}, \& {Rich}}]{font2006}
{Font}, A.~S., {Johnston}, K.~V., {Guhathakurta}, P., {Majewski}, S.~R., \& {Rich}, R.~M. 2006, \aj, 131, 1436

\bibitem[{{Gajda} {et~al.}(2021){Gajda}, {Gerhard}, {Bla{\~n}a}, {Zhu}, {Shen}, {Saglia}, \& {Bender}}]{Gajda2021}
{Gajda}, G., {Gerhard}, O., {Bla{\~n}a}, M., {et~al.} 2021, \aap, 647, A131

\bibitem[{{Garcia} {et~al.}(2025){Garcia}, {Torrey}, {Bhagwat}, {Wright}, {Chen}, {Grasha}, {Ridolfo}, {Hemler}, {Sarkar}, {Chakraborty}, {Nelson}, {Sanders}, {Costa}, {Vogelsberger}, {Kewley}, {Ellison}, \& {Hernquist}}]{Garcia}
{Garcia}, A.~M., {Torrey}, P., {Bhagwat}, A., {et~al.} 2025, arXiv e-prints, arXiv:2503.03804

\bibitem[{{Geehan} {et~al.}(2006){Geehan}, {Fardal}, {Babul}, \& {Guhathakurta}}]{Geehan}
{Geehan}, J.~J., {Fardal}, M.~A., {Babul}, A., \& {Guhathakurta}, P. 2006, \mnras, 366, 996

\bibitem[{{Gilbert} {et~al.}(2007){Gilbert}, {Fardal}, {Kalirai}, {Guhathakurta}, {Geha}, {Isler}, {Majewski}, {Ostheimer}, {Patterson}, {Reitzel}, {Kirby}, \& {Cooper}}]{Gilbert2007}
{Gilbert}, K.~M., {Fardal}, M., {Kalirai}, J.~S., {et~al.} 2007, \apj, 668, 245

\bibitem[{{Gilbert} {et~al.}(2009){Gilbert}, {Guhathakurta}, {Kollipara}, {Beaton}, {Geha}, {Kalirai}, {Kirby}, {Majewski}, \& {Patterson}}]{gilbert09}
{Gilbert}, K.~M., {Guhathakurta}, P., {Kollipara}, P., {et~al.} 2009, \apj, 705, 1275

\bibitem[{{Gilbert} {et~al.}(2019){Gilbert}, {Kirby}, {Escala}, {Wojno}, {Kalirai}, \& {Guhathakurta}}]{Gilbert2019}
{Gilbert}, K.~M., {Kirby}, E.~N., {Escala}, I., {et~al.} 2019, \apj, 883, 128

\bibitem[{{Gilbert} {et~al.}(2020){Gilbert}, {Wojno}, {Kirby}, {Escala}, {Beaton}, {Guhathakurta}, \& {Majewski}}]{Gilbert2020}
{Gilbert}, K.~M., {Wojno}, J., {Kirby}, E.~N., {et~al.} 2020, \aj, 160, 41

\bibitem[{{Gregersen} {et~al.}(2015){Gregersen}, {Seth}, {Williams}, {Lang}, {Dalcanton}, {Girardi}, {Skillman}, {Bell}, {Dolphin}, {Fouesneau}, {Guhathakurta}, {Hamren}, {Johnson}, {Kalirai}, {Lewis}, {Monachesi}, \& {Olsen}}]{Gregersen2015}
{Gregersen}, D., {Seth}, A.~C., {Williams}, B.~F., {et~al.} 2015, \aj, 150, 189

\bibitem[{{Guhathakurta} {et~al.}(2006){Guhathakurta}, {Rich}, {Reitzel}, {Cooper}, {Gilbert}, {Majewski}, {Ostheimer}, {Geha}, {Johnston}, \& {Patterson}}]{Guhathakurta06}
{Guhathakurta}, P., {Rich}, R.~M., {Reitzel}, D.~B., {et~al.} 2006, \aj, 131, 2497

\bibitem[{Hahsler \& Piekenbrock(2025)}]{dbscan}
Hahsler, M. \& Piekenbrock, M. 2025, dbscan: Density-Based Spatial Clustering of Applications with Noise (DBSCAN) and Related Algorithms, r package version 1.2.2

\bibitem[{{Hammer} {et~al.}(2005){Hammer}, {Flores}, {Elbaz}, {Zheng}, {Liang}, \& {Cesarsky}}]{hammer05}
{Hammer}, F., {Flores}, H., {Elbaz}, D., {et~al.} 2005, \aap, 430, 115

\bibitem[{{Hammer} {et~al.}(2009){Hammer}, {Flores}, {Puech}, {Yang}, {Athanassoula}, {Rodrigues}, \& {Delgado}}]{hammer09}
{Hammer}, F., {Flores}, H., {Puech}, M., {et~al.} 2009, \aap, 507, 1313

\bibitem[{{Hammer} {et~al.}(2018){Hammer}, {Yang}, {Wang}, {Ibata}, {Flores}, \& {Puech}}]{hammer}
{Hammer}, F., {Yang}, Y.~B., {Wang}, J.~L., {et~al.} 2018, \mnras, 475, 2754

\bibitem[{{Harris} {et~al.}(2020){Harris}, {Millman}, {van der Walt}, {Gommers}, {Virtanen}, {Cournapeau}, {Wieser}, {Taylor}, {Berg}, {Smith}, {Kern}, {Picus}, {Hoyer}, {van Kerkwijk}, {Brett}, {Haldane}, {del R{\'\i}o}, {Wiebe}, {Peterson}, {G{\'e}rard-Marchant}, {Sheppard}, {Reddy}, {Weckesser}, {Abbasi}, {Gohlke}, \& {Oliphant}}]{numpy}
{Harris}, C.~R., {Millman}, K.~J., {van der Walt}, S.~J., {et~al.} 2020, \nat, 585, 357

\bibitem[{{Helmi}(2020)}]{Helmi2020}
{Helmi}, A. 2020, \araa, 58, 205

\bibitem[{{Hendel} \& {Johnston}(2015)}]{Hendel2015MNRAS.454.2472H}
{Hendel}, D. \& {Johnston}, K.~V. 2015, \mnras, 454, 2472

\bibitem[{{Hopkins} {et~al.}(2009){Hopkins}, {Cox}, {Younger}, \& {Hernquist}}]{hopkins09}
{Hopkins}, P.~F., {Cox}, T.~J., {Younger}, J.~D., \& {Hernquist}, L. 2009, \apj, 691, 1168

\bibitem[{{Hopkins} {et~al.}(2008){Hopkins}, {Hernquist}, {Cox}, {Younger}, \& {Besla}}]{Hopkins08}
{Hopkins}, P.~F., {Hernquist}, L., {Cox}, T.~J., {Younger}, J.~D., \& {Besla}, G. 2008, \apj, 688, 757

\bibitem[{{Hubble}(1929)}]{Hubble1929}
{Hubble}, E.~P. 1929, \apj, 69, 103

\bibitem[{{Hunter}(2007)}]{matplotlib}
{Hunter}, J.~D. 2007, Computing in Science and Engineering, 9, 90

\bibitem[{{Ibata} {et~al.}(2004){Ibata}, {Chapman}, {Ferguson}, {Irwin}, {Lewis}, \& {McConnachie}}]{ibata2004}
{Ibata}, R., {Chapman}, S., {Ferguson}, A.~M.~N., {et~al.} 2004, \mnras, 351, 117

\bibitem[{{Ibata} {et~al.}(2005){Ibata}, {Chapman}, {Ferguson}, {Lewis}, {Irwin}, \& {Tanvir}}]{Ibata2005}
{Ibata}, R., {Chapman}, S., {Ferguson}, A.~M.~N., {et~al.} 2005, \apj, 634, 287

\bibitem[{{Ibata} {et~al.}(2001){Ibata}, {Irwin}, {Lewis}, {Ferguson}, \& {Tanvir}}]{ibata2001}
{Ibata}, R., {Irwin}, M., {Lewis}, G., {Ferguson}, A. M.~N., \& {Tanvir}, N. 2001, \nat, 412, 49

\bibitem[{{Kalirai} {et~al.}(2006){Kalirai}, {Guhathakurta}, {Gilbert}, {Reitzel}, {Majewski}, {Rich}, \& {Cooper}}]{Kalirai2006}
{Kalirai}, J.~S., {Guhathakurta}, P., {Gilbert}, K.~M., {et~al.} 2006, \apj, 641, 268

\bibitem[{{Kirihara} {et~al.}(2014){Kirihara}, {Miki}, \& {Mori}}]{kirihara2014}
{Kirihara}, T., {Miki}, Y., \& {Mori}, M. 2014, \pasj, 66, L10

\bibitem[{{Kirihara} {et~al.}(2017){Kirihara}, {Miki}, {Mori}, {Kawaguchi}, \& {Rich}}]{Kirihara2017}
{Kirihara}, T., {Miki}, Y., {Mori}, M., {Kawaguchi}, T., \& {Rich}, R.~M. 2017, \mnras, 464, 3509

\bibitem[{{Kobayashi} {et~al.}(2023){Kobayashi}, {Bhattacharya}, {Arnaboldi}, \& {Gerhard}}]{Kobayashi2023}
{Kobayashi}, C., {Bhattacharya}, S., {Arnaboldi}, M., \& {Gerhard}, O. 2023, \apjl, 956, L14

\bibitem[{{Kwitter} \& {Henry}(2022)}]{Kwitter2022PASP..134b2001K}
{Kwitter}, K.~B. \& {Henry}, R.~B.~C. 2022, \pasp, 134, 022001

\bibitem[{{Lewis} {et~al.}(2015){Lewis}, {Dolphin}, {Dalcanton}, {Weisz}, {Williams}, {Bell}, {Seth}, {Simones}, {Skillman}, {Choi}, {Fouesneau}, {Guhathakurta}, {Johnson}, {Kalirai}, {Leroy}, {Monachesi}, {Rix}, \& {Schruba}}]{Lewis2015ApJ...805..183L}
{Lewis}, A.~R., {Dolphin}, A.~E., {Dalcanton}, J.~J., {et~al.} 2015, \apj, 805, 183

\bibitem[{{Longobardi} {et~al.}(2015){Longobardi}, {Arnaboldi}, {Gerhard}, \& {Mihos}}]{longobardi15}
{Longobardi}, A., {Arnaboldi}, M., {Gerhard}, O., \& {Mihos}, J.~C. 2015, \aap, 579, L3

\bibitem[{{Maiolino} \& {Mannucci}(2019)}]{maiolino}
{Maiolino}, R. \& {Mannucci}, F. 2019, \aapr, 27, 3

\bibitem[{{Martig} {et~al.}(2014){Martig}, {Minchev}, \& {Flynn}}]{Martig2014}
{Martig}, M., {Minchev}, I., \& {Flynn}, C. 2014, \mnras, 443, 2452

\bibitem[{{McConnachie} {et~al.}(2018){McConnachie}, {Ibata}, {Martin}, {Ferguson}, {Collins}, {Gwyn}, {Irwin}, {Lewis}, {Mackey}, {Davidge}, {Arias}, {Conn}, {C{\^o}t{\'e}}, {Crnojevic}, {Huxor}, {Penarrubia}, {Spengler}, {Tanvir}, {Valls-Gabaud}, {Babul}, {Barmby}, {Bate}, {Bernard}, {Chapman}, {Dotter}, {Harris}, {McMonigal}, {Navarro}, {Puzia}, {Rich}, {Thomas}, \& {Widrow}}]{mcconnachie}
{McConnachie}, A.~W., {Ibata}, R., {Martin}, N., {et~al.} 2018, \apj, 868, 55

\bibitem[{{McConnachie} {et~al.}(2009){McConnachie}, {Irwin}, {Ibata}, {Dubinski}, {Widrow}, {Martin}, {C{\^o}t{\'e}}, {Dotter}, {Navarro}, {Ferguson}, {Puzia}, {Lewis}, {Babul}, {Barmby}, {Bienaym{\'e}}, {Chapman}, {Cockcroft}, {Collins}, {Fardal}, {Harris}, {Huxor}, {Mackey}, {Pe{\~n}arrubia}, {Rich}, {Richer}, {Siebert}, {Tanvir}, {Valls-Gabaud}, \& {Venn}}]{mcconnachie09nature}
{McConnachie}, A.~W., {Irwin}, M.~J., {Ibata}, R.~A., {et~al.} 2009, \nat, 461, 66

\bibitem[{{McConnachie} {et~al.}(2003){McConnachie}, {Irwin}, {Ibata}, {Ferguson}, {Lewis}, \& {Tanvir}}]{McConnachie03}
{McConnachie}, A.~W., {Irwin}, M.~J., {Ibata}, R.~A., {et~al.} 2003, \mnras, 343, 1335

\bibitem[{{Merrett} {et~al.}(2003){Merrett}, {Kuijken}, {Merrifield}, {Romanowsky}, {Douglas}, {Napolitano}, {Arnaboldi}, {Capaccioli}, {Freeman}, {Gerhard}, {Evans}, {Wilkinson}, {Halliday}, {Bridges}, \& {Carter}}]{Merrett2003}
{Merrett}, H.~R., {Kuijken}, K., {Merrifield}, M.~R., {et~al.} 2003, \mnras, 346, L62

\bibitem[{{Merrett} {et~al.}(2006){Merrett}, {Merrifield}, {Douglas}, {Kuijken}, {Romanowsky}, {Napolitano}, {Arnaboldi}, {Capaccioli}, {Freeman}, {Gerhard}, {Coccato}, {Carter}, {Evans}, {Wilkinson}, {Halliday}, \& {Bridges}}]{Merrett2006}
{Merrett}, H.~R., {Merrifield}, M.~R., {Douglas}, N.~G., {et~al.} 2006, \mnras, 369, 120

\bibitem[{{Merrifield} \& {Kuijken}(1998)}]{Merrifield}
{Merrifield}, M.~R. \& {Kuijken}, K. 1998, \mnras, 297, 1292

\bibitem[{{Milo{\v{s}}evi{\'c}} {et~al.}(2022){Milo{\v{s}}evi{\'c}}, {Mi{\'c}i{\'c}}, \& {Lewis}}]{milosevic2022}
{Milo{\v{s}}evi{\'c}}, S., {Mi{\'c}i{\'c}}, M., \& {Lewis}, G.~F. 2022, \mnras, 511, 2868

\bibitem[{{Milo{\v{s}}evi{\'c}} {et~al.}(2024){Milo{\v{s}}evi{\'c}}, {Mi{\'c}i{\'c}}, \& {Lewis}}]{milosevic2024}
{Milo{\v{s}}evi{\'c}}, S., {Mi{\'c}i{\'c}}, M., \& {Lewis}, G.~F. 2024, \mnras, 527, 4797

\bibitem[{{Moll{\'a}} {et~al.}(2019){Moll{\'a}}, {D{\'\i}az}, {Cavichia}, {Gibson}, {Maciel}, {Costa}, {Ascasibar}, \& {Few}}]{molla2019MNRAS.482.3071M}
{Moll{\'a}}, M., {D{\'\i}az}, {\'A}.~I., {Cavichia}, O., {et~al.} 2019, \mnras, 482, 3071

\bibitem[{{Mori} \& {Rich}(2008)}]{Mori2008}
{Mori}, M. \& {Rich}, R.~M. 2008, \apjl, 674, L77

\bibitem[{{Nordstr{\"o}m} {et~al.}(2004){Nordstr{\"o}m}, {Mayor}, {Andersen}, {Holmberg}, {Pont}, {J{\o}rgensen}, {Olsen}, {Udry}, \& {Mowlavi}}]{nordstrom2004}
{Nordstr{\"o}m}, B., {Mayor}, M., {Andersen}, J., {et~al.} 2004, \aap, 418, 989

\bibitem[{{Ogami} {et~al.}(2025){Ogami}, {Tanaka}, {Komiyama}, {Chiba}, {Guhathakurta}, {Kirby}, {Wyse}, {Filion}, {Gilbert}, {Escala}, {Mori}, {Kirihara}, {Tanaka}, {Ishigaki}, {Hayashi}, {Lee}, {Sharma}, {Kalirai}, \& {Lupton}}]{ogami}
{Ogami}, I., {Tanaka}, M., {Komiyama}, Y., {et~al.} 2025, \mnras, 536, 530

\bibitem[{{Okamoto} {et~al.}(2005){Okamoto}, {Eke}, {Frenk}, \& {Jenkins}}]{Okamoto2005}
{Okamoto}, T., {Eke}, V.~R., {Frenk}, C.~S., \& {Jenkins}, A. 2005, \mnras, 363, 1299

\bibitem[{{Pillepich} {et~al.}(2014){Pillepich}, {Vogelsberger}, {Deason}, {Rodriguez-Gomez}, {Genel}, {Nelson}, {Torrey}, {Sales}, {Marinacci}, {Springel}, {Sijacki}, \& {Hernquist}}]{Pillepich2014M}
{Pillepich}, A., {Vogelsberger}, M., {Deason}, A., {et~al.} 2014, \mnras, 444, 237

\bibitem[{{Richardson} {et~al.}(2008){Richardson}, {Ferguson}, {Johnson}, {Irwin}, {Tanvir}, {Faria}, {Ibata}, {Johnston}, {Lewis}, {McConnachie}, \& {Chapman}}]{Richardson2008}
{Richardson}, J.~C., {Ferguson}, A.~M.~N., {Johnson}, R.~A., {et~al.} 2008, \aj, 135, 1998

\bibitem[{{Roca-F{\`a}brega} {et~al.}(2021){Roca-F{\`a}brega}, {Kim}, {Hausammann}, {Nagamine}, {Lupi}, {Powell}, {Shimizu}, {Ceverino}, {Primack}, {Quinn}, {Revaz}, {Vel{\'a}zquez}, {Abel}, {Buehlmann}, {Dekel}, {Dong}, {Hahn}, {Hummels}, {Kim}, {Smith}, {Strawn}, {Teyssier}, {Turk}, \& {AGORA Collaboration}}]{Roca-Fabrega2021}
{Roca-F{\`a}brega}, S., {Kim}, J.-H., {Hausammann}, L., {et~al.} 2021, \apj, 917, 64

\bibitem[{{Rodrigues} {et~al.}(2008){Rodrigues}, {Hammer}, {Flores}, {Puech}, {Liang}, {Fuentes-Carrera}, {Nesvadba}, {Lehnert}, {Yang}, {Amram}, {Balkowski}, {Cesarsky}, {Dannerbauer}, {Delgado}, {Guiderdoni}, {Kembhavi}, {Neichel}, {{\"O}stlin}, {Pozzetti}, {Ravikumar}, {Rawat}, {di Serego Alighieri}, {Vergani}, {Vernet}, \& {Wozniak}}]{rodriguez}
{Rodrigues}, M., {Hammer}, F., {Flores}, H., {et~al.} 2008, \aap, 492, 371

\bibitem[{{Rubin} \& {Ford}(1970)}]{Rubin1970ApJ}
{Rubin}, V.~C. \& {Ford}, W.~Kent, J. 1970, \apj, 159, 379

\bibitem[{{Sadoun} {et~al.}(2014){Sadoun}, {Mohayaee}, \& {Colin}}]{Sadoun2014}
{Sadoun}, R., {Mohayaee}, R., \& {Colin}, J. 2014, \mnras, 442, 160

\bibitem[{{Saglia} {et~al.}(2018){Saglia}, {Opitsch}, {Fabricius}, {Bender}, {Bla{\~n}a}, \& {Gerhard}}]{Saglia2018}
{Saglia}, R.~P., {Opitsch}, M., {Fabricius}, M.~H., {et~al.} 2018, \aap, 618, A156

\bibitem[{Salaris \& Cassisi(2005)}]{salaris}
Salaris, M. \& Cassisi, S. 2005, Evolution of stars and stellar populations (John Wiley \& Sons)

\bibitem[{{Savino} {et~al.}(2025){Savino}, {Weisz}, {Dolphin}, {Durbin}, {Kallivayalil}, {Wetzel}, {Anderson}, {Besla}, {Boylan-Kolchin}, {Brown}, {Bullock}, {Cole}, {Collins}, {Cooper}, {Deason}, {Dotter}, {Fardal}, {Ferguson}, {Fritz}, {Geha}, {Gilbert}, {Guhathakurta}, {Ibata}, {Irwin}, {Jeon}, {Kirby}, {Lewis}, {Mackey}, {Majewski}, {Martin}, {McConnachie}, {Patel}, {Rich}, {Skillman}, {Simon}, {Sohn}, {Tollerud}, \& {van der Marel}}]{savino2025}
{Savino}, A., {Weisz}, D.~R., {Dolphin}, A.~E., {et~al.} 2025, \apj, 979, 205

\bibitem[{{Stinson} {et~al.}(2010){Stinson}, {Bailin}, {Couchman}, {Wadsley}, {Shen}, {Nickerson}, {Brook}, \& {Quinn}}]{mugs2010MNRAS.408..812S}
{Stinson}, G.~S., {Bailin}, J., {Couchman}, H., {et~al.} 2010, \mnras, 408, 812

\bibitem[{{Tamura} {et~al.}(2016){Tamura}, {Takato}, {Shimono}, {Moritani}, {Yabe}, {Ishizuka}, {Ueda}, {Kamata}, {Aghazarian}, {Arnouts}, {Barban}, {Barkhouser}, {Borges}, {Braun}, {Carr}, {Chabaud}, {Chang}, {Chen}, {Chiba}, {Chou}, {Chu}, {Cohen}, {de Almeida}, {de Oliveira}, {de Oliveira}, {Dekany}, {Dohlen}, {dos Santos}, {dos Santos}, {Ellis}, {Fabricius}, {Ferrand}, {Ferreira}, {Golebiowski}, {Greene}, {Gross}, {Gunn}, {Hammond}, {Harding}, {Hart}, {Heckman}, {Hirata}, {Ho}, {Hope}, {Hovland}, {Hsu}, {Hu}, {Huang}, {Jaquet}, {Jing}, {Karr}, {Kimura}, {King}, {Komatsu}, {Le Brun}, {Le F{\`e}vre}, {Le Fur}, {Le Mignant}, {Ling}, {Loomis}, {Lupton}, {Madec}, {Mao}, {Marrara}, {Mendes de Oliveira}, {Minowa}, {Morantz}, {Murayama}, {Murray}, {Ohyama}, {Orndorff}, {Pascal}, {Pereira}, {Reiley}, {Reinecke}, {Ritter}, {Roberts}, {Schwochert}, {Seiffert}, {Smee}, {Sodre}, {Spergel}, {Steinkraus}, {Strauss}, {Surace}, {Suto}, {Suzuki}, {Swinbank}, {Tait}, {Takada}, {Tamura}, {Tanaka}, {Tresse}, {Verducci},
  {Vibert}, {Vidal}, {Wang}, {Wen}, {Yan}, \& {Yasuda}}]{Subarupfs}
{Tamura}, N., {Takato}, N., {Shimono}, A., {et~al.} 2016, in Society of Photo-Optical Instrumentation Engineers (SPIE) Conference Series, Vol. 9908, Ground-based and Airborne Instrumentation for Astronomy VI, ed. C.~J. {Evans}, L.~{Simard}, \& H.~{Takami}, 99081M

\bibitem[{{Tanaka} {et~al.}(2010){Tanaka}, {Chiba}, {Komiyama}, {Guhathakurta}, {Kalirai}, \& {Iye}}]{tanaka2010}
{Tanaka}, M., {Chiba}, M., {Komiyama}, Y., {et~al.} 2010, \apj, 708, 1168

\bibitem[{{The pandas development Team}(2024)}]{pandas}
{The pandas development Team}. 2024, {pandas-dev/pandas: Pandas}

\bibitem[{{Tissera} {et~al.}(2019){Tissera}, {Rosas-Guevara}, {Bower}, {Crain}, {del P Lagos}, {Schaller}, {Schaye}, \& {Theuns}}]{Tissera2019MNRAS.482.2208T}
{Tissera}, P.~B., {Rosas-Guevara}, Y., {Bower}, R.~G., {et~al.} 2019, \mnras, 482, 2208

\bibitem[{{Tornatore} {et~al.}(2007){Tornatore}, {Borgani}, {Dolag}, \& {Matteucci}}]{Tornatore2007}
{Tornatore}, L., {Borgani}, S., {Dolag}, K., \& {Matteucci}, F. 2007, \mnras, 382, 1050

\bibitem[{{Tremonti} {et~al.}(2004){Tremonti}, {Heckman}, {Kauffmann}, {Brinchmann}, {Charlot}, {White}, {Seibert}, {Peng}, {Schlegel}, {Uomoto}, {Fukugita}, \& {Brinkmann}}]{tremonti}
{Tremonti}, C.~A., {Heckman}, T.~M., {Kauffmann}, G., {et~al.} 2004, \apj, 613, 898

\bibitem[{{Watkins} {et~al.}(2013){Watkins}, {Evans}, \& {van de Ven}}]{Watkins2013}
{Watkins}, L.~L., {Evans}, N.~W., \& {van de Ven}, G. 2013, \mnras, 430, 971

\bibitem[{{White} \& {Rees}(1978)}]{White1978}
{White}, S.~D.~M. \& {Rees}, M.~J. 1978, \mnras, 183, 341

\bibitem[{{Wiersma} {et~al.}(2009){Wiersma}, {Schaye}, {Theuns}, {Dalla Vecchia}, \& {Tornatore}}]{Wiersma2009}
{Wiersma}, R. P.~C., {Schaye}, J., {Theuns}, T., {Dalla Vecchia}, C., \& {Tornatore}, L. 2009, \mnras, 399, 574

\bibitem[{{Williams} {et~al.}(2017){Williams}, {Dolphin}, {Dalcanton}, {Weisz}, {Bell}, {Lewis}, {Rosenfield}, {Choi}, {Skillman}, \& {Monachesi}}]{williams17}
{Williams}, B.~F., {Dolphin}, A.~E., {Dalcanton}, J.~J., {et~al.} 2017, \apj, 846, 145

\bibitem[{{Wojno} {et~al.}(2023){Wojno}, {Gilbert}, {Kirby}, {Escala}, {Guhathakurta}, {Beaton}, {Kalirai}, {Chiba}, \& {Majewski}}]{Wojno2023}
{Wojno}, J.~L., {Gilbert}, K.~M., {Kirby}, E.~N., {et~al.} 2023, \apj, 951, 12

\bibitem[{{Wyse}(2001)}]{Wyse}
{Wyse}, R.~F.~G. 2001, in Astronomical Society of the Pacific Conference Series, Vol. 230, Galaxy Disks and Disk Galaxies, ed. J.~G. {Funes} \& E.~M. {Corsini}, 71--80

\bibitem[{{Yin} {et~al.}(2009){Yin}, {Hou}, {Prantzos}, {Boissier}, {Chang}, {Shen}, \& {Zhang}}]{yin}
{Yin}, J., {Hou}, J.~L., {Prantzos}, N., {et~al.} 2009, \aap, 505, 497

\bibitem[{{Zinchenko} {et~al.}(2015){Zinchenko}, {Berczik}, {Grebel}, {Pilyugin}, \& {Just}}]{Zinchenko2015ApJ...806..267Z}
{Zinchenko}, I.~A., {Berczik}, P., {Grebel}, E.~K., {Pilyugin}, L.~S., \& {Just}, A. 2015, \apj, 806, 267

\end{thebibliography}

\begin{appendix} \label{appendix}

\section{Adopted values for the initial metallicity gradient in the two progenitors}\label{gradient appendix}

To establish the initial metallicity gradient of the two disc progenitors, we relied on constraints derived from observations of galaxies at comparable redshifts. \citet{curtigrad}, utilizing emission line measurements from gravitationally lensed galaxies in the KLEVER survey, determined oxygen abundance gradients within approximately one to two effective radii (R$\rm_{e}$). However, we needed to set gradient values that are suitable for the entire radial extent of the discs out to a few tens of kiloparsecs, thus extending beyond the inner $\sim$1$-$2 R$\rm_e$ region.

To address this, we considered predictions from chemical evolution models that determine metallicity gradients across the entire radial range of galactic discs. \citet[see their Figure 5]{molla2019MNRAS.482.3071M} present the radial distribution of 12 + log(O/H) for MW-like galaxies across redshifts 0 to 4.0. These models exhibit a step-like metallicity profile: a relatively flat gradient in the inner disc regions (up to $\sim$10 kpc at redshift z~$=$~1), followed by a rapid decline at intermediate radii, and a subsequent flattening in the outer disc (beyond $\sim$15 kpc at redshift z~$=$~1), where oxygen abundances reach approximately 20\% of the solar value.
Given our assumption of a linear gradient for the initial discs, we adopt a value of –0.1 dex/kpc. This choice, which is slightly steeper than the near-flat observational gradients reported by \citet{curtigrad}, accounts for the steeper decline predicted by chemical evolution models over the entire radial range covered by the simulated exponential discs in the simulations. Additional support for the adopted gradient value comes from simulated radial metallicity profiles of star-forming galaxies in recent cosmological simulations (see TNG; \citealt{Garcia} and MUGS; \citealt{mugs2010MNRAS.408..812S} cosmological simulations).

\section{The star formation history of the M31 stellar disc} \label{sfh appendix}

\begin{figure}[htp]
    \centering
    \includegraphics[width=0.5\textwidth]{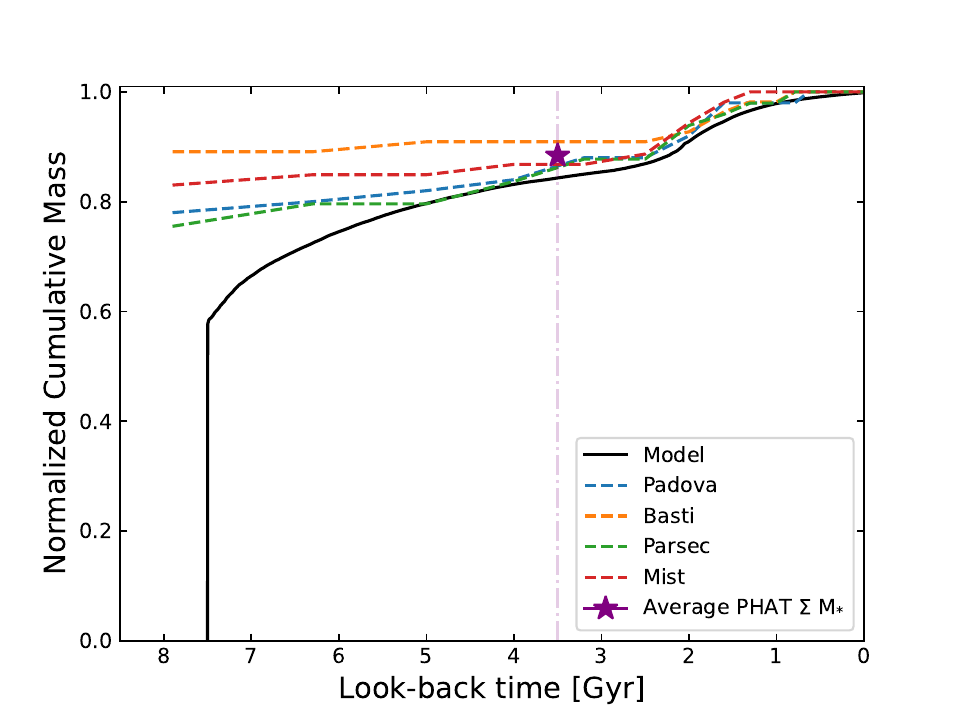}
    \caption{Cumulative stellar mass formed in the north-east quadrant of the disc of M31 from \citet{williams17}, as probed by four different stellar evolution models (each model is denoted with a dashed coloured line). The purple star indicates the average value for the four models computed 3.5 Gyr ago. The vertical dotted line indicates the separation of the two age groups. The cumulative stellar mass formed during the simulation of the M31 merger event is plotted as a black, solid curve.} 
    \label{fig:sfh}
\end{figure}

\citet{williams17} used deep multi-band observations from HST's PHAT survey in the M31 disc to constrain the SFH of its stars. In Figure~\ref{fig:sfh}, we show the SFH using four different stellar evolution models as reproduced in \citet{williams17}. These results show that $\sim$80\% of the stellar mass in the disc is formed at a lookback time of $\sim$8~Gyr. At $\sim$3.5~Gyr ago, a second star-forming episode started in the observed part of the disc, peaking at $\sim$2~$-$~3 Gyr ago and providing $\sim$20\% of the total stellar mass in the M31 disc.
 
 A simplified approach to describing the SFH is to distinguish the M31 disc stars into two age groups. One group includes stars older than 3.5 Gyr, which account for the bulk of the stellar mass in the disc, that is $\sim$80\%. The second group includes stars born in the subsequent burst of star formation. Stars of this second younger group have ages in the range of 2 to 4 Gyr and younger.

 The SFH which results from the \citetalias{hammer} simulations deviates somewhat from the M31 SFHs published by \citet{williams17} in the PHAT area; the \citetalias{hammer} SFH is shown by the black continuous curve in Figure~\ref{fig:sfh}. In the \citetalias{hammer} simulations, stars are being formed in the intermediate time range 7.5 $-$ 3.5 Gyr also.  We address such a discrepancy by including all stars older than 3.5 Gyr in the old age group and all stars younger than 3.5 Gyr in the young age group. 
 
 We then compare the cumulative mass fraction values for ages $<$3.5~Gyr and $>$3.5~Gyr in the simulations with the measured values in the PHAT area.  The observed constrained value for the cumulative stellar mass formed for ages older than 3.5~Gyr is $\sim$88\% from the average of the four modelled SFHs in \citet{williams17}. Such value is fully comparable with the cumulative stellar mass faction value generated in the \citetalias{hammer} simulation, which amounts to $\sim$84\%. 

Considering these two age families, older and younger than 3.5~Gyr, from the PHAT SFH, we then associate the group of old PNe with the progenitor stars, which are older than 3.5~Gyr. The group of young PNe is then associated with the progenitor stars younger than 3.5~Gyr, which are linked to the recent episode of star formation in the M31 disc. 

The age milestone of 3.5~Gyr is in line with the star formation history of the M31's disc as traced by PNe which show two distinct kinematical and chemical populations  \citep{paperiv} and with the results from the chemical evolution models in \citet{arna22}.

    \section{DBSCAN parameters} 
    \label{appendix:dbscan}
    The parameters that must be specified for the implementation of DBSCAN are the $eps$ and $min samples$. $Eps$ is the maximum distance between two samples for one to be considered as being in the neighborhood of the other, and $minsamples$ is the number of samples in a neighborhood for a point to be considered as a core point \citep{dbscan}. The parameters utilised to distinguish the components of each substructure are separately discussed below.
    
    \subsection{The Giant Stellar Stream}
    The GSS in the model consists of regions with diverse densities. The regions that lie in the vicinity of the disc (roughly the area denoted as GSS-component [3] in Figure ~\ref{fig:3d_gs}) exhibit a significantly higher density than the rest. DBSCAN clustering does not work efficiently for data sets with varying densities. Therefore, to properly distinguish the constituent components of the GSS, we resort to two consecutive runs of DBSCAN with different values for $eps$ and $min samples$. 

    For the first run, the values of the parameters were $eps$~=~3.9 and $min samples$~=~250. These values are appropriate for the identification of the most dense component, near the disc of the galaxy; GSS-component [3]. We mask this region and resort to the next run of DBSCAN.

    For the second run, the values of the parameters were $eps$~=~4.1 and $min samples$~=~78. Three more clusters are identified within the premises of the GSS, which, together with the disc component, reveal the GSS's multi-component, complicated intrinsic structure in the model.

    \subsection{The North-East Shelf}
    Stars within the NE shelf region selected in the model are almost uniformly spread in density and the DBSCAN parameters used were $eps$~=~2 and $min samples$~=~100.
    
\subsection{The Western Shelf}
The parameters for the identification of the two components of the (uniformly spread) W shelf are $eps$~=~2.6 and $min samples$~=~137.

    \section{Line-of-sight modification of the M31 model} \label{different kinetic angle}
\citetalias{hammer} chose a specific temporal output during an evolving sequence, which optimises the reproduction of a considerable amount of observations that were previously available, as detailed in Section~\ref{section general concepts}. Since the substructures in the inner halo are time-dependent, they may be found closer or further away in the projected distance from their observed counterparts. For instance, \citetalias{hammer} reported that some shells can be found at the same radial distance as in M31, but perhaps not simultaneously with other features.

The second limitation is related to the parameters of the final rotation of the simulated analogue at z~$=$~0,  utilised to place the galaxy in the observed reference frame (PA and inclination of the disc, relative positions of inner halo substructures, etc.). There is some rotation uncertainty in the model since the accuracy of both the PA and the disc inclination is up to a few degrees. Specifically, the third angle that is used to adjust the relative position between the disc and the halo in the simulation is that of the rotation around the disc axis in the plane of the disc. This angle is the least constrained ($\sim$ $\pm$10$\degree$) in reproducing the observational data. Variations in angles impact the V$\rm_{LOS}$  to the MW, as the M31 disc is seen almost edge-on. 

The third limitation has to do with the large number of initial parameters (to mention only some of them: the in-fall angle of the secondary galaxy, the value of the first pericentre passage, and the internal light distribution in the progenitors) associated with a major merger simulation. A variation in the initial parameters may provide better agreement with observations; nonetheless, the fine-tuning of the simulation is beyond the scope of this work. 

To compare the final temporal output with the observed LOS of the M31 system from our vantage point in the MW, the coordinate system must be fixed.  To achieve this, the modelled disc is aligned with the observed inclination and PA of the M31 disc. However, there is a third angle of rotation, the rotation along the disc plane. This angle is termed the kinetic angle. The kinetic angle is accountable for the relative arrangement between the disc and the inner halo substructures. Hence, the chosen kinetic angle is the one that aptly reproduces the observed morphology (i.e. surface brightness) of the GSS and the two shells. However, there is some leeway of approximately $\pm$10$\degree$ in the choice of this kinetic angle while maintaining the GSS (and other substructures) surface brightness similar to the observed. 

\begin{figure*}[h!]
    \centering
    \includegraphics[width=\textwidth]{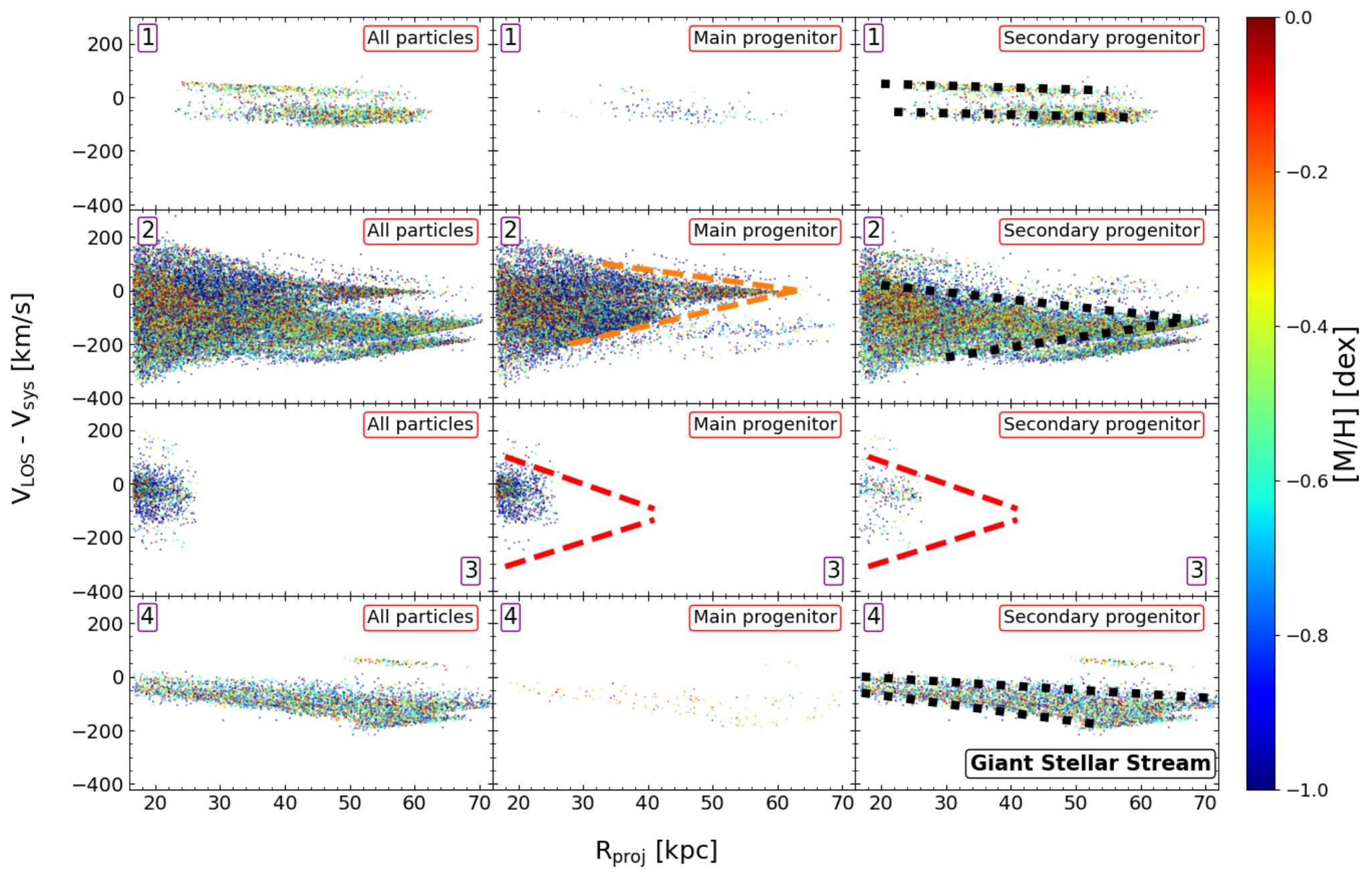}
    \caption{Same as Figure \ref{fig:desi_gs}, but for a different kinetic angle. Delineated ridges are the ones previously identified for the GSS with the default kinetic angle, to explicitly show the differences that emerge from choosing a different kinetic angle.}
    \label{fig:gs_kin_angle_137}
\end{figure*}

To illustrate the extent of the impact on the phase space diagram introduced from choosing a different kinetic angle, Figure~\ref{fig:gs_kin_angle_137} depicts the R$\rm_{proj}$ versus V$\rm_{LOS}$ diagram of stars in the GSS for a slightly different kinetic angle (+7$\degree$), which still reproduces the observed morphology of the inner halo features. The density of stars in the identified clusters is significantly altered compared to that observed for the default kinetic angle (compare with Figure~\ref{fig:desi_gs}). This is explicitly noticeable in the reduced density of GSS-component [3], which impacts the identification of coherent features in the phase space. 

Therefore, the adjustment of the LOS alignment of the simulated galaxy at z~$=$~0 may provide a viable mechanism to account for discrepancies between the observed and modelled phase space features in M31's substructures.
\FloatBarrier
\FloatBarrier

\section{Comparing photometric metallicity of substructures with younger model population} \label{metallicity young stars}

\begin{figure*}[htp]
    \centering
    \includegraphics[width=\textwidth]{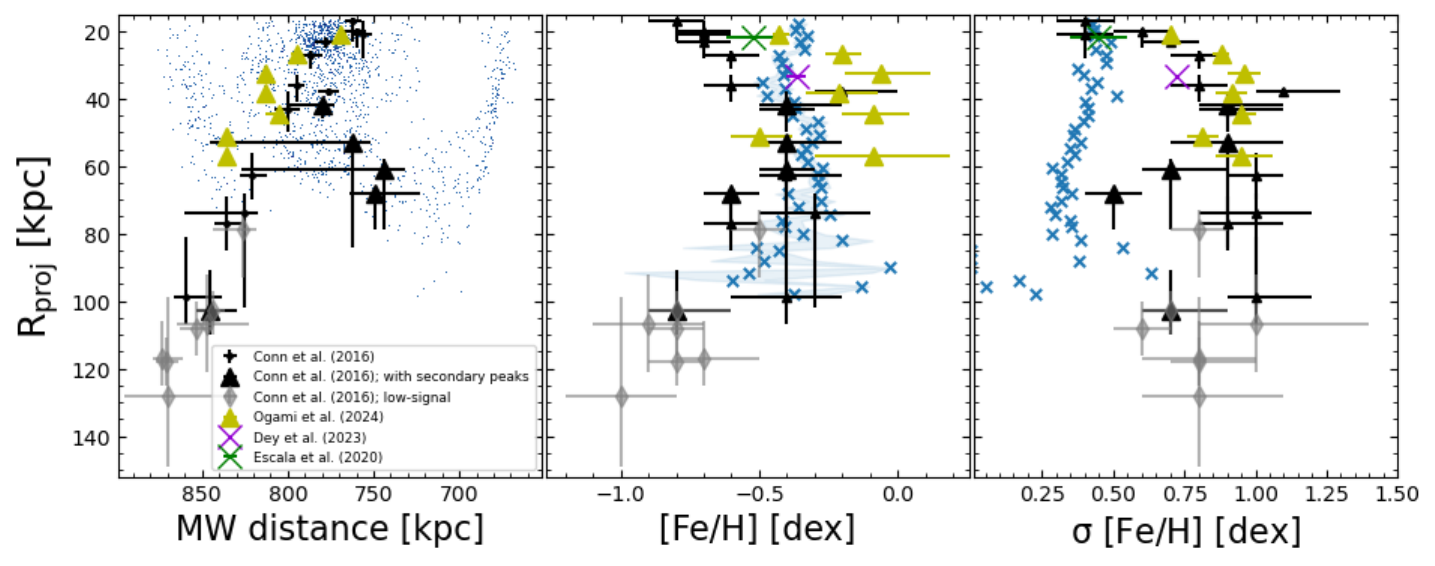}
    \caption{Same as Figure \ref{fig:ogami_gs} but for modelled stars younger than 3.5~Gyr.}
    \label{fig:ogami_gs_young}
    \end{figure*}

\begin{figure}[htp]
    \centering
    \includegraphics[width=0.5\textwidth]{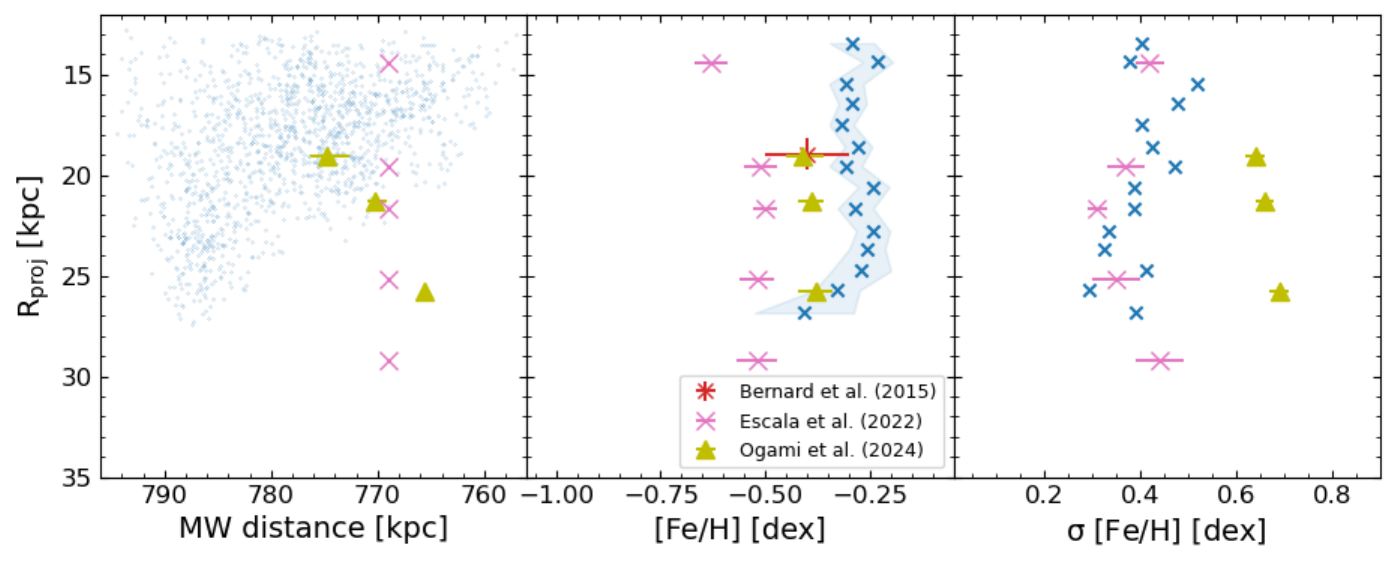}
    \caption{Same as Figure \ref{fig:ogami_ne} but for modelled stars younger than 3.5~Gyr.}
    \label{fig:ogami_ne_young}
    \end{figure}

\begin{figure}[htp]
    \centering
    \includegraphics[width=0.5\textwidth]{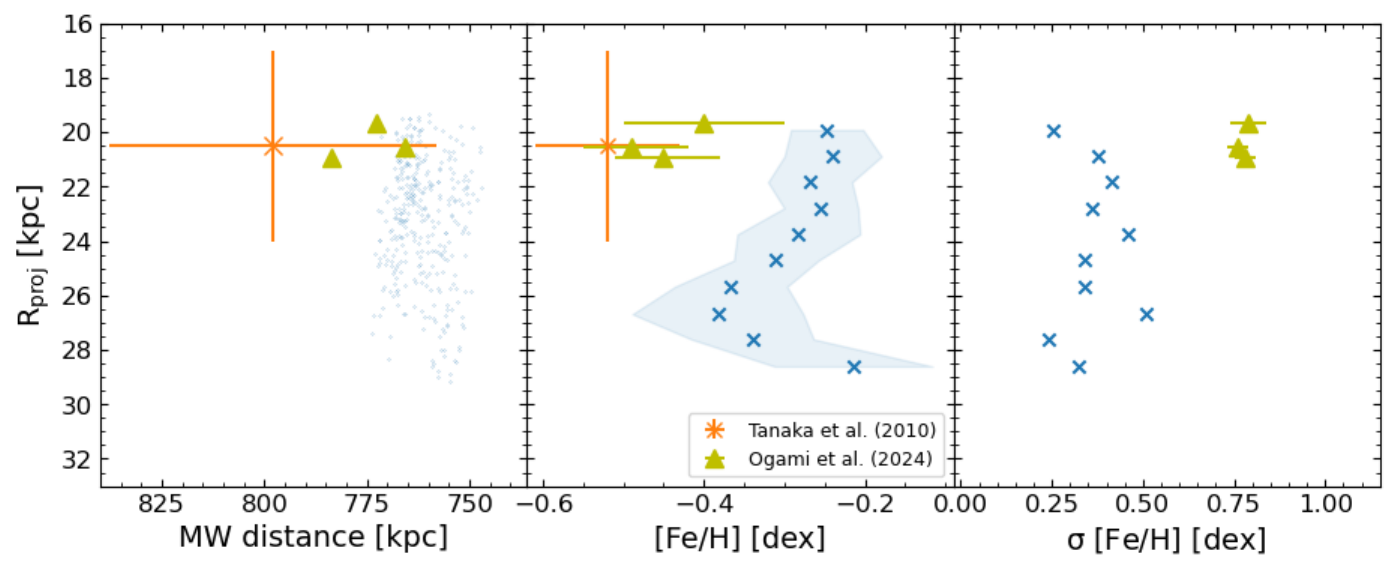}
    \caption{Same as Figure \ref{fig:ogami_ws} but for modelled stars younger than 3.5~Gyr.}
    \label{fig:ogami_ws_young}
    \end{figure}

There is a degeneracy inherent in determining the age and metallicity of a stellar population from photometric observations. A single value of the age of the stellar population has been assumed to obtain the corresponding metallicity and distance from the MW for the M31 inner halo substructures from their CMDs \citep{conn,ogami}. As shown in Figure~\ref{fig:ogami_gs}, the GSS spans a distance of $\sim$160~kpc along the LOS. Hence, determining a unique distance modulus for the GSS may bias its determined metallicity. Furthermore, a photometric survey from the entirety of stars stretching along $\sim$160~kpc LOS distance will preferentially sample the brightest resolved giants, therefore, those near the tip of the RGB. These stars, however, are in principle the younger RGBs. Therefore, photometric studies to determine stellar metallicity in the GSS may entail a possible bias towards younger RBG stars. To examine such a potential bias, we replicate Figure~\ref{fig:ogami_gs} but selecting only model stars younger than 3.5~Gyr (Figure~\ref{fig:ogami_gs_young}). We find that selecting only young ($<$3.5~Gyr old) stars in the model results in a better agreement with photometric metallicity observations in the GSS. For the NE and W shelves, selecting younger stars that are  $<$3.5~Gyr old in the model leads to average metallicity values that are slightly more metal-rich than the metallicity values determined from photometric observations for these substructures, obtained assuming an age of 12 $-$ 13 Gyr (see Figures~\ref{fig:ogami_ne_young} and \ref{fig:ogami_ws_young}).
\FloatBarrier
\FloatBarrier

\FloatBarrier
\FloatBarrier

 \end{appendix}

\end{document}